\documentclass[aps,rmp,reprint,amsmath,amssymb,graphicx,longbibliography]{revtex4-2}

\usepackage{amsmath,amssymb,amsfonts,mathtools} 

\usepackage{amsmath,amsfonts,amssymb}

\usepackage{verbatim}
\usepackage{mathrsfs}
\usepackage{hyperref}
\usepackage{graphicx}
\usepackage{bm}
\usepackage{lineno}
\usepackage{color}

\begin{document}
\title{Quantum Geometry Phenomena in Condensed Matter Systems}

\author{\footnotesize Anyuan Gao}
\email{anyuangao@sjtu.edu.cn}
\affiliation{\footnotesize Tsung-Dao Lee Institute $\And$ School of Physics and Astronomy, Shanghai Jiao Tong University, Shanghai 200240, China.}

\author{\footnotesize Naoto Nagaosa}
\email{nnagaosa.naoto@googlemail.com}
\affiliation{\footnotesize Fundamental Quantum Science Program (FQSP), TRIP Headquarters, RIKEN, Wako 351-0198, Japan.}
\affiliation{\footnotesize RIKEN Center for Emergent Matter Science (CEMS), Wako 351-0198, Japan.}

\author{\footnotesize Ni Ni}
\email{nini@physics.ucla.edu}
\affiliation{\footnotesize Department of Physics and Astronomy and California NanoSystems Institute, University of California, Los Angeles, Los Angeles, CA 90095, USA.}

\author{\footnotesize Su-Yang Xu}
\email{suyangxu@fas.harvard.edu}
\affiliation{\footnotesize Department of Chemistry and Chemical Biology, Harvard University, Massachusetts 02138, USA.}

\date{\today}

\begin{abstract}

Quantum geometry, which describes the geometry of Bloch wavefunctions in solids, has become a cornerstone of modern quantum condensed matter physics. The quantum geometrical tensor encodes this geometry through two fundamental components: the quantum metric (real part) and the Berry curvature (imaginary part). While the Berry curvature gained prominence through its manifestation in the intrinsic anomalous Hall effect, recent advances have revealed equally significant effects arising from the quantum metric. This includes its signatures in nonlinear transport, superfluid density of flat-band superconductors, and nonlinear optical responses. These advances underscore how quantum geometry is reshaping our understanding of condensed matter systems, with far-reaching implications for future technologies. In this review, we survey recent progress in the field, focusing on both foundational concepts and emergent phenomena in transport and optics—with particular emphasis on the pivotal role of the quantum metric.

\end{abstract}

\maketitle

\tableofcontents

\section{Introduction}
\label{sec:intro}
 From the discovery of the Berry phase to the emergence of topological materials and related phenomena in recent decades, quantum geometry has become a foundational framework for understanding a wide spectrum of effects in condensed matter physics. Quantum geometry is a local property described by differential geometry in a parameter space, which, in condensed matter systems, corresponds to the momentum space. Going beyond conventional band theory, the concept of quantum geometry, formalized in the 1980s, is captured by the quantum geometric tensor (QGT), which describes how the Bloch wavefunction evolves locally with respect to the change of momentum in Hilbert space \cite{Vanderbilt_2018}. Due to its quantum nature, the QGT consists of two parts: the real part, known as the quantum metric \cite{provost1980riemannian}, and the imaginary part, known as the Berry curvature \cite{RevModPhys.82.1959}.

\begin{figure}[h]
\centering
\includegraphics[width=8cm]{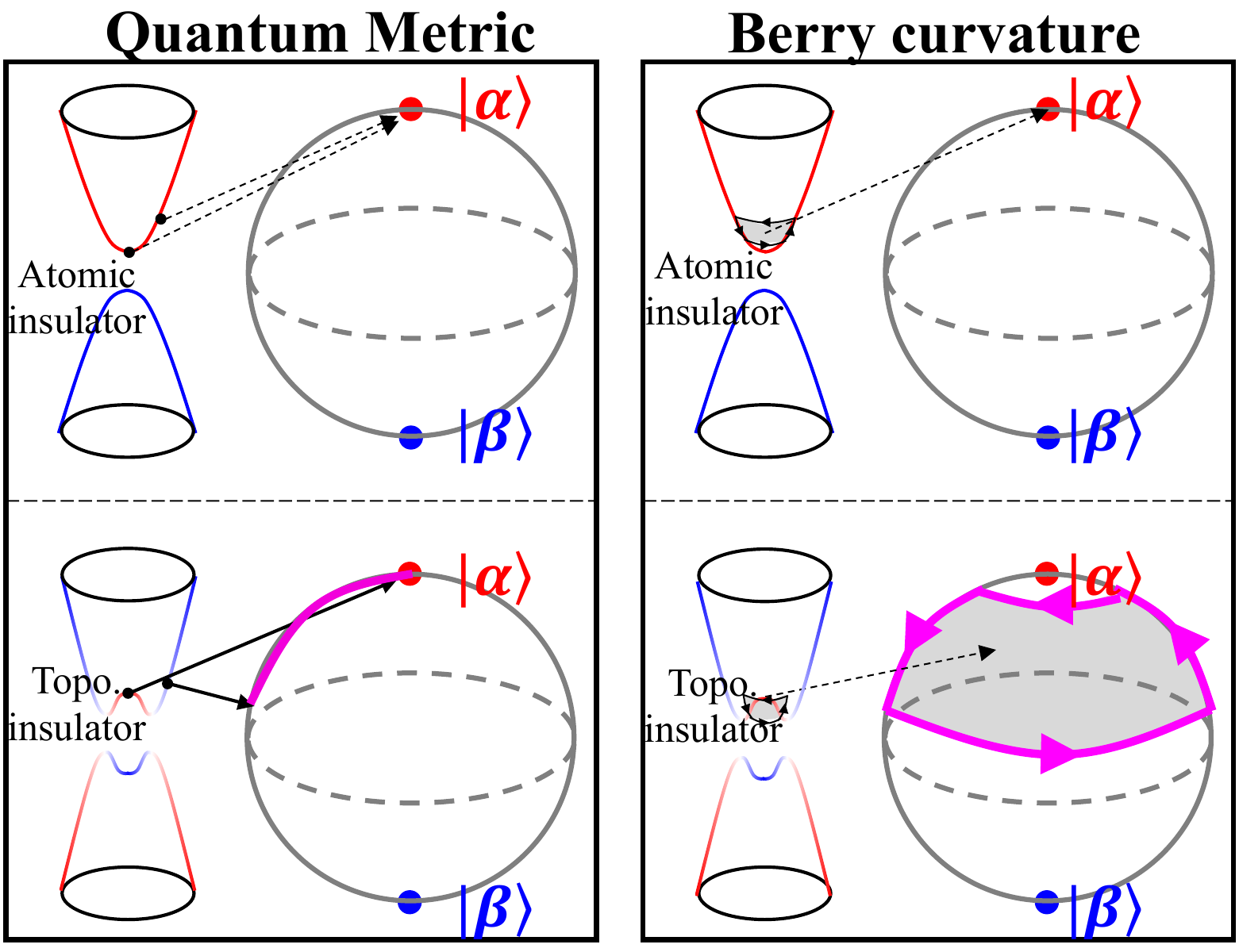}
 \caption{\small An intuitive physical picture for quantum metric and Berry curvature by mapping a 2D two-level system onto the Bloch sphere, where quantum metric characterizes the quantum distance between two nearby states on the Bloch sphere and Berry curvature corresponds to the area mapped on the Bloch sphere. }
\label{QM_BC}
\end{figure}

 The quantum metric captures the distance between nearby wavefunctions, \textit{i.e.}, how distinguishable they are. For instance, if two nearby wavefunctions are orthogonal, their overlap is zero, and the quantum metric becomes large, indicating a significant local change in the wavefunction; if two nearby wavefunctions are identical, the quantum metric is zero, reflecting no local change. Since the quantum metric is not immediately intuitive, we begin by elucidating its physics picture through a simple two-band model consisting of two orthogonal states, $\alpha$ and $\beta$. Within a unit cell, the electronic wave may occupy purely the $\alpha$ state or the $\beta$ state, or spread over both states. This means that the Bloch wavefunction at any $k$ point will be a superposition of the two orthogonal basis, $ \arrowvert u_n^{\textrm{c(v)}}\rangle=C_\alpha^{\textrm{c(v)}}\arrowvert \alpha \rangle + C_\beta^{\textrm{c(v)}}\arrowvert \beta \rangle$, which equivalently corresponds to a distinct point on the Bloch sphere, where the north and south poles are pure $\alpha$ and $\beta$ states, respectively. First, consider a trivial atomic insulator, as shown in the top left panel of Fig.~\ref{QM_BC}, where the entire conduction (valence) band is composed of the $\alpha$ ($\beta$) orbital. We can choose two adjacent momenta $k_1$ and $k_2$, and map their Bloch wavefunctions onto the Bloch sphere. The quantum metric determines the distance between them on the Bloch sphere. In this case, because of the atomic insulator nature, both Bloch wavefunctions map onto the same point of the Bloch sphere, the north pole. As such, the quantum metric is zero. Second, consider an insulator with band inversion, as shown in the bottom left panel of Fig.~\ref{QM_BC}. Because of the band inversion, the orbital composition of the band changes quickly in momentum space. We choose the same momenta $k_1$ and $k_2$ and carry out the same mapping procedure on the Bloch sphere. We clearly obtain a large distance on the Bloch sphere, indicating a nonzero quantum metric.

The quantum metric is closely related to the quantum fluctuation and uncertainty relation \cite{PhysRevLett.65.1697}, which determines the minimum size of the Wannier wavefunction \cite{PhysRevB.56.12847}. For example, in Sec. II and Sec. III, we show that both the quantum distance and the Wannier orbital size involve terms reminiscent of the energy-time uncertainty relation. The generic state, i.e., the ``wavepacket" constructed in the real space, thus feels the local nontrivial geometry such as connection,
metric, and curvature.  
In the momentum space, since a larger quantum metric implies a faster change of the wavefunction with respect to momentum, this results in a greater shift of the Wannier center in real space under momentum changes, ultimately leading to quantum-metric-driven electron motion. Since this type of electron motion is typically much weaker than the Drude response, enhancing the effect of quantum metric requires suppressing the Drude contribution. This may be achieved by using
flat band systems, or by investigating the higher-order nonlinear responses.

 The Berry curvature characterizes the local curvature (twisting or rotation of the phase) of the wavefunction. Berry curvature equips the momentum space with a local gauge structure, analogous to an effective magnetic field.  When integrated over a closed loop in momentum space, it gives rise to the Berry phase; when integrated over the whole manifold, i.e., all occupied states in the Brillouin zone, one can define the global geometry, i.e., topology, with topological invariance such as the Chern number. Using the similar setup, we can also visualize Berry curvature, as shown in the right panel of Fig.~\ref{QM_BC}. We draw a loop in $k$ space and map every Bloch wavefunction on the loop to the Bloch sphere. The Berry curvature is then the area mapped on the Bloch sphere, which is zero in a trivial insulator and nonzero in an insulator with a band inversion.

Historically, studies on the intrinsic anomalous Hall effect (AHE) \cite{nagaosa2010anomalous} in ferromagnetic metals have highlighted the significance of Berry curvature, providing the first example of how quantum geometry can profoundly influence electron motion in a solid \cite{PhysRevB.59.14915, Chen:2014PRL, Nakatsuji:2015Nature}. Specifically, the integral of Berry curvature over the occupied states or the Berry connection around the Fermi surface 
\cite{PhysRevLett.93.206602} determines the intrinsic anomalous Hall conductivity. Since then, it has been increasingly evident that the influence of quantum geometry extends far beyond the AHE in ferromagnetic metals. Berry curvature can lead to many novel Hall phenomena in quantum materials beyond ferromagnets, such as the valley Hall \cite{Xiao:2007PRL, Mak:2014Science} spin Hall \cite{Hiesch:1999PRL, kato:2004Science} and layer Hall effects \cite{Gao:2021Nature}. The Berry curvature dipole can lead to the nonlinear Hall effect \cite{sodemann2015quantum, Ma:2019Nature, Kang2019nonlinear}, the circular photogalvanic effect \cite{Ma:2017direct, Yuan:2014generation} and optical activity \cite{Xu:2020Spontaneous}. 

Although traditionally, Berry curvature has received more attention due to its connection with the Hall phenomena and its central role in topological materials, very recently, the effects of quantum metric are getting increasing interest \cite{Wang2021Intrinsic, gao2023quantum, wang2023quantum}. Theoretical proposals show that even in systems where the Berry curvature vanishes, quantum metric plays a profound role in optical absorption and nonlinear transport \cite{PhysRevLett.112.166601}, as well as the superfluid density of flat band superconductors \cite{peotta2015superfluidity, tian2023evidence}. New experimental advances include the observation of third-order nonlinear Hall effects \cite{Lai2021third}, the measurement of quantum metric tensors in high-finesse planar microcavities \cite{Gianfrate2020measurement}, and the quantum-metric-dipole-induced intrinsic second-harmonic nonlinear Hall effect \cite{gao2023quantum, wang2023quantum}. Beyond the quantum geometric tensor, i.e., Berry curvature and quantum metric, there are many quantum geometric phenomena such as the shift current in noncentrosymmetric matter due to Berry connection \cite{Baltz}, and various nonlinear optical responses formulated by Riemannian geometry \cite{Ahn2022}. These exciting developments in quantum geometry open a new era in condensed matter systems and potential technological breakthroughs in areas such as classical and quantum nonlinearity, rectification and energy harvesting, classical and quantum sensing, as well as next generation logic and memory devices.

Here we provide a comprehensive overview of the role of quantum geometry in condensed matter systems, covering both its theoretical foundations and recent experimental developments. The topology in electronic states in solids has been extensively 
studied, with numerous references and review articles available 
\cite{kane-mele2,bernevig2006quantum,hasan-kane10,qi2011topological,
chiu2016classification,bernevig2013topological}. Therefore, although we touch on 
this topics below, in this article, we rather focus on the physical consequences of the local quantum geometry. Excellent review articles on the quantum geometry have appeared recently 
\cite{yu2025quantumgeometryquantummaterials, jiang2025revealingquantumgeometrynonlinear, verma2025quantumgeometryrevisitingelectronic}, and we aim at being complementary to those. In particular, we tried to be self-contained in the theory part of this article.

The review is organized as follows. In Section II, we introduce the theoretical framework of quantum geometry in parameter space by defining the quantum geometric tensor and the quantum distance between states. We then explore how quantum distance relates to both Fisher information and entanglement entropy, followed by a discussion of various types of time evolutions, including adiabatic evolution, first-order non-adiabatic corrections, non-adiabatic dynamics treated within the Born–Oppenheimer approximation, and fully generic non-adiabatic processes, highlighting the role of quantum geometry in modifying the effective Hamiltonian. In Section III, we apply the framework of quantum geometry to momentum space, focusing on how local quantum geometry gives rise to a variety of physical phenomena. We begin by analyzing its impact on the real-space wavepacket dynamics, Wannier localization, and the semiclassical equations of motion. We then illustrate how it enters conductivity through linear response theory, appears in Landau levels and the fractional quantum Hall effect, contributes to quantum-metric-induced geometric superfluid density in flat-band systems, underlies nonlinear Hall effects driven by both Berry curvature dipole and quantum metric dipole, and plays an important role in various nonlinear optical responses. In Section IV and Section V, we review the recent experimental observations related to quantum geometry. In Section IV, we focus on the transport and optical phenomena arising from Berry curvature and Berry connection, including the intrinsic anomalous Hall effect, the quantum anomalous Hall effect and and fractional quantum Hall effect in intrinsic topological insulators and TMD Moire superlattices; the spin-Hall, valley-Hall, and layer-Hall effects; the nonlinear Hall effect induced by the Berry curvature dipole; the orbital magnetoelectric effect in A-type intrinsic magnetic topological insulators; the Edelstein effect and natural optical activity, the Berry-connection-driven shift current and circular photogalvanic effect in semiconductors, topological insulators and Weyl semimetals. In Section V, we focus on the quantum phenomena related to the quantum metric. We first provide an intuitive picture on how quantum metric can lead to an electronic motion, and then discuss the recent experimental progress in quantum-metric-related phenomena, including the flat band superconductivity, the intrinsic quantum-metric-dipole nonlinear Hall effect and the momentum-resolved photoemission spectroscopy measurement of quantum geometry. Finally, in Section VI, we conclude our review with an outlook on the challenges and future directions.

\section{Theoretical Framework of Quantum Geometry}
\label{sec:sec1}

\subsection{Definition of Quantum Geometric Tensor}
We start with the Hilbert space of wavefunction where the inner-product
$\langle u | v\rangle$ between the two wavefunctions satisfies \cite{provost1980riemannian,Stusy1905Kurzeste,doi:10.1142/0613}
\begin{align}
& \langle u | a v\rangle = a \langle u |v\rangle   \nonumber \\
& \langle u | a_1 v_1 + a_2 v_2\rangle = a_1 \langle u |v_1\rangle + a_2 \langle u | v_2\rangle  \nonumber \\
& \langle u | v\rangle = \langle v | u\rangle^*  \nonumber \\
& || u ||^2  =\langle u | u\rangle \geq 0  \nonumber \\
\end{align}
where $a_1,a_2$ are complex c-numbers and $^*$ means the complex conjugate.
On might define the distance between the two wavefunctions $|u\rangle$ and $|v \rangle$ as
$|| u - v ||^2 = \langle u - v | u - v\rangle$.
However, this does not work because the two wavefunctions which differs only by 
the phase factor should be identified as the same wavefunction. 
Mathematically this means that the quantum states become `Rays'.
In other words, the wavefunction has the gauge degrees of freedom.
Suppose $|u (\lambda)\rangle$ depends on a set of real parameters 
$\lambda = ( \lambda_1, \lambda_2, \cdot \cdot \cdot, \lambda_n)$, and 
consider the two neighboring wavefunctions 
$|u (\lambda)\rangle$ and $|u (\lambda+ d \lambda)\rangle$.
First one can define the inner product of the two up to the first-order in 
$\lambda_i$, i.e.,
\begin{align}
\langle u (\lambda) |u (\lambda+ d \lambda)\rangle = 1 + \sum_i \langle u (\lambda) | \partial_i | u (\lambda)\rangle d \lambda_i.
\end{align}
Here we used the fact that $|u(\lambda)\rangle$ is normalized, i.e., $\langle u (\lambda) |u (\lambda)\rangle = 1$,
and introduced the notation $\partial_i = \partial/\partial \lambda_i$.
By taking the derivative of the normalization condition, one obtains
\begin{align}
\langle \partial_i u | u\rangle + \langle u | \partial_i u\rangle = 2{\rm Re}[\langle u | \partial_i u\rangle] = 0
\end{align}
where we omitted $\lambda$ hereafter if no confusion occurs.
Therefore, 
$\langle u | \partial_i u\rangle$ is pure-imaginary and we define Berry connection $a_i(\lambda)$ as \cite{doi:10.1098/rspa.1984.0023,simon1983holonomy,WOS:A1987G882400002}
\begin{align}
a_i(\lambda) = i \langle u (\lambda) | \partial_i | u (\lambda)\rangle.
\end{align}
and one can write up to the linear order in $d \lambda_i$ 
\begin{align}
\langle u (\lambda) |u (\lambda+ d \lambda)\rangle = \exp[ - i \sum_i a_i(\lambda) d \lambda_i].
\end{align}
It is evident that the Berry connection is gauge dependent, and
for the gauge transformation $|u(\lambda)\rangle \to |u(\lambda)\rangle e^{i \phi(\lambda)}$, $a_i(\lambda)$ 
transforms as 
\begin{align}
a_i(\lambda) \to a_i(\lambda) - \partial_i \phi(\lambda).
\end{align}
One can define the Berry curvature tensor $F_{ij}(\lambda)$ as
\begin{align}
F_{ij}(\lambda) = \partial_i a_j(\lambda) -  \partial_j a_i(\lambda).
\end{align}
which is gauge invariant.

\begin{figure}[htb]
\begin{center}
\includegraphics[width =3.3in]{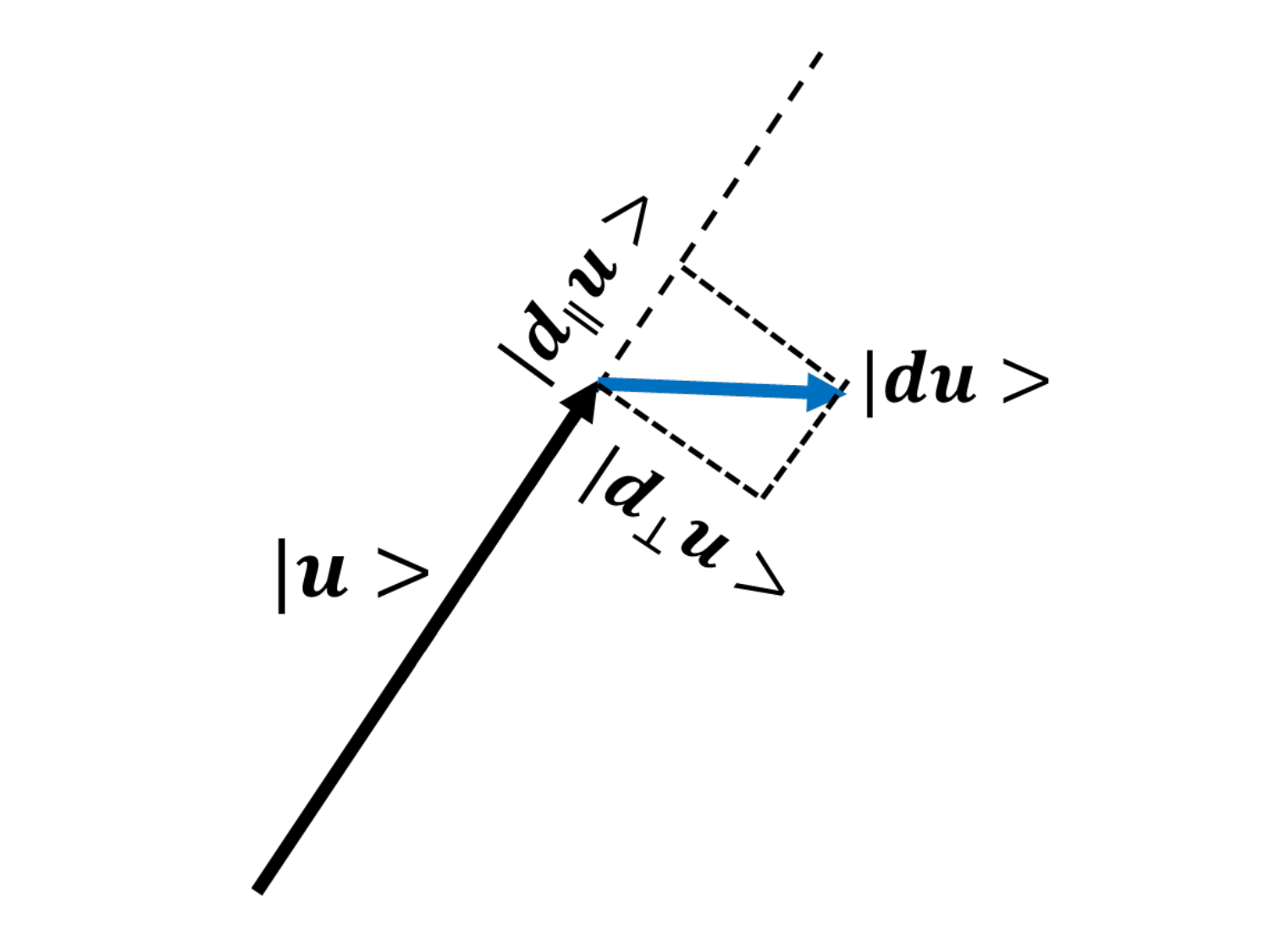} 
\caption{
{\bf Wavefunction in Hilbert space.}
The wavefunction $|u \rangle$ is defined as the
ray in the Hilbert space where the broken line corresponds to 
the $U(1)$ phase of the wavefunction.
Therefore, the small change $|d u \rangle$ of the wavefunction 
can be decomposed into the parallel $|d_\parallel u \rangle$ and
perpendicular  $|d_\perp u \rangle$ parts. Note that the 
broken line extended beyond the arrow for $| u \rangle$ 
represents the ray representation corresponding to the U(1) gauge
degree of freedom.
}
\label{fig:geometry}
\end{center}
\end{figure}

Another way to get rid of this gauge dependence is to decompose 
$| d u (\lambda)\rangle = |u (\lambda + d \lambda)\rangle -|u (\lambda)\rangle$ to
longitudinal and transverse parts as
\begin{align}
| d_\parallel u (\lambda)\rangle &=  |u (\lambda)\rangle \langle u (\lambda) | d u (\lambda)\rangle, \nonumber \\
| d_\perp u (\lambda)\rangle &= | d u (\lambda)\rangle - |u (\lambda)\rangle \langle u (\lambda) | d u (\lambda)\rangle.
\end{align}
Then the proper distance $d s$ between $|u (\lambda)\rangle$ and $|u (\lambda+ d \lambda)\rangle$ is defined as
\begin{align}
d s^2 = || d_\perp u (\lambda)\rangle ||^2 = \sum_{i,j} g_{ij}(\lambda) d \lambda_i \lambda_j ,
\end{align}
where the metric tensor $g_{ij}$ is explicitly given by
\begin{align}
g_{ij}(\lambda) = {\rm Re} [ \langle \partial_i u(\lambda) | \partial_j u(\lambda)\rangle] - a_i(\lambda)a_j(\lambda).
\end{align}

Now we define the
quantum geometric tensor \cite{doi:10.1142/0613}  (QGT) $G_{ij}(\lambda)$ as 
\begin{align}
G_{ij} (\lambda) = {\rm Tr} [ P(\lambda) \partial_i P(\lambda) \partial_j P(\lambda) ]
\end{align}
where $P(\lambda)= |u(\lambda)\rangle\langle u(\lambda)|$ is the projection operator.
One can easily confirm 
\begin{align}
G_{ij} (\lambda) = \langle\partial_i u(\lambda) |\partial_j u(\lambda)\rangle - a_i(\lambda)a_j(\lambda), 
\end{align}
and 
\begin{align}
{\rm Re}G_{ij} (\lambda) &= g_{ij}(\lambda) = g_{ji}(\lambda), \nonumber \\
{\rm Im}G_{ij} (\lambda) &= - \frac{1}{2} F_{ij}(\lambda) = \frac{1}{2} F_{ji}(\lambda)  . 
\end{align}
Namely, 
\begin{align}
G_{ij} (\lambda) = g_{ij}(\lambda) - \frac{i}{2} F_{ij}(\lambda). 
\end{align}

The QGT has some properties. First, it is a hermitian matrix.
This originates from the fact that $P(\lambda)^\dagger = P(\lambda)$ as
\begin{align}
G_{ij} (\lambda)^* & = {\rm Tr} [ \partial_j P(\lambda)^\dagger  \partial_i P(\lambda)^\dagger P(\lambda)^\dagger] \nonumber \\
&=  {\rm Tr} [ \partial_j P(\lambda) \partial_i P(\lambda) P(\lambda)] \nonumber \\
&=  {\rm Tr} [P(\lambda) \partial_j P(\lambda) \partial_i P(\lambda) ] \nonumber \\
&= G_{ji} = [G^\dagger]_{ij}^*.
\end{align}
Second, it does not have negative eigenvalues. To prove it,
let us consider the quadratic form by introducing the 
complex numbers $\eta_i$
and $T(\lambda,\eta) = \sum_i \eta_i \partial_i P(\lambda)$,
\begin{align}
I(\eta) = \sum_{i,j} \eta_i^* G_{ij}(\lambda) \eta_j  
= {\rm Tr} [ P(\lambda) T^\dagger(\lambda,\eta) T(\lambda,\eta) )]
\label{eq:positive}
\end{align}
cannot be negative. Therefore, the eigenvalue of $G_{ij}(\lambda)$ 
is real and non-negative.

There are other quantities related to the QGT defined here.
One is 
\begin{align}
B_{ij} (\lambda) & = {\rm Tr} [ \partial_i P(\lambda) \partial_j P(\lambda)] \nonumber \\
&=  G_{ij}(\lambda) + G_{ji}(\lambda).
\end{align}
The other is 
\begin{align}
\tilde{G}_{ij} (\lambda) & = {\rm Tr} [ \partial_i P(\lambda) (1 - P(\lambda) )\partial_j P(\lambda)] \nonumber \\
&=  B_{ij}(\lambda) - G_{ji}(\lambda) = G_{ij}(\lambda).
\end{align}

The distance between the two generic states 
$|u\rangle$ and $|v\rangle$ can be defined
\begin{align}
s^2 = 1 - |\langle u|v \rangle|^2
\end{align}
which is evidently gauge invariant.
For $|u\rangle = |u(\lambda)\rangle$ and $|v\rangle = |u(\lambda + d\lambda)\rangle$,
we need to expand $|v\rangle$ up to the second order in $d\lambda$ as
\begin{align}
\langle u(\lambda) |u(\lambda + d\lambda)\rangle &= 1 + \sum_i \langle u(\lambda) |\partial_i |u(\lambda)\rangle d\lambda_i  \nonumber \\
&+ \frac{1}{2} \sum_{i,j} \langle u(\lambda) |\partial_i \partial_j |u(\lambda)\rangle d\lambda_i d\lambda_j 
\end{align}

We need the first order term in $d\lambda_i$ for the imaginary part, while
the second order term in the real part.
\begin{align}
{\rm Im}\langle u(\lambda |u(\lambda + d\lambda)\rangle = - \sum_i a_i(\lambda) d \lambda_i
\end{align}
and 
\begin{align}
& {\rm Re}\langle u(\lambda) |u(\lambda + d\lambda)\rangle   \nonumber \\
&= 1 - 
\frac{1}{2} \sum_{i,j} {\rm Re}\langle\partial_i u(\lambda) | \partial_j u(\lambda)\rangle d\lambda_i d\lambda_j
\end{align}
Here we have used the equation $\partial_i \partial_j \langle u(\lambda) |u(\lambda)\rangle = 0$.

Therefore, up to second order in $d \lambda_i$, we obtain 
\begin{align}
& |\langle u(\lambda) |u(\lambda + d\lambda)\rangle|^2  \nonumber \\
&= 1 - 
\sum_{i,j} {\rm Re}\langle\partial_i u(\lambda) | \partial_j u(\lambda)\rangle d\lambda_i d\lambda_j \nonumber \\
&  + \sum_{i,j} a_i(\lambda) a_j(\lambda)d\lambda_i d\lambda_j
\end{align}
and we obtain $d s^2 = \sum_{i,j} g_{ij} (\lambda) d\lambda_i d\lambda_j$.
The distance between the general two states $| u_1 \rangle$ and $| u_2 \rangle$
is defined as
\begin{align}
    s_{12}^2 = \frac{1}{2} Tr[( \hat{P}_1 - \hat{P}_2)^2]
\end{align}
where $\hat{P}_i = | u_i \rangle \langle u_i |$ is the projection 
operator to $| u_1 \rangle$.
Evidently this is gauge invariant, and explicitly given by
\begin{align}
    s_{12}^2 = 1 - |\langle u_1 | u_2 \rangle |^2.
\end{align}
For the more general case, the projectors are replaced by
the density matrices to define the distance between the 
two states, which will be relevant to the discussion on 
the quantum Fisher information in the next subsection. 

Non-Abelian generalization of the QGT is given by the following. 
Let us define $|u \rangle = ( |u_1 \rangle, |u_2 \rangle \cdots |u_N \rangle )$.
Then the QGT ($N \times N$ matrix) is defined by $G_{ij}(\lambda)$ 
\begin{align}
G_{ij}(\lambda) = \langle \partial_i u | [ 1 - P ] | \partial_j u \rangle,
\end{align}
or its matrix element ($n,m=1,2, \cdots N)$
\begin{align}
G_{ij}^{nm}(\lambda) = \langle \partial_i u_n | [ 1 - P ] | \partial_j u_m \rangle,
\end{align}
where $P= \sum_l |u_l \rangle \langle u_l |$ is the projection operator.
The quantum metric tensor and the non-Abelian Berry curvature tensor is 
given by \cite{PhysRevB.81.245129}
\begin{align}
g_{ij}(\lambda) = \frac{1}{2} ( G_{ij} + G^\dagger_{ij} )
\end{align}
and 
\begin{align}
F_{ij}(\lambda) = i ( G_{ij} - G^\dagger_{ij} ).
\end{align}
Note here that the meaning of $\dagger$ is to take the Hermitian conjugate 
for the indices $n,m$ and not $i,j$.
The non-Abelian Berry connection ($N \times N$ matrix) is
defined by 
\begin{align}
a_{i}^{nm}(\lambda) = i \langle u_n | \partial_j u_m \rangle,
\end{align}
and the corresponding Berry curvature defined above is 
\begin{align}
F_{ij}(\lambda) = \partial_i a_j - \partial_j a_i - i [ a_i, a_j].
\end{align}

For non-interacting fermion systems,
the ground state is given by 
$(\Pi_{n: {\rm occupied}}\hat{a}_n^\dagger) | {\rm vacuum \rangle}$,
and the QGT are defined by the projector $\hat{P}$ given by
\begin{align}
\hat{P}(\lambda) = \sum_{ n: {\rm occupied}} | u_n \rangle
\langle u_n |. 
\end{align}

Here some remarks are in order. 
First, the wavefunction $u(\lambda)$ depends on the parameter 
set $\lambda$ which can be characterized as the eigenstate of the Hamiltonian 
$H(\lambda)$ which depends on the parameter
set $\lambda$, i.e., 
\begin{align}
     H(\lambda)|u_k(\lambda)\rangle=E_k(\lambda)|u_k(\lambda)\rangle.
     \label{eq:eigen}
\end{align}
As we will discuss later in section II.C, the state
stays on the same $k$ eigenstate as long as the time dependence of 
$\lambda(t)$ is much slower than the $\hbar/\Delta E$ where $\Delta E$
is the energy gap between the eigenvalues. This is the situation considered
by Berry in his original paper on Berry phase \cite{doi:10.1098/rspa.1984.0023}. 
(Note that here the no degeneracy of the eigenvalues is assumed.)
In this adiabatic limit, the quantum geometric tensor is 
given in terms of Feynman-Hellman theorem as follows.
By taking the derivative of eq.(\ref{eq:eigen}) with respect to $\lambda_i$,
\begin{align}
    &[\partial_i H(\lambda)]|u_k(\lambda)\rangle + H(\lambda) |\partial_i u_k(\lambda)\rangle
    \nonumber \\
    &= [\partial_i E_k(\lambda)]|u_k(\lambda)\rangle 
    + E_k(\lambda) |\partial_i u_k(\lambda)\rangle 
\end{align}
Taking the inner product with 
$\langle u_k(\lambda) |$, we obtain
\begin{align}
\langle u_k(\lambda) |\partial_i H(\lambda) |u_k(\lambda)\rangle = \partial_i E_k(\lambda)
\end{align}
while with 
$\langle u_l(\lambda) |$ $(l \ne k)$,
$\langle u_k(\lambda) |$, we obtain
\begin{align}
\langle u_l (\lambda) |\partial_i u_k(\lambda) \rangle = 
\frac{\langle u_l(\lambda) | \partial_i H(\lambda) |u_k(\lambda)\rangle }{E_k(\lambda)-E_l(\lambda) }
\label{eq:FH2}
\end{align}
In this representation, the QGT can be written as
\begin{align}
    G_{k,ij}(\lambda) &= \langle \partial_i u_k | ( 1 - |u_k \rangle \langle u_k| ) |\partial_j u_k \rangle \nonumber \\
    &= \sum_{m(\ne k)} \langle \partial_i u_k | u_m \rangle \langle u_m |\partial_j u_k \rangle \nonumber \\
     &= \sum_{m(\ne k)} \frac{\langle u_k| \partial_i H |u_m\rangle \langle u_m| \partial_j H |u_k\rangle }{(E_k-E_m)^2 }.  
     \label{eq:Gkij}
\end{align}

Now let us take a look at an example. We consider the two-state model whose Hamiltonian is given by
\begin{align}
H = h_0(\lambda) + \vec{h}(\lambda) \cdot \vec{\sigma}
\end{align}
where $\vec{\sigma} = (\sigma_1, \sigma_2, \sigma_3)$ are Pauli matrices 
which satisfy the anti-commutation relation $ \{ \sigma_a, \sigma_b \} = 2 \delta_{a,b}$ and 
commutation relation $[ \sigma_a, \sigma_b] = 2i \varepsilon_{abc} \sigma_c$
where $\varepsilon_{abc}$ is the totally antisymmetric tensor.
Defining the unit vector $\vec{n}(\lambda) =  \vec{h}(\lambda)/|\vec{h}(\lambda)|$,
the projection operators  $P_{\pm} (\lambda)$ to the eigenstates with the 
eigenenergy $E_\pm(\lambda) = h_0 \pm |\vec{h}|$ is given by
\begin{align}
P_{\pm}(\lambda) = \frac{1}{2} ( 1 \pm \vec{n}(\lambda) \cdot \vec{\sigma} )
\end{align}
which satisfy
\begin{align}
HP_{\pm}(\lambda) = E_\pm(\lambda) P_{\pm}(\lambda). 
\end{align}
Therefore, one can calculate $G^\pm_{ij}(\lambda)$ by using the identities
${\rm Tr}[\sigma_a \sigma_b] = 2 \delta_{ab}$ and 
${\rm Tr}[\sigma_a \sigma_b \sigma_c] = 2i \varepsilon_{abc}$ as
\begin{align}
G^\pm_{ij} = \frac{1}{4} \partial_i \vec{n} \cdot \partial_j \vec{n} \pm \frac{i}{4} \vec{n} \cdot (\partial_i \vec{n} \times \partial_j \vec{n}). 
\end{align}

When we introduce the polar coordinates $\theta$ and $\varphi$ as
\begin{align}
   \vec{n} = ( \sin \theta \cos \varphi, \sin \theta \sin \varphi, 
   \cos \theta ), 
\end{align}
the spinor wavefunction 
\begin{align}
  | \vec{n} \rangle =  
  e^{i \chi} ( \cos(\theta/2), e^{i \varphi}\sin(\theta/2) ) 
\end{align}
satisfies
\begin{align}
   \vec{n} \cdot \vec{\sigma} | \vec{n} \rangle = | \vec{n} \rangle.
\end{align}
So the distance $s_{12}$ between 
$| \vec{n}_1 \rangle$ and $| \vec{n}_2 \rangle $ can be shown to be
\begin{align}
   s_{12}^2 = 1 -   |\langle \vec{n}_1 | \vec{n}_2 \rangle|^2 =  \frac{1}{2} ( 1 - \vec{n}_1 \cdot \vec{n}_2 )
\end{align}

\subsection{Information and Quantum Geometry}

The quantum geometry is closely related to quantum information. 
First, we show that the quantum metric tensor is closely related
to the quantum Fisher information. 
To elucidate it, let us start with the classical Fisher information
\cite{doi:10.1142/9789814338745_0015}.
Suppose the probability distribution $p_m(\lambda)$ depends on 
a real parameter $\lambda$,
the Fisher information $f_\lambda$
is defined by 
\begin{align}
    f_\lambda = \sum_m \frac{(\partial_\lambda p_m( \lambda))^2}{p_m(\lambda)}
    = - \sum_m p_m( \lambda) \partial_\lambda^2 \log p_m( \lambda), 
\end{align}
where we used the fact $\partial_\lambda^2 \sum_m p_m(\lambda) = \partial_\lambda^2 1 = 0$.
Define $\langle A\rangle_\lambda = \sum_m p_m(\lambda) A_m$ as the expectation value of 
$A$, and $\langle(\Delta A)^2\rangle_\lambda = \langle A^2\rangle_\lambda 
- (\langle A\rangle_\lambda)^2$.
Also defining $L_m(\lambda) = \partial_\lambda \log p_m(\lambda)$, we can show 
$\langle L_\lambda \rangle_\lambda=0$, $\langle (\Delta L_\lambda)^2 \rangle_\lambda= 
\langle L_\lambda^2 \rangle_\lambda = f_\lambda$.
Therefore, using the Cauchy-Schwartz inequality, one obtains
\begin{align}
    \langle(\Delta A)^2\rangle_\lambda \cdot \langle (L_\lambda)^2\rangle \geq ( \langle \Delta A \cdot L_\lambda \rangle_\lambda)^2.
\end{align}
Lastly, one can show
\begin{align}
    \langle(\Delta A) L_\lambda \rangle_\lambda  &= \sum_m ( A_m - \langle A \rangle_\lambda ) \partial_\lambda p_m(\lambda) \nonumber \\
    &= \partial_\lambda \sum_m A_m p_m(\lambda) = \partial_\lambda \langle A \rangle_\lambda    
\end{align}
and we finally reach the Cramer-Rao inequality
\begin{align}
    \langle(\Delta A)^2\rangle_\lambda f_\lambda \geq (\partial_\lambda \langle A\rangle_\lambda)^2.
\end{align}
which gives a very useful bound for the statistical uncertainty for the 
estimation of $A$.

When the parameter set $\lambda_i$($i = 1,2, \cdot \cdot \cdot ,n$) is considered,
the Fisher information becomes a matrix defined by 
\begin{align}
f_{ij}(\lambda) = \sum_m \frac{ \partial_i p_m \partial_j p_m}{p_m}.
\end{align}

Now we relate this Fisher information to quantum mechanics
\cite{PhysRevLett.72.3439}. 
Let the wavefunction $|u\rangle$ be the linear combination of 
some base $|m\rangle$ as
\begin{align}
|u\rangle = \sum_m c_m |m\rangle = \sum_m \sqrt{p_m} e^{i\phi_m} |m\rangle
\end{align}
where $p_m = |c_m|^2$ is the probability to find the state $|m\rangle$.
Consider another state $|\tilde{u}\rangle$ which is the neighbor of 
$|u\rangle$ and is characterized by $p_m + d p_m$ and $\phi + d \phi$, then, let us calculate the difference between  $|u\rangle$  and  $|\tilde{u}\rangle$.
First we have 
\begin{align}
\langle \tilde{u} |u\rangle = \sum_m \sqrt{(p_m + d p_m)p_m} e^{- i d\phi_m} |m \rangle
\end{align}
and expand it to second order in $d p_m$ and $d \phi_m$ to obtain 
\begin{align}
|\langle \tilde{u} |u \rangle|^2 &= 1 - \frac{1}{4} \sum_m \frac{d p_m^2}{p_m}  
\nonumber \\
&- \sum_m p_m d \phi_m^2 + ( \sum_m p_m d \phi_m )^2.
\label{eq:overlap}
\end{align} 
One can show that 
$p_m ( d \phi_m - \sum_k p_k d \phi_k) = {\rm Im}(\langle u|m\rangle \langle m | d_\perp u\rangle$
where $|d_\perp u\rangle = |d u\rangle - |u\rangle \langle u | d u\rangle$.
$ {\rm Im}(\langle u|m\rangle \langle m | d_\perp u\rangle )$ vanishes, which can be seen if one chooses
the base $|m \rangle$ such that $|m_1\rangle = |u\rangle$ and $|m_2\rangle \propto  |d_\perp u\rangle$.
(Note that $\langle u |d_\perp u\rangle =0$.)
Therefore, the second line of eq.(\ref{eq:overlap}) vanishes,
and we obtain the distance as 
\begin{align}
1 - |\langle \tilde{u} |u\rangle|^2 = \frac{1}{4} \sum_m \frac{d p_m^2}{p_m} = d s^2  
\label{eq:distance}
\end{align} 
Because $p_m$ is the probability to find the state at $|m\rangle$,
the r.h.s. of eq.(\ref{eq:distance}) has the meaning of 
Fisher information, i.e.,
\begin{align}
\sum_m \frac{d p_m^2}{p_m} = \sum_{i,j} \sum_m \frac{\partial_i p_m \partial_j p_m}{p_m} d \lambda_i d \lambda_j = \sum_{i,j} f_{ij}(\lambda) d \lambda_i d \lambda_j  
\label{eq:distance}
\end{align} 
and we obtain the relation $g_{ij}(\lambda)= f_{ij}(\lambda)/4$, i.e.,
the metric tensor is nothing but (1/4 of) Fisher information matrix.

Now we define the quantum Fisher information for the general density 
matrix $\hat{\rho}(\lambda)$ characterized by the parameter $\lambda$. 
For that purpose, we introduce a complete 
set of non-negative Hermitian operator $\hat{E}(\xi)$ 
satisfying
\begin{align}
    \int d \xi \hat{E}(\xi) = \hat{1},
\end{align}
which describes the measurement. 
Then the probability density for measurement result $\xi$ for 
the parameter $\lambda$ is 
\begin{align}
    p(\xi | \lambda) = {\rm tr}[\hat{E}(\xi) \hat{\rho}(\lambda)]
\end{align}
and we are interested in $\partial_\lambda \ln p(\xi | \lambda)$.
In the operator level, we want to know the operator $\hat{L}(\lambda)$
which satisfies
\begin{align}
    \partial_\lambda \hat{\rho}(\lambda) = \hat{L}(\lambda) \hat{\rho}(\lambda),
    \label{eq:logderi}
\end{align}
which roughly corresponds to 
$\hat{L}(\lambda) =\partial_\lambda \ln \hat{\rho}(\lambda)$.
However, $\hat{L}$ and $\hat{\rho}$ are not commuting in general,
and in that case the right hand side of eq.(\ref{eq:logderi}) 
should be replaced by 
\begin{align}
    \mathcal{R}_{\hat{\rho}} (\hat{L}) 
    = \frac{1}{2} (\hat{L} \hat{\rho} + \hat{\rho} \hat{L} ),
    \label{eq:logderi2}
\end{align}
Introducing the basis $|n \rangle$ which diagonalizes the density matrix $\hat{\rho}$,
one can write in general 
\begin{align} 
    \hat{\rho} = \sum_n p_n |n \rangle \langle n |
\end{align}
with the probability $p_n$ for the pure state $| n \rangle$.
Then the ``super operator"  $\mathcal{R}_{\hat{\rho}} (\hat{L})$, 
which transforms the operator to
another operator, is explicitly written as
\begin{align}
      \mathcal{R}_{\hat{\rho}} (\hat{L}) 
    = \frac{1}{2} \sum_{n,m}
     (p_n + p_m) (\hat{L})_{n m} | n \rangle \langle m |.    
     \label{eq:logderi3}
\end{align}
Therefore, the inverse super operator is given by
\begin{align}
      \mathcal{R}^{-1}_{\hat{\rho}} (\hat{L}) 
    = \frac{2 (\hat{L})_{n m}}{p_n + p_m}  | n \rangle \langle m |.    \label{eq:logderi4}
\end{align}
Using this  $\mathcal{R}^{-1}_{\hat{\rho}}$, one can show a useful 
identity for the general Hermitian operators $\hat{A}$, $\hat{B}$:
\begin{align}
     {\rm tr}(\hat{A} \hat{B}) = 
     {\rm Re}[ {\rm tr}( \hat{\rho} \hat{A} \mathcal{R}^{-1}_{\hat{\rho}} (\hat{B} ) ]
     \label{eq:ident}
\end{align}
This can be proven by using the expression
\begin{align}
      {\rm tr}( \hat{\rho} \hat{A} \mathcal{R}^{-1}_{\hat{\rho}} (\hat{B} )) 
      = \sum_{n,m} \frac{ p_n A_{nm} B_{mn} + p_m A_{mn} B_{nm} }{p_n + p_m}. 
     \label{eq:rhs}
\end{align}

Now the quantum Fisher information $F(\lambda)$ is given by
the following expression with 
$\hat{\rho}' = \partial \hat{\rho}/\partial \lambda$:
\begin{align}
      F(\lambda) = \int d \xi 
      \frac{[ {\rm tr}( \hat{E}(\xi) \hat{\rho}'(\lambda) )]^2}{{\rm tr}[ \hat{E}(\xi)  \hat{\rho}(\lambda) ]}  
      \label{eq:qfisher}
\end{align}

Let us proceed to evaluate the integrand in eq.(\ref{eq:qfisher})
as
\begin{align}
     \bigg| \frac{[ {\rm tr}( \hat{E} \hat{\rho}')]^2}{{\rm tr}[\hat{E}\hat{\rho}]} \bigg|^2  &\leq
      \bigg| {\rm tr} \biggl( \frac{\hat{\rho}^{1/2} \hat{E}^{1/2}}{\sqrt{ 
      {\rm tr}(\hat{E} \hat{\rho}) } } \cdot
      \hat{E}^{1/2} \hat{\rho}^{1/2} \mathcal{R}^{-1}_{\hat{\rho}} 
      (\hat{\rho}') \hat{\rho}^{1/2} \biggr) \bigg|^2  \nonumber \\
      &\leq
      {\rm tr}\biggl[  \frac{\hat{\rho}^{1/2} \hat{E}^{1/2}}{\sqrt{ 
      {\rm tr}(\hat{E} \hat{\rho}) } } \cdot
      \frac{\hat{E}^{1/2} \hat{\rho}^{1/2}}{\sqrt{ 
      {\rm tr}(\hat{E} \hat{\rho}) } }  \biggr] \nonumber \\
      &\times 
      {\rm tr}[ \hat{E}^{1/2} \hat{\rho}^{1/2} \mathcal{R}^{-1}_{\hat{\rho}} 
      (\hat{\rho}') \hat{\rho}^{1/2} \cdot \hat{\rho}^{1/2}  \mathcal{R}^{-1}_{\hat{\rho}} 
      (\hat{\rho}') \hat{E}^{1/2}     ]  \nonumber \\
      &= \frac{{\rm tr}[ \hat{E} \hat{\rho} ]}{{\rm tr}[ \hat{E} \hat{\rho} ]}
         \cdot {\rm tr}[ \hat{E} \mathcal{R}^{-1}_{\hat{\rho}}(\hat{\rho}')
          \hat{\rho} \mathcal{R}^{-1}_{\hat{\rho}}(\hat{\rho}' ]
      \label{eq:qfisher2}
\end{align}
Note that we have used the Schwarz inequality
$|{\rm tr}(\hat{A}^\dagger \hat{B} )|^2 \leq 
{\rm tr}(\hat{A}^\dagger \hat{A} ) {\rm tr}(\hat{B}^\dagger \hat{B} )
$.
Inserting this inequality to eq.(\ref{eq:qfisher}), we
obtain 
\begin{align}
      F(\lambda) &\leq \int d \xi {\rm tr}[ \hat{E}(\xi) \mathcal{R}^{-1}_{\hat{\rho}}(\hat{\rho}')
          \hat{\rho} \mathcal{R}^{-1}_{\hat{\rho}}(\hat{\rho}') ]
          \nonumber \\
        &=   {\rm tr}[\mathcal{R}^{-1}_{\hat{\rho}}(\hat{\rho}')
          \hat{\rho} \mathcal{R}^{-1}_{\hat{\rho}}(\hat{\rho}') ]
      \label{eq:qfisher3}
\end{align}

 By using eq.(\ref{eq:ident}) with $\hat{A} = \mathcal{R}^{-1}_{\hat{\rho}}(\hat{\rho}')  $ and $\hat{B} = \hat{\rho}'$, and that
  ${\rm tr}[\mathcal{R}^{-1}_{\hat{\rho}}(\hat{\rho}')
          \hat{\rho} \mathcal{R}^{-1}_{\hat{\rho}}(\hat{\rho}')$
  is real,
  one obtains
\begin{align}
       {\rm tr}[\mathcal{R}^{-1}_{\hat{\rho}}(\hat{\rho}')
          \hat{\rho} \mathcal{R}^{-1}_{\hat{\rho}}(\hat{\rho}') ]
        = {\rm tr}[ \hat{\rho}' \mathcal{R}^{-1}_{\hat{\rho}}(\hat{\rho}') ]
      \label{eq:qfisher4}
\end{align}

Let us here analyze the content of $\hat{\rho}'$.
\begin{align}
      \hat{\rho}  + d \lambda \hat{\rho}' &= \sum_n ( p_n + d p_n) 
      | n' \rangle \langle n' | \nonumber \\
       &= \sum_n d p_n | n \rangle \langle n| + 
          e^{i d \lambda \hat{h}}  \hat{\rho} e^{- i d \lambda \hat{h}}      
      \label{eq:qfisher5}
\end{align}
where the change of the eigenstate $|n \rangle \to |n' \rangle$
is driven by the operator $\hat{h}$ as 
$|n' \rangle = e^{i d \lambda \hat{h}} |n \rangle  $.

Therefore, one obtains 
\begin{align}
       \mathcal{R}^{-1}_{\hat{\rho}}(\hat{\rho}') 
        = \sum_{n,m} \frac{2}{p_n + p_m} 
         \biggl( \delta_{nm} \frac{d p_n}{d \lambda}   + i (p_n - p_m) h_{nm}
          \biggr) | n \rangle \langle m| 
      \label{eq:qfisher6}
\end{align}
and 
\begin{align}
       &{\rm tr}[ \hat{\rho}' \mathcal{R}^{-1}_{\hat{\rho}}(\hat{\rho}') ]
       \nonumber \\
       &= \sum_n \frac{1}{p_n} \biggl(\frac{d p_n}{d \lambda} \biggr)^2
       + 2 \sum_{n,m} \frac{(p_n - p_m)^2}{p_n + p_m} {|h_{nm}|^2} 
       \label{eq:qfisher7}
\end{align}
which is the upper bound of quantum Fisher information $F_Q(\lambda)$.

It is remarkable that $F_Q$ can be measured experimentally
\cite{Hauke2016,Yu2022}. 
Let us consider the case that the probability $p_n$ does not change
and is given by the equilibrium distribution $p_n = e^{- \beta E_n}/Z$
where $\beta= 1/(k_B T)$ is the inverse of the temperature and
$Z$ is the partition function.
Therefore, we have the expression 
\begin{align}
      F_Q(\lambda) &= 2 \sum_{n,m} \frac{(p_n - p_m)^2}{p_n + p_m} |h_{nm}|^2
      \nonumber \\
       &= 2 \sum_{n,m} \tanh{\frac{\beta(E_m - E_n)}{2}} (p_n - p_m) |h_{nm}|^2. 
       \label{eq:linearres}
\end{align}
It can be related to the linear response function 
of the observable $\hat{h}$ to the field conjugate to $\hat{h}$, which is 
given by
\begin{align}
      \chi(\omega, T) = \frac{i}{\hbar} \int_0^{\infty} d t 
      e^{ i \omega t} {\rm tr}( \hat{\rho} [ \hat{h}(t), \hat{h} ] )
             \label{eq:linearres2}
\end{align}
In terms of the energy eigenstate $| n \rangle$, one obtains
\begin{align}
      \chi(\omega, T) = \sum_{n,m} (p_m - p_n) 
      \frac{| h_{nm}|^2}{ \omega + i \delta - E_m + E_n}              \label{eq:linearres3}
\end{align}
and its imaginary part $\chi''(\omega, T)$ as
\begin{align}
      \chi''(\omega, T) = \pi \sum_{n,m} (p_m - p_n) 
      | h_{nm}|^2 \delta(\omega + - E_m + E_n)
      \label{eq:linearres4}
\end{align}

Using eqs.(\ref{eq:linearres}) and (\ref{eq:linearres4}),
one obtains the useful relation 
\begin{align}
      F_Q(T) = \frac{4}{\pi} \int_0^{\infty} d \omega  
      \chi''(\omega, T)  \tanh{\frac{\beta \omega}{2}} 
             \label{eq:linearres5}
\end{align}

For a quantum spin system, the Fisher information gives a measure on how many
spins are entangled, which provides a useful clue to its ground state
\cite{PhysRevA.85.022321,PhysRevLett.102.100401}.
Recently, the experimental detection of quantum Fisher information
in terms of Resonant Inelastic X-ray Scattering (RIXS) has been reported 
for Ba$_3$CeIr$_2$O$_9$ where the entanglement of the 
electronic orbitals between neighboring Ir sites is evaluated.
\cite{ren2024witnessingquantumentanglementusing}. 

Quantum entanglement is one of the most important concepts in 
quantum physics \cite{RevModPhys.80.517}. 
The simplest example is the spin singlet,
where the wavefunction is given by
\begin{align}
    | \Psi_0 \rangle = \frac{1}{2} 
    ( | \uparrow \rangle_1 | \downarrow \rangle_2 
     -  | \downarrow \rangle_1 | \uparrow \rangle_2 ).
\end{align}
This state is distinct from the product state
\begin{align}
    | \Psi_1 \rangle = | \uparrow \rangle_1 | \uparrow \rangle_2.
\end{align}
To distinguish between the two, one can define the
entanglement entropy as follows.
Taking the trace over the system 2, one can obtain the
reduced density matrix for system 1 as
\begin{align}
   \rho_1 = {\rm Tr}_2 [ \rho_{1+2}] 
\end{align}
where $\rho_{1+2} = | \Psi \rangle \langle \Psi |$
is the density matrix for the total system (pure state).
The entanglement entropy $S_{\rm ent.}$ is defined by
\begin{align}
   S_{\rm ent.} = - {\rm Tr}_1 [ \rho_1 \ln \rho_1 ] 
\end{align}
For $|\Psi_0 \rangle$, $\rho_1 = (| \uparrow \rangle_1  \langle_1 \uparrow |
+| \downarrow \rangle_1  \langle_1 \downarrow |)/2$ and
$ S_{\rm ent.} = \ln 2$, while for $|\Psi_0 \rangle$,
 $\rho_1 = |\uparrow \rangle_1  \langle_1 \uparrow |$ and
$ S_{\rm ent.} = 0$.

For more general situation, we divide the total system into 
$A$ and $\bar{A}$. 
For a non-interacting fermion system on
$N$-sites described by \cite{Peschel_2003}
\begin{align}
   H_{\rm tot.} = - \sum_{n,m=1}^N t_{nm} c^\dagger_n c_m, 
\end{align}
where $c^\dagger_n$ ($c_n$) is the creation (annihilation) operator
of an electron on site $n$.
Suppose that $N_e$ electrons occupy the energy eigenstates
$\psi_k (i)$ ($k=1,2, \cdots, N_e)$ of $ H_{\rm tot.}$.
The ground state of this system is completely characterized by
the correlation 
$C_{n,m} = \langle c^\dagger_n c_m \rangle = \sum_{k=1}^{N_e} \psi^*_k (n) \psi_k(m)$.
Note that $C_{n,m}^* = C_{m,n}$, i.e., the $N \times N$ matrix $C$ is hermitian.
Let $i$, $j$ be the two sites in the subsystem $A$ with dimension $M ( < N) $, and the
$M \times M$ matrix $C^A_{i,j} = C_{m=i, n=j}$ is the subset of $C$.
Starting from $\rho_{\rm total} =| \Psi \rangle \langle \Psi |$
with $| \Psi \rangle$ being the ground state of $H_{\rm tot.}$,
the reduced density matrix
$\rho_A = {\rm Tr}_{\bar{A}} [\rho_{\rm total}]$ is expressed by the
effective Hamiltonian $H_{\rm eff.}$ as
\begin{align}
   \rho_A = \kappa e^{- H_{\rm eff.}}  
\end{align}
where 
\begin{align}
   H_{\rm eff.} = \sum_{i,j=1}^M h_{ij} c^\dagger_i c_j = \sum_\lambda \epsilon_\lambda a^\dagger_\lambda a_\lambda  
   \label{eq:effctiveH}
\end{align}
is again hermitian and $\kappa$ is the normalization factor. 
Here we introduced the unitary transformation 
\begin{align}
   c_i &= \sum_{\lambda =1}^M \phi_\lambda(i) a_\lambda
   \nonumber \\
   a_\lambda &= \sum_{i=1}^M \phi^*_\lambda(i) c_i
\end{align}
This can be understood when one considers the path integral representation 
of the density matrix in terms of the Grassmann variables.
With $\rho_A$, the expectation value of $c_i^\dagger c_j$ is
\begin{align}
    \langle c_i^\dagger c_j \rangle  &= \sum_{\lambda =1}^M 
   \phi^*_\lambda(i) \phi_\lambda(j)  \langle a^\dagger_\lambda
   a_\lambda \rangle 
   \nonumber \\
    &= \sum_{\lambda =1}^M 
   \phi^*_\lambda(i) \phi_\lambda(j)  \frac{1}{ e^{\varepsilon_\lambda} + 1}
   \label{eq:cij}
\end{align}
From eq.(\ref{eq:effctiveH}), one can express $h_{ij}$ as
\begin{align}
    h_{ij} = \sum_{\lambda =1}^M \varepsilon_\lambda  
   \phi_\lambda(i) \phi^*_\lambda(j).
   \label{eq:hij}
\end{align}
It is easy to relate eq.(\ref{eq:cij}) and eq.(\ref{eq:hij})
to obtain 
\begin{align}
    h^T = \ln \biggl( \frac{1-C}{C} \biggr ).
   \label{eq:hT}
\end{align}
where $^T$ means the transpose of the matrix.

From $\rho_A$ one can define the generalized entropy called
R\'{e}nyi entropy by
 \begin{align}
    S^{(\alpha)}_A = \frac{1}{1-\alpha} \ln {\rm Tr} \rho_A^\alpha
   \label{eq:Renyi},
\end{align}
which is reduced to the von Neumann entropy in the limit of $\alpha \to 1$.
Now let $p_i$ ($i=1,2, \cdots, M$) be the eigenvalue and
$| 0 \rangle_j$ and $| 1 \rangle_j$ be the empty and occupied states
of that eigenstate of the matrix $C$, 
and then \cite{PhysRevB.109.085146}
\begin{align}
    \rho_A = \prod_{j=1}^M [ (1 - p_j) | 0 \rangle_j \langle_j  0 | + 
    p_j | 0 \rangle_j \langle_j  0 | ].    
\end{align}
Therefore,
 \begin{align}
    S^{(\alpha)}_A = \sum_{j=1}^M e_\alpha( p_j)
   \label{eq:Renyi2},
\end{align}
with
\begin{align}
    e_\alpha(p) &= \frac{1}{1- \alpha} \ln[(1-p)^\alpha + p^\alpha] 
    \nonumber \\
    &\to - (1-p) \ln(1-p) - p \ln p  \ \ \ \ \ {\rm as} \ \ \ \ \alpha \to 1
   \label{eq:Renyi3},
\end{align}

Now we define the following quantity
\begin{align}
    D_A (p) = {\rm det} [ p 1 - C^A]
\end{align}
and we can write $S^{(\alpha)}_A$
as
\begin{align}
    S^{(\alpha)}_A = \oint_C \frac{d p}{2 \pi i} e_\alpha (p) 
    \frac{d \ln D_A(p)}{d p} 
   \label{eq:Renyi4}, 
\end{align}
where the contour $C$ includes the segment of real axis $[0,1]$.
This equality can be easily understood because
$\ln D_A(p) = \sum_{j=1}^M \ln (p - p_j)$. 
Now a key relation is that 
\begin{align}
{\rm Tr}O^r = {\rm Tr} (C^A)^r
\label{eq:equal}
\end{align}
with 
\begin{align}
    O_{k,k'} = \langle \psi_k | \psi_{k'} \rangle_A 
    =\sum_{j \in A} \psi^\dagger_k (j) \psi_{k'} (j) 
    \label{eq:Renyi5}, 
\end{align}
being the overlap integral of occupied states restricted to region $A$.
Note that $C^A$ is $M \times M$ matrix, while $O$ is $N_e \times N_e$
matrix.
The proof of eq.(\ref{eq:equal}) goes as follows.

\begin{align}
{\rm Tr} O^q &=
\sum_{k_1, \cdots, k_q = 1}^{N_e} \sum_{j_1 \in A} \psi_{k_1}^*(j_1) \psi_{k_2} (j_1)
\nonumber \\
&\times\sum_{j_2 \in A} \psi_{k_2}^*(j_2) \psi_{k_3} (j_2)
    \cdots \times\sum_{j_q \in A} \psi_{k_q}^*(j_q) \psi_{k_1} (j_q)
\nonumber \\
&= \sum_{j_1, \cdots, j_q \in A} C^A (j_1, j_q) C^A( j_q, j_{q-1}) 
   \cdots C^A( J_2, j_1) 
\nonumber \\
&= {\rm Tr} (C^A)^q
\end{align}
Therefore,
\begin{align}
    \ln D_A (p) &= \ln p - \sum_{q=1}^\infty \frac{{\rm Tr} (C^A)^q}{q p^q} 
    \nonumber \\
    &= \ln p - \sum_{q=1}^\infty \frac{{\rm Tr} O^q}{q p^q} 
    \nonumber \\
    &= \ln {\rm det} [ p 1 - O]
\end{align}
Therefore, the  R\'{e}nyi entropy is given by 
\begin{align}
    S^{(\alpha)}_A = \sum_{r=1}^{N_e} e_\alpha(o_r)
\end{align}
with $o_r$ being the eigenvalue of matrix $O$.
Note that the overlap integral, i.e., the inner product, is 
related to the quantum distance, and hence the quantum metric is 
closely related to the entanglement entropy, and several authors
studied this relation extensively 
\cite{PhysRevB.73.245115,PhysRevB.88.115114}. 

The quantum geometry tensor shows the critical scaling
at the quantum phase transition 
\cite{CAROLLO20201,PhysRevLett.99.095701,PhysRevA.73.010303,
zanardi2007differentialinformationgeometryquantumphase}.
Generally, the ground state is expected to be very sensitive to 
the change of the parameter $\lambda$'s near the quantum
critical point, but it is not trivial that the
QGT diverges or not there.
Let the QGT $G_{k,ij}$ in eq.(\ref{eq:Gkij})
for the ground state $k=0$ be $Q_{ij}$, which   
can be rewritten as \cite{PhysRevLett.99.095701}
\begin{align}
    Q_{ij} = \int_0^\infty d \tau  \tau G_{ij}(\tau),
\end{align}
where 
\begin{align}
    G_{ij}(\tau) = \langle 0 | \partial_i H(\tau) \partial_j H(0) | 0 \rangle
    -  \langle 0 | \partial_i H(0) | 0 \rangle
    \langle 0 | \partial_j H(0) | 0 \rangle
\end{align}
is the imaginary time correlation function 
$(A(\tau)= e^{H \tau} A e^{- H \tau})$, 
with $\partial_i = \partial/\partial \lambda_i$.
Define the QGT per unit volume $q_{ij}= Q_{ij}/L^d$
with $L$ and $d$ are the linear dimension 
and dimensionality of the system.
The critical property of $q_{ij}$ near the critical 
point $\lambda=\lambda_c$
is determined by the
scaling dimension $\Delta_i$ of $\partial_i H$ and the 
dynamical critical exponent $\zeta$ as \cite{PhysRevLett.99.095701}
\begin{align}
    q ( \lambda \to \lambda_c) \sim | \lambda - \lambda_c |^{\nu \Delta_Q}
\end{align}
with $\Delta_Q = \Delta_i + \Delta_j - 2 \zeta - d$
and $\nu$ is the critical exponent for the correlation length, i.e.,
$\xi= |\lambda - \lambda_c |^{-\nu} $.
The exponent $\Delta_Q$ can be either positive or negative, 
which depends on the model.
In \cite{PhysRevLett.99.095701}, $S=1/2$ $XXZ$ Heisenberg model 
is analyzed and $\Delta_Q=1$ is concluded, which indicates that
$q$ does not diverge. 

\subsection{Adiabatic and Non-Adiabatic Time Evolution and Quantum Geometry}

In section II.A, we considered the eigenstates and eigenvalues of 
Hamiltonian $H(\lambda)$ which depends on the parameter $\lambda$.
However, we did not consider the time-evolution of the 
state when $\lambda(t)$ depends on the time $t$.
We start with the time-dependent Schr\"{o}dinger equation given by 
\begin{align}
      i \hbar \frac{\partial}{\partial t} | \psi(t) \rangle =
      H(\lambda(t)) | \psi(t) \rangle 
             \label{eq:schroedinger}
\end{align}
For each $\lambda$, the eigenstate problem of $H(\lambda)$ is solved
as
\begin{align}
      H(\lambda) | \phi_n(\lambda) \rangle =
      E_n(\lambda) | \phi_n(\lambda) \rangle 
             \label{eq:eigenschroedinger}
\end{align}
We consider the situation where the state is at $|\phi_k (\lambda)\rangle$ 
at $t = - \infty$, and consider its time-evolution \cite{KOLODRUBETZ20171,doi:10.1098/rspa.1984.0023,doi:10.1142/0613}.
Expanding $| \psi (t) \rangle$ in terms of $| \phi_n(\lambda) \rangle $ as
\begin{align}
      | \psi (t) \rangle = \sum_n c_n(t) | \phi_n(\lambda(t) ) \rangle,
      \label{eq:eigenschroedinger2}
\end{align}
one can obtain the differential equation for $c_n(t)$
as
\begin{align}
      &i \hbar \frac{d c_n(t) }{d t} - E_n(\lambda(t)) c_n(t)
      \nonumber \\
      &=
      - i \hbar \sum_m c_m(t) \langle \phi_n(\lambda(t)) |
      \partial_j \phi_m(\lambda(t)) \rangle \frac{d \lambda_j}{d t},
      \label{eq:timeev}
\end{align}
where $\partial_j = \partial/\partial \lambda_j$.
First, let us consider the fully adiabatic case.
In the original paper by Berry, it was assumed that the
adiabatic time-evolution does not allow the transition 
from the state $|\phi_k \rangle$ to other state $|\phi_n \rangle$ ($n \ne k)$,
and hence only $k$ is relevant both in the left and right hand sides of 
eq.(\ref{eq:timeev}).
Therefore, defining
\begin{align}
      \tilde{c}_k(t) = c_k(t) \exp\biggl( \frac{i}{\hbar} 
      \int_{-\infty}^t d t' E_k(\lambda(t')) \biggr), 
      \label{eq:timeev2}
\end{align}
we obtain 
\begin{align}
      \frac{d \tilde{c}_k(t) }{d t} = - \tilde{c}_k(t)
      \langle \phi_k(\lambda(t)) |
      \partial_j \phi_k(\lambda(t)) \rangle \frac{d \lambda_j}{d t}.
      \label{eq:timeev3}
\end{align}
Equation (\ref{eq:timeev3})
can be easily solved as
\begin{align}
      \tilde{c}_k(t) &= 
      \exp\biggl(  
      - \int_{-\infty}^t d t'  \langle \phi_k(\lambda(t')) |
      \partial_j \phi_k(\lambda(t')) \rangle \frac{d \lambda_j}{d t'} \biggr)
     \nonumber \\
     &=  \exp\biggl(  
      - \int_{\lambda(-\infty)}^{\lambda(t)} \langle \phi_k(\lambda) |
      \partial_j \phi_k(\lambda) \rangle d \lambda_j \biggr)
\nonumber \\
       &= \exp \biggl( i  
      \int_{\lambda(-\infty)}^{\lambda(t)} a_{k, j} d \lambda_j \biggr)
      \nonumber \\
       &= \exp(  i\gamma_k(t) )  
      \label{eq:timeev4}
\end{align}
where $\gamma_k(t)$ is the celebrated Berry phase,
and the Berry connection $a_{k,j}(\lambda)$ is defined by 
\begin{align}
a_{k,j} = \langle \phi_k(\lambda) | i
      \partial_j \phi_k(\lambda) \rangle. 
\end{align}
which is real because 
$\langle \phi_k(\lambda) | \partial_j \phi_k(\lambda) \rangle
+ \langle \partial_j \phi_k(\lambda) | \phi_k(\lambda) \rangle
= \langle \phi_k(\lambda) | \partial_j \phi_k(\lambda) \rangle
+ \langle \phi_k(\lambda) | \partial_j \phi_k(\lambda) \rangle^* =0.
$

Up to now, we assume that no transition is allowed. 
Second, let us consider the case with first-order non-adiabatic correction. 
This treatment is typically used in situations where the system is
subject to external perturbations that vary slowly in time, 
such as magnetic fields, electric fields, strain, or laser pulses.
More explicitly, 
in eq.(\ref{eq:timeev}), $c_n(t)$ with $n \ne k$ can be induced 
first order in $\frac{d\lambda}{d t}$.
To calculate the non-adiabatic correction up to first order in 
$\frac{d\lambda}{d t}$, we regard $\lambda(t)$ as if it
is time-independent compared with the dynamical
time-evolution $e^{- i E_n(\lambda) t/ \hbar}$ in 
 eq.(\ref{eq:timeev}).
 Therefore, eq.(\ref{eq:timeev}) is reduced for $c_n(t)$ ($n \ne k$)
 to
\begin{align}
      &i \hbar \frac{d c_n(t) }{d t} - E_n(\lambda) c_n(t)
      \nonumber \\
      &=
      - i \hbar c_k(t) \langle \phi_k(\lambda) |
      \partial_j \phi_n(\lambda) \rangle \frac{d \lambda_j}{d t},
      \label{eq:timeevn}
\end{align}
 which is solved as
 \begin{align}
      c_n(t) = 
       \frac{i \hbar \exp ( - \frac{i}{\hbar} E_k t + i \gamma_k )}
       { E_k(\lambda) - E_n(\lambda) + i \delta} 
        \langle \phi_n(\lambda) |
      \partial_j \phi_k(\lambda) \rangle \frac{d \lambda_j}{d t},
       \label{eq:timeevn}
\end{align}
 where we omitted the explicit $\lambda$-dependence for brevity,
 and $\delta$ is the infinitesimal positive convergence factor
 which can be neglected for $n \ne k$.
Therefore, 
\begin{align}
      &| \psi(t) \rangle = | \phi_k \rangle e^{- i E_k t/\hbar + i \gamma_k}
      \nonumber \\
      &+ \sum_{n (\ne k)} e^{- i E_k t/\hbar + i \gamma_k} 
      \frac{i \hbar}{ E_k- E_n} 
      | \phi_n \rangle \langle \phi_n|
      \partial_j \phi_k \rangle \frac{d \lambda_j}{d t}.
       \label{eq:timeevn2}
\end{align}
With this wavefunction, one can calculate the 
expectation value of $\frac{\partial H(\lambda)}{\partial \lambda_i}$ as
\begin{align}
      \langle \psi(t) | \frac{\partial H}{\partial \lambda_i}|
      \psi(t) \rangle &= 
      \langle \phi_k(t) | \frac{\partial H}{\partial \lambda_i}|
      \phi_k(t) \rangle
      \nonumber \\
      + \sum_{n (\ne k)}  
      \frac{i \hbar}{ E_k - E_n} 
      &\biggl[
      \langle \phi_k|\frac{\partial H}{\partial \lambda_i} |
      \phi_n\rangle \langle \phi_n|
      \partial_j \phi_k\rangle ]
      \nonumber \\
      &-  \langle \partial_j \phi_k| \phi_n \rangle 
      \langle \phi_n |\frac{\partial H(\lambda)}{\partial \lambda_i} |
      \phi_k \rangle \biggr] \frac{d \lambda_j}{d t}.
       \label{eq:timeevn2}
\end{align}
Using Feynman-Hellmann theorem, this expression is reduced to
\cite{PhysRevLett.75.697}
\begin{align}
      &\langle \psi(t) | \frac{\partial H}{\partial \lambda_i}|
      \psi(t) \rangle = 
      \frac{\partial E_k}{\partial \lambda_i}
      \nonumber \\
      &+ \sum_{n (\ne k)}  
    i \hbar  \biggl[
  \langle \partial_i \phi_k|  \phi_n\rangle \langle \phi_n|
      \partial_j \phi_k\rangle 
      - \langle \partial_j \phi_k|  \phi_n\rangle \langle \phi_n|
      \partial_i \phi_k\rangle  \biggr] \frac{d \lambda_j}{d t}.
          \nonumber \\
      &=  \frac{\partial E_k}{\partial \lambda_i} + \hbar \Omega^k_{ij}  
      \frac{d \lambda_j}{d t}.
       \label{eq:timeevn3}
\end{align}
Here the second term is the non-adiabatic correction
which is proportional to the Berry curvature 
$\Omega^k_{ij}$ given by
\begin{align}
    \Omega^k_{ij} = \partial_i a_{k,j} - \partial_j a_{k,i}.
\end{align}

Third, we now consider situations of non-adiabatic effect where
$\lambda(t)$ itself is the dynamical variable 
which changes slowly in time compared with the
rapid system described by $H(\lambda(t))$ \cite{PhysRevLett.121.020401}.
Examples include the electron-phonon coupled system,
where the Born-Oppenheimer approximation (BOA) is justified
\cite{doi:10.1142/0613,RevModPhys.64.51,doi:10.1143/JPSJ.76.015003}.
Let us write the Hamiltonian as
\begin{align}
H = \frac{1}{2M} \hat{P}^2 + h(\hat{\xi}, \hat{X})
\end{align}
where $\hat{P}$ $(\hat{X})$ is the momentum (coordinate) operator of phonon,
while $\hat{\xi} = ( \hat{x}, \hat{p})$ is those of electrons.

In BOA, the wavefunction of the coupled system is given by the
product of electronic and phononic parts as
\begin{align}
\Psi(x,X) = \phi_n(x, X) \Phi(X) 
\end{align}
where
\begin{align}
h(\hat{\xi}, X) \phi_n(x,X) = E_n (X) \phi_n(x, X). 
\end{align}
Here, the phonon coordinate $X$ is regarded as a parameter, and not
the dynamical variable. 
In the bra-ket formalism, this corresponds to $| m(X) \rangle \otimes | X \rangle $
which forms the complete set as
\begin{align}
\int d \mu(X) |X \rangle \langle X| &= 1,  \nonumber \\
\sum_m |m(X) \rangle \langle m(X)| &= 1.
\end{align}
Inserting these expression, one obtain the following 
path integral representation of time-evolution:
\begin{align}
& K(T) = {\rm Tr} [ \exp( - i H T/\hbar) ]  \nonumber \\
     &= \sum_n \int d \mu(X_0) \langle n(X_0), X_0 | \exp( - i H T/\hbar) |  n(X_0), X_0 \rangle 
                      \nonumber \\
     &= \sum_n \int d \mu(X_0) \langle n(X_0), X_0 | [\exp( - i H \epsilon/ \hbar)]^L |  n(X_0), X_0 \rangle 
                      \nonumber \\
    &= \sum_n \sum_{m_1} \sum_{m_2} \cdots \sum_{m_{L-1}} 
     \int d \mu(X_0) \biggl( \Pi_{k=1}^{L-1} \int d \mu(X_k) \biggr)  
     \nonumber \\ 
    &\times \langle n(X_0), X_0 | \exp( - i H \epsilon/\hbar) | m_{L-1} (X_{L-1}), X_{L-1} \rangle \times \cdots \nonumber \\
    &\times \langle m_1(X_1), X_1 | \exp( - i H \epsilon/\hbar) | n(X_{0}), X_{0} \rangle, 
    \label{eq:path}
\end{align}
where $T = L \epsilon$ and we consider the limit $L \to \infty$.
Up to now, the expression is exact, and we introduce here the ``adiabatic approximation"
by restricting the sum over $m_1, m_2, \cdots , m_{L-1}$ to the initial $n$.
Namely, the transitions between the different electronic eigenstates are forbidden.
In this case, eq.(\ref{eq:path}) is reduced to 
\begin{align}
K(T) = \sum_n  \Pi_{k=1}^{L} \int d \mu(X_k) \omega_k 
\end{align}
with $X_L = X_0$, and
\begin{align}
\omega_k = \langle n(X_k), X_k | \exp( - i H \epsilon/\hbar) | n(X_{k-1}), X_{k-1} \rangle. 
\end{align}
For small $\epsilon$, 
\begin{align}
\omega_k &\cong  \langle n(X_k), X_k | \biggl( 1  - \frac{i}{\hbar} H \biggr) | n(X_{k-1}), X_{k-1} \rangle \nonumber \\
 &= \biggl( 1 -  \frac{i}{\hbar} E_n (X_{k-1}) \biggr) 
 \langle n(X_k)| n(X_{k-1})\rangle \langle X_k | X_{k-1} \rangle
 \nonumber \\
  &- i \frac{\epsilon}{\hbar} \frac{1}{2M} 
  \biggl[ - \hbar^2 \langle n(X_k)| \nabla_k^2 n(X_{k-1})\rangle \langle X_k | X_{k-1} \rangle
  \nonumber \\
  &+ 2 \frac{\hbar}{i} \langle n(X_k)| \nabla_k n(X_{k-1})\rangle \langle X_k | \hat{P} | X_{k-1} \rangle  \nonumber \\
  &+ \langle n(X_k)| n(X_{k-1})\rangle \langle X_k | \hat{P}^2 | X_{k-1} \rangle  
  \biggr]
    \label{eq:element}
\end{align}
where $\nabla_k = \partial/\partial X_k$.
One can easily show
\begin{align}
\langle X_k | \hat{P}^n | X_{k-1} \rangle = 
\int d \mu(P_k) P_k^n e^{i P_k ( X_k - X_{k-1} )/\hbar}
\end{align}
and $\omega_k$ is written up to the first order in $\epsilon$ as
\begin{align}
\omega_k &\cong  \int d \mu(P_k) e^{i P_k ( X_k - X_{k-1} )/\hbar} 
\langle n(X_k) | n(X_{k-1})\rangle 
\nonumber \\
&\times \biggl[ 1 -  \frac{i \epsilon}{\hbar} (E_n (X_{k-1}) +H_{kin} ) \biggr]
    \label{eq:element2}
\end{align}
where $H_{kin}$ is given by 
\begin{align}
H_{kin} = \frac{1}{2M} &\biggl( P_k^2 + 2 \frac{\hbar}{i} P_k 
 \langle n(X_k)| \nabla_k | n(X_{k-1})\rangle \nonumber \\
&- \hbar^2  \langle n(X_k)| \nabla_k^2 |n(X_{k-1})\rangle \biggr).     
\end{align}
Eventually,
\begin{align}
K(T) = \sum_n \int \mathcal{D} X \mathcal{D} P 
  \langle n(X_T) | n(X_0) \rangle_C
  \exp \biggl( \frac{i}{\hbar} S_n^{ad} \biggr)
    \label{eq:path2}
\end{align}
where  
\begin{align}
  \langle n(X_T) | n(X_0) \rangle_C &= \exp \biggl( 
  - i\oint_C a_n (X) d X  \biggr)  \nonumber \\
   &= 
  \exp  \biggl(  - i \int b_n (X) \cdot d S  \biggr)
\end{align}
is the Berry phase with the Berry connection 
$a_n(X) = i \langle n(X) | \nabla_X | n(X) \rangle$
and the Berry curvature $b_n(X) = \nabla_X \times a_n (X)$.
On the other hand, $S_n^{ad}$ is given by 
\begin{align}
  S_n^{ad} = \int_0^T d t (P \dot{X} - H_n^{ad} )
\end{align}
with the effective Hamiltonian
\begin{align}
  H_n^{ad} = \frac{1}{2M} \biggl( \frac{\hbar}{i} \nabla_X - a_n(X) 
  \biggr)^2 + V_n(X) + E_n(X) 
\end{align}
where 
\begin{align}
  V_n(X) = \frac{\hbar^2}{2M} \langle \nabla_X n| ( 1 - | n \rangle \langle n |)
  | \nabla_X  n \rangle 
\end{align}
is the correction of the potential to the BOA 
represented by the quantum metric tensor.
In summary, the corrections to the BOA are 
represented by the QGT.

Finally, we consider the generic non-adiabatic process
without assuming adiabatic process addressed 
by Anandan-Aharanov \cite{PhysRevLett.65.1697}.
In this case, the parameter $\lambda$ is 
replaced by the real time $t$, and the time evolution from 
$t$ to $t+ dt$ is given by the Hamiltonian $H(t)$ up to the 
second order in $d t$ as,
\begin{align}
|u(t+\mathrm{d}t)\rangle=|u(t)\rangle+\frac{d}{dt}|u(t)\rangle dt+\frac12\frac{d^2}{dt^2}|u(t)\rangle dt^2+\cdots
\end{align}
The time-dependent Schr\"{o}dinger equation is,
\begin{align}
\frac{d}{dt}|u(t)\rangle =- \frac i\hbar H(t)|u(t)\rangle, 
\end{align}
which leads to 
\begin{align}
\frac{d^2}{dt^2}|u(t)\rangle &= -\frac i\hbar \frac{d H(t)}{dt}|u(t)\rangle-\frac1{\hbar^2}H(t)^2|u(t)\rangle    
\end{align}
Therefore, we can calculate the distance between 
$|u(t) \rangle$ and $| u(t + dt)\rangle$
as
\begin{align}
ds^2 &= 1 - |\langle u(t)| u(t+dt) \rangle |^2  \nonumber \\
&=
\frac{1}{\hbar^2} ( \langle u(t)|H(t)^2 |u(t) \rangle - \langle u(t)|H(t) |u(t) \rangle^2 ) dt^2
\label{eq:ds2}
\end{align}
Note that the term with $d H(t)/dt$ contributes only to $O(dt^3)$.
Here the fluctuation of the energy $\Delta E$ is defined as
\begin{align}
(\Delta E)^2 =  \langle u(t)|H(t)^2 |u(t) \rangle - \langle u(t)|H(t) |u(t) \rangle^2, 
\end{align}
and hence eq.(\ref{eq:ds2}) becomes $ds = \Delta E dt/\hbar$,
i.e., the quantum distance corresponding to the quantum evolution in time is given by the product of the energy uncertainty 
$\Delta E$ times the elapsed time $t$.
Anandan-Aharanov argued that the shortest distance between orthogonal states, 
is $2s \geq \pi$ along a geodesic, which leads to the uncertainty relation between the
time and energy as
\begin{align}
\Delta E \Delta t \gtrsim \frac{h}{4} 
\end{align}

\section{Quantum Geometric Tensor for Bloch States}

Now we consider the Bloch wavefunction for solids where the
parameter $\lambda$ corresponds to the crystal wavenumber
$k$ \cite{Vanderbilt_2018}. Namely, the wavefunction is written as
\begin{align}
    |\psi_{nk} \rangle = e^{i k \cdot \hat{x} } | u_{nk} \rangle
    \label{eq:Bloch1}
\end{align} 
where $\hat{x}$ is the position operator.  
$| u_{n,k} \rangle$ is the periodic part satisfying
$\langle x + a | u_{n,k} \rangle = \langle x | u_{n,k} \rangle $
with $a$ being the lattice vector. 

Since the Bloch wavefunction $| \psi_{nk} \rangle$ is the eigenstate of the 
Hamiltonian, i.e., 
\begin{align}
    \hat{H} | \psi_{nk} \rangle = \varepsilon_{nk} | \psi_{nk} \rangle,
    \label{eq:Bloch}
\end{align}

Here we take the normalization of the Bloch wavefunction as
\begin{align}
    \langle \psi_{mk} | \psi_{nk'} \rangle = \delta_{m,n} \delta_{k,k'},
    \label{eq:innerproduct}
\end{align}
where the wavenumber $k$ is discretized as 
$k = \frac{2 \pi}{La}( n_1,\cdots,n_d)$ ($d$: dimension of the system)
with $La$ being the linear dimension of the
system and $n_1, \cdots,n_d$ are integers. The volume of the system is 
$V= N a^d = N V_{\rm cell}$ with $N= L^d$, and $V_{\rm cell}=a^d$ is the volume
of the unit cell. 
In terms of these quantities, the integral and sum over the momentum
$k$ is related by 
\begin{align}
\frac{1}{N} \sum_k =  \int [ dk],       
\end{align}
where $[dk] = \frac{V_{\rm cell}}{(2 \pi)^d} d^d k$
and $\delta^d(k-k') = \frac{V}{(2 \pi)^d}\delta_{k,k'}$.
The advantage of this convention is that 
$\frac{1}{N} \sum_k 1=  \int [ dk] 1 = 1$ because  the volume of the
first Brillouin zone is $\frac{(2\pi)^d}{V_{\rm cell}}$, and the number of 
$k$-points in it is $N$.
Using the relation 
\begin{align}
\sum_R e^{i (k - k') \cdot R} = N \delta_{k,k'} = \frac{(2 \pi)^d}{V_{\rm cell}} \delta^d(k-k'),
\end{align}
where $R$ is the position of the unit cell.

From eq.(\ref{eq:Bloch}), the periodic part $| u_{n,k} \rangle$ satisfies 
the relation
\begin{align}
    \hat{H}_k | u_{nk} \rangle = \varepsilon_{nk} | u_{nk} \rangle,
    \label{eq:Bloch2}
\end{align}
with 
\begin{align}
    \hat{H}_k  = e^{-i k \cdot \hat{x} } \hat{H} e^{i k \cdot \hat{x} } ,
    \label{eq:Hk}
\end{align}
In this representation, the velocity operator $\hat{v}_k$ is
given by
\begin{align}
    \hat{v}_k  = \frac{\partial \hat{H}_k }{\partial k}.
    \label{eq:vk}
\end{align}
Here we derive the Feynman-Hellmann theorem in the context of 
Bloch wavefunction.
Differentiating eq.(\ref{eq:Bloch2}) with respect to $k$,
one obtains 
\begin{align}
    \frac{\partial \hat{H}_k}{\partial k} | u_{nk} \rangle  +
    \hat{H}_k | \frac{\partial}{\partial k} u_{nk} \rangle = 
    \frac{\partial \varepsilon_{nk}}{\partial k} | u_{nk} \rangle
    +  \varepsilon_{nk} | \frac{\partial} {\partial k} u_{nk} \rangle 
    \label{eq:FH} 
\end{align}
Taking the inner product with $\langle u_{nk} |$, one obtains
\begin{align}
    \langle u_{nk} | \hat{v}_k  | u_{nk} \rangle  
    = \frac{\partial \varepsilon_{nk}}{\partial k}  
    \label{eq:gv} 
\end{align}
which is nothing but the group velocity.
With $\langle u_{mk} |$ ($m \ne n$),
one obtains the 
\begin{align}
    \langle u_{mk} | \hat{v}_k  | u_{nk} \rangle  
    = (\varepsilon_{nk} - \varepsilon_{mk} )  \langle u_{mk} | 
    \frac{\partial}{\partial k}  u_{nk} \rangle   
    \label{eq:gv2} 
\end{align}

Now we consider the matrix elements of the position operator $\hat{x}$ between
the Bloch wavefunctions. In this case, one needs to consider the general 
case of $k$ and $k'$.

\begin{align}
 &  \langle \psi_{mk} | \hat{x} | \psi_{nk'} \rangle \nonumber \\ 
 &=  \langle u_{mk}|  e^{-i k \hat{x}}  \hat{x}  e^{i k' \hat{x}} | u_{nk'} \rangle
    \nonumber \\
 &= -i \frac{\partial}{\partial k'}  \langle u_{mk} |  e^{-i k \hat{x}} e^{i k' \hat{x}} | u_{nk'} \rangle+ 
 i \langle u_{mk} |  e^{-i k \hat{x}}  e^{i k' \hat{x}} 
 | \frac{\partial}{\partial k'}u_{nk'} \rangle  \nonumber \\
 &= - i  \delta_{m,n} \frac{\partial \delta_{k,k'}}{\partial k'} 
    + i  \langle u_{mk} |  \frac{\partial}{\partial k'}u_{nk'} \rangle \delta_{k,k'}  
    \nonumber \\
 &=  i  \delta_{m,n} \frac{\partial \delta_{k,k'}}{\partial k} 
    + i  \langle u_{mk} |  \frac{\partial}{\partial k}u_{nk} \rangle \delta_{k,k'}  
  \label{eq:xhat}
\end{align}

It is instructive to see the relation between eq.(\ref{eq:xhat}) and 
eqs.(\ref{eq:gv} and (\ref{eq:gv2}).
\begin{align}
    & \langle \psi_{mk} | \hat{v} | \psi_{nk'} \rangle =
     \langle \psi_{mk} | \frac{d \hat{x}}{ d t }  | \psi_{nk'} \rangle \nonumber \\
     &= \langle \psi_{mk} | \frac{1}{i \hbar} [ \hat{x}, H] | \psi_{nk'} \rangle
      = \frac{1}{i \hbar}  \langle \psi_{mk} |\hat{x}| \psi_{nk'} \rangle 
      (\varepsilon_{n k'}- \varepsilon_{mk})
      \nonumber \\
      &= \frac{1}{\hbar}  \biggl[ 
        \delta_{m,n} \frac{\partial \delta_{k,k'}}{\partial k} 
    +   \langle u_{mk} |  \frac{\partial}{\partial k'}u_{nk'} \rangle \delta_{k,k'} 
    \biggr] (\varepsilon_{n k'}- \varepsilon_{mk})
      \label{eq:xdot} 
\end{align}
Now we use the identity
\begin{align}
    \frac{\partial}{\partial k}[ \delta_{k,k'} ( \varepsilon_{n k}- \varepsilon_{n k'}) ] =
    \frac{\partial}{\partial k}[ 0 ] =0,
\end{align}
and 
\begin{align}
    \frac{\partial  \delta_{k,k'} }{\partial k}( \varepsilon_{n k}- \varepsilon_{n k'})  =
    - \frac{\partial \varepsilon_{n k}} {\partial k}  \delta_{k,k'}.
\end{align}
Therefore, it is easily seen that eq.(\ref{eq:xhat}) and 
eqs.(\ref{eq:gv}), (\ref{eq:gv2}) are consistent with each other.

The QGT is defined for this periodic part (not for $|\psi_{n,k} \rangle$) because $\langle \psi_{m k'}|\psi_{n,k} \rangle = 0$ for 
$k \ne k'$. 
The Berry connection is defined by 
\begin{align}
    A_{nk} = i \langle u_{nk}| \nabla_k |u_{nk} \rangle 
\end{align}
and QGT is
\begin{align}
    G_{n,\mu \nu}(k)  = \langle \partial_\mu  u_{n k}|
    ( 1 - \hat{P}) |\partial_\nu u_{n,k} \rangle. 
\end{align}
where $\hat{P}$ is the projection operator 
$\hat{P} = \sum_l |u_{l k} \rangle \langle u_{l k} |$ with the sum taken
over the bands of interests.

The non-Abelian generalization is 
\begin{align}
    A_{n,m k} = i \langle u_{nk}| \nabla_k |u_{mk} \rangle 
\end{align}
and QGT is
\begin{align}
    G_{n,m: \mu \nu}(k)  = \langle \partial_\mu  u_{n k}|
    ( 1 - | u_{n,k} \rangle \langle u_{n k}|) |\partial_\nu u_{m,k} \rangle. 
\end{align}

Using the matrix elements discussed above, 
one can obtain the following identity
\begin{align}
& \sum_{m (\ne n)} \langle u_{nk} | \hat{x}_i| u_{mk} \rangle 
 \langle u_{mk} | \hat{x}_j| u_{nk} \rangle  
 \nonumber \\
 &= 
 - \sum_{m (\ne n)} \langle u_{nk} | \frac{\partial}{\partial k_i} u_{mk} \rangle 
 \langle u_{mk} | \frac{\partial}{\partial k_j}  u_{nk} \rangle
 \nonumber \\
 &= \sum_{m (\ne n)} \langle \frac{\partial}{\partial k_i} u_{nk} | u_{mk} \rangle 
 \langle u_{mk} | \frac{\partial}{\partial k_j}  u_{nk} \rangle
 \nonumber \\
 &= \langle \frac{\partial}{\partial k_i} u_{nk} | \frac{\partial}{\partial k_j}  u_{nk} \rangle - \langle  \frac{\partial}{\partial k_i} u_{nk} |  u_{nk} \rangle
 \langle u_{nk} | \frac{\partial}{\partial k_j} u_{nk} \rangle
 \nonumber \\
 &= G_{n,ij}
 \label{eq:xxg}
\end{align}

Now some remarks about the symmetry of the QGT are in order.
Of particular importance are the time-reversal $\cal{T}$ 
and spatial inversion $\cal{P}$ symmetries.
One can show that 
$G_{\mu \nu} (k) = G^*_{\mu \nu}(-k)$
when the system is $\cal{T}$-symmetric, while
$G_{\mu \nu} (k) = G_{\mu \nu}(-k)$
when the system is $\cal{P}$-symmetric.
When the system is $\cal{PT}$-symmetric, 
$G_{\mu \nu} (k) = G_{\mu \nu}^*(k)$, and
the imaginary part, i.e., Berry curvature $F_{\mu \nu}(k)=0$.
In this case, the Kramers degeneracy occurs at each $k$-point
and the non-Abelian QGT occurs and the 
Euler class an be defined \cite{PhysRevX.9.021013}.

\subsection{Wavepacket Dynamics and Semiclassical Treatment}
The semiclassical treatment in terms of the wavepacket made from the
Bloch wavefunctions provide transparent and useful physical pictures
\cite{PhysRevB.59.14915,RevModPhys.82.1959}.
Suppose the perturbation of the Hamiltonian 
is characterized by the slowly varying set of parameters 
$\beta(x,t)=\beta_1(x,t), \cdots , \beta_r(x, t)$ as
$\hat{H}(\hat{x},\hat{p}: \beta(x,t))$.

The meaning of the Berry connection can be understood when 
one considers the wavefunction of the wavepacket as
\begin{align}
 | \Psi \rangle = \sum_{k} a(k,t) | \psi_k (x_c, t) \rangle,
\end{align}
where $| \psi_k (x_c, t) \rangle$ is the Bloch eigenstate of the local Hamiltonian 
$\hat{H}_c(x_c, t) = \hat{H}(\hat{x},\hat{p}: \beta(x_c,t) )$
at the center of the wavepacket $x_c$ with the energy eigenvalue 
$\varepsilon(k,t)$, i.e.,
\begin{align}
 \hat{H}_c (x_c,t)  | \psi_k (x_c, t) \rangle = \varepsilon(k,t) | \psi_k (x_c, t) \rangle.
\end{align}
$|a(k)|$ has a sharp peak near $k_c$ with the
normalization condition 
\begin{align}
 \sum_k |a(k)|^2  = 1
\end{align}
so that both the position and momentum can be specified
within the accuracy consistent with the uncertainty principle,
i.e., $\Delta x \Delta k > 1$.
The center of the momentum $k_c$ is given by
\begin{align}
    k_c = \sum_k  k |a(k)|^2. 
\end{align}
(Note that the band index $n$ is dropped in this section 
assuming that the transition between the bands are forbidden in the
low energy phenomena.)
One can define the phase $\gamma(k,t)$ as
\begin{align}
    a(k,t) = |a(k,t)| \exp(- i \gamma (k,t) ). 
\end{align}

The Bloch wavefunction 
$| \psi_k (x_c, t) \rangle$ is represented by 
$| \psi_k (x_c, t) \rangle = e^{i k \hat{x}} | u_k (x_c, t) \rangle$
in terms of the periodic part $| u_k (x_c, t) \rangle$ as given in 
eq.(\ref{eq:Bloch1}). 

Using eq.(\ref{eq:xhat}), the expectation value of $\hat{x}$ for the
wavepacket is
\begin{align}
 x_c &=  \langle \Psi(x_c, t) | \hat{x} | \Psi(x_c, t) \rangle \nonumber \\ 
 &= \int d^3 k \int d^3 k' a^*(k,t) a(k',t)\langle \psi_k | \hat{x} 
  | \psi_{k'} \rangle
    \nonumber \\
 &= \int d^3 k \biggl[ a^*(k) i \frac{\partial}{\partial k}a(k) + 
 i \langle u_k | \frac{\partial}{\partial k}u_{k} \rangle |a(k)|^2  \biggr]
\end{align}
where the partial integration with respect to $k'$ has been done.
Now considering that $a(k,t)$ is peaked such that both $x$ and $k$ 
are sharply peaked, the wavepacket center $x_c$ is given by
\begin{align}
 x_c &= i a^*(k_c)  \frac{\partial a(k_c)}{\partial k_c} + 
 i \biggl\langle u_{k_c} \bigg| \frac{\partial}{\partial k_c}u_{k_c} \biggr\rangle 
 \nonumber \\
 &=  \frac{\partial \gamma (x_c,t)}{\partial k_c} + A(k_c)
 \label{eq:xc}
\end{align}
thus Berry connection
$A(k_c) =  i \langle u_{k_c} | \frac{\partial}{\partial k_c}u_{k_c} \rangle $
has the meaning of intra-cell coordinates which gives the center of the 
wavepacket measured from the Wannier center \cite{ADAMS1959286}. 

The dynamics of the wavepacket can be described by the effective 
Lagrangian given $L$ defined by
\begin{align}
    L = \langle \Psi | \frac{d}{dt} - \hat{H} | \Psi \rangle
\end{align}
which was discussed in details in \cite{PhysRevB.59.14915,RevModPhys.82.1959,
PhysRevB.88.104412};
\begin{align}
    L &= - \varepsilon + k_c \cdot \dot{x}_c + 
    \dot{k}_c \cdot \biggl\langle u \bigg| i \frac{\partial u}{\partial k_c} 
    \biggr\rangle \nonumber \\
    &+
    \dot{x}_c \cdot \biggl\langle u \bigg| i \frac{\partial u}{\partial x_c} 
    \biggr\rangle +
    \biggl\langle u \bigg| i \frac{\partial u}{\partial t} 
    \biggr\rangle,
    \nonumber \\
    &=  - \varepsilon + \frac{1}{2}\sum_{i,j=1}^6 z_i  K_{ij} \dot{z}_j 
    k_c \cdot \dot{x}_c + \sum_{i=1}^6 a_i \dot{z}_i + a_t
    \end{align}
where we introduced $z_i=((x_c)_1,(x_c)_2,(x_c)_3,(k_c)_1,(k_c)_2,(k_c)_3)$,
and $K_{ij}$ by 
\begin{align}
    \begin{pmatrix}
   0 & -I \\
   I & 0
\end{pmatrix}
\end{align}
where $I$ is the $3 \times 3$ unit matrix.
The equation of motion is derived by taking the variation with respect to
$\delta z_i$, i.e.,
\begin{align}
    \delta \int dt L &= \int dt \delta z_i \biggl( K_{ij} \dot{z}_j 
    - \frac{\partial \epsilon}{\partial z_i} + \dot{z}_j \partial_i a_j 
    - \frac{d a_i}{d t} + \partial_i a_t \biggr)
    \nonumber \\
    &= \int dt \delta z_i \biggl( K_{ij} \dot{z}_j  
    - \frac{\partial \epsilon}{\partial z_i} + \dot{z}_j \Omega_{ij} + \Omega_{it}     \biggr) = 0
    \label{eq:Lag}
    \end{align}
where 
\begin{align}
    \Omega_{ij} = \partial_i a_j - \partial_j a_i =
      2 {\rm Im} \langle \partial_i u | \partial_j u \rangle.
    \end{align}
is the Berry curvature.
The equation of motion obtained from eq.(\ref{eq:Lag}) is
\begin{align}
   \Gamma_{ij} \dot{z}_j  &= \frac{\partial \epsilon}{\partial z_i} + \Omega_{it}
    \label{eq:Lag2}
    \end{align}
where $\Gamma_{ij} = K_{ij} + \Omega_{ij}$.

 Here we consider the canonical transformation from $z_i$ to 
 $\bar{z}_i(z)$ and $H(z) = \bar{H}(\bar{z})$ where  $\bar{z}_i$ satisfies the usual 
 canonical equation of motion \cite{RevModPhys.70.467}, i.e.,
\begin{align}
    \dot{\bar{z}}_i = (J_c)_{ij} \frac{\partial \bar{H}(\bar{z})}{\partial \bar{z}_j }
                    = [ \bar{z}_i, \bar{H}(\bar{z})]
                    \label{eq:can1}
\end{align}
where $J_c= K^{-1}$.
From eq.(\ref{eq:can1}), 
\begin{align}
    \dot{z}_i &= \frac{\partial z_i}{\partial \bar{z}_j} (J_c)_{j l}
    \frac{\partial \bar{H}}{\partial \bar{z}_l}
    \nonumber \\
    &= \frac{\partial z_i}{\partial \bar{z}_j} (J_c)_{j l}
    \frac{\partial H}{\partial z_k} \frac{\partial z_k}{\partial \bar{z}_l}
    \nonumber \\
    &= J_{ij} \frac{\partial H(z)}{\partial z_j}
   \label{eq:can2}
\end{align}
Therefore, $J= \Gamma^{-1}$ is related to $J_c$
by
\begin{align}
  J_{ik}= \frac{\partial z_i}{\partial \bar{z}_j } (J_c)_{j l}
    \frac{\partial z_k}{\partial \bar{z}_l}
   \label{eq:can3}
\end{align}
Therefore, the phase space integral is given by
\begin{align}
    \int \prod_{i = 1}^6 d \bar{z}_i =
    \int  \prod_{j=1}^6 d z_j  \frac{\partial(\bar{z})}{\partial(z)} 
    \label{eq:can4}
\end{align}
where the Jacobian 
\begin{align}
    \frac{\partial(\bar{z})}{\partial (z)} = {\rm det} \biggl( \frac{\partial \bar{z}_i}{\partial z_j} \biggr) 
    \label{eq:Jac}
\end{align}
On the other hand, taking the determinant of eq.(\ref{eq:can3}),
we obtain 
\begin{align}
    {\rm det}J = 
    \frac{\partial( z)}{\partial (\bar{z})} {\rm det}J_c \frac{\partial( z)}{\partial (\bar{z})} = \biggl(\frac{\partial( z)}{\partial (\bar{z})} \biggr)^2 
    \label{eq:Jac2}
\end{align}
because ${\rm det}J_c = 1$.
Therefore, the Jacobian 
${ \cal J}=\frac{\partial(\bar{z})}{\partial(z)}$
is given by
\begin{align}
    {\cal J} =  
    \biggl(\frac{\partial( z)}{\partial (\bar{z})} \biggr)^{-1} 
    = (\sqrt{{\rm det} J})^{-1} = \sqrt{{\rm det} \Gamma}
    \label{eq:Jac3}
\end{align}

Explicitly, eq.(\ref{eq:Lag2}) is given by 
\begin{align}
    \dot{x}_c &= \frac{\partial \varepsilon}{\partial k_c}
      - (\Omega_{k x} \cdot \dot{x}_c + \Omega_{k k} \cdot \dot{k}_c )
      + \Omega_{tk}   \nonumber \\
      \dot{k}_c &= - \frac{\partial \varepsilon}{\partial x_c}
      + (\Omega_{x x} \cdot \dot{x}_c + \Omega_{x k} \cdot \dot{k}_c )
      - \Omega_{tx} 
      \label{eq:SN}
\end{align}

Equations (\ref{eq:SN}) contain a lot of physics and can be the
starting point of various nontrivial physical phenomena.
One example is the Hall effect which originates from 
$\Omega_{k_1 k_2}$, i.e., the Berry curvature in momentum space.
A heuristic argument to interpret eq.(\ref{eq:SN}) is given using eq.
(\ref{eq:xc}). Namely the position operator $\hat{x}$ is roughly
given by the derivative with respect to $k$, i.e., $i \partial/\partial k$ 
in the space of $a(k)$, while the gauge covariance requires the 
connection $A(k)$. Therefore, $x_\mu$ and $x_\nu$ are not commuting anymore due to the $k$ dependence of $A(k)$.
For example, 
\begin{align}
    [x_1, x_2] &= \biggl[i \frac{\partial}{\partial k_1} + A_1(k) , 
    i \frac{\partial}{\partial k_2}  + A_2(k) \biggr] \nonumber \\
    &= i \biggl(\frac{\partial A_2}{\partial k_1} 
     - \frac{\partial A_2}{\partial k_1} \biggr) = i \Omega_{k_1 k_2}, 
\end{align}
which leads to the Heisenberg equation 
\begin{align}
 \frac{ d x_1}{dt} &= \frac{1}{i \hbar} [x_1, H(x,p)] \nonumber \\ 
 &= \frac{1}{\hbar} \frac{\partial H}{\partial k_1} 
    + \frac{1}{i \hbar} [x_1, x_2] \frac{\partial H}{\partial x_2}
    \nonumber \\
 &= v_1(k)  
    - \frac{1}{\hbar} \Omega_{k_1 k_2} \frac{\partial H}{\partial x_2}.  
    \label{eq:eom}
\end{align}
The first term $v_1(k)$ in the right hand side of eq.(\ref{eq:eom}) is the 
group velocity of the wavepacket, while the second term is called
anomalous velocity.
Note that this second term can be finite when integrated over the
occupied state in thermal equilibrium, in sharp contrast to 
the first term.
For example, when some bands are totally occupied in the whole
first Brillouin zone, i.e., band insulators, usually there is no
current because the integral of the group velocity
$\frac{1}{\hbar} \frac{\partial H}{\partial k_1}$ over the first Brillouin zone
vanished due to the periodicity of the energy dispersion $\varepsilon_{k}$.
However, the second term is given by the integral of 
$\Omega_{k_1 k_2}$ over the first Brillouin zone, which results in 
the TKNN formula to the Hall conductivity \cite{PhysRevLett.49.405,KOHMOTO1985343}
\begin{align}
 \sigma_H  = \frac{e^2}{\hbar} \int_{\rm 1st BZ} \frac{d^2 k}{(2 \pi)^2} 
\Omega_{k_1 k_2}
    \label{eq:TKNN}
\end{align}
By using the Stokes theorem, 
\begin{align}
\int_{\rm 1st BZ} \frac{d^2 k}{(2 \pi)^2} \Omega_{k_1 k_2}
= \frac{1}{(2 \pi)^2} \int_{C = \partial \rm 1st BZ} d k \cdot A(k)
\end{align}
is the contour integral along the boundary of the first Brillouin zone
and is quantized as $n/(2 \pi)$ ($n$: integer called Chern number).
This connects the topological properties of the Bloch wavefunction
in momentum space to the physical response, i.e., Hall effect,
providing the theoretical basis for its quantization, i.e.,
\begin{align}
 \sigma_H  = \frac{e^2}{h} n.
\end{align}

Next, we focus on the term with $\Omega_{tk}$ in eq.(\ref{eq:SN}).
Suppose we have the adiabatic change of the atomic positions
from the centrosymmetric ones to the noncentrosymmetric one during
the long time period $T$ in a band insulator.
Let the displacements parametrized by $\lambda(t)$. 
Therefore, the integrated total current during this adiabatic change,
i.e., the change in the polarization $\Delta P$ is
given by \cite{PhysRevB.47.1651,RevModPhys.66.899,Vanderbilt_2018}(consider here one-dimensional case)
\begin{align}
\Delta P &= - e \int_0^T d t  \int_{\rm 1st BZ} \frac{ dk}{ 2 \pi} \dot{x}_c 
\nonumber \\
&= - e \int_0^T d t \int_{\rm 1st BZ} \frac{ dk}{ 2 \pi} \Omega_{tk} \nonumber \\  
&= - e \int_0^T d t \int_{\rm 1st BZ} \frac{ dk}{ 2 \pi} \frac{d \lambda}{d t} 
\Omega_{\lambda k} \nonumber \\  
&= - e \int_0^\lambda \int_{\rm 1st BZ} \frac{ dk}{ 2 \pi} 
\Omega_{\lambda k}  
\end{align}

Writing $\Omega_{k_i k_j} = \varepsilon_{ijl} b_l$, 
$\Omega_{x_i x_j} = \varepsilon_{ijl} \frac{e}{c} B_l$,
$\frac{\partial \varepsilon}{\partial x_c} = e E$
and assuming $\Omega_{x_i, k_j} = \Omega_{k_i, x_j}=0$, 
eq.(\ref{eq:SN}) can be rewritten as
\begin{align}
    \dot{x}_c &= \biggl( 1 + e B \cdot b \biggr)^{-1}
    \biggl[ v(k) + e E \times b + e ( b \cdot v(k)) B \biggr]
    \nonumber \\
     \dot{k}_c &= \biggl( 1 + e B \cdot b \biggr)^{-1}
    \biggl[ e E + e ( v(k) \times B )
    + e^2 ( E \cdot B) b \biggr].
    \label{eq:SN2}
\end{align}
Next, we calculate the Jacobian ${\cal J}$ discussed in 
eq.(\ref{eq:Jac3})
\cite{RevModPhys.82.1959,PhysRevB.75.035114,xiao2005berry}.
\begin{align}
    {\cal J}^2 &= {\rm det}( K+ \Omega) 
    = {\det}
    \begin{pmatrix}
   \Omega_{xx} & -I \\
   I & \Omega_{kk}
\end{pmatrix}
 \nonumber \\
 &= {\det}
    \begin{pmatrix}
   \Omega_{xx} & -I \\
   I & \Omega_{kk}
\end{pmatrix}
  \begin{pmatrix}
   I & \Omega_{xx}^{-1} \\
   0 & I
\end{pmatrix}
    \nonumber \\
 &= {\det}
    \begin{pmatrix}
   \Omega_{xx} & 0 \\
   I & \Omega_{xx}^{-1} + \Omega_{kk}
\end{pmatrix}
    \nonumber \\
 &= {\det}
[\Omega_{xx}( \Omega_{xx}^{-1} + \Omega_{kk} )]
   \nonumber \\
 &= {\det}[1 + \Omega_{xx}\Omega_{kk}].
    \label{eq:Jac4}
\end{align}
Here the matrix element of $1 + \Omega_{xx}\Omega_{kk}$ is
\begin{align}
    &[1 + \Omega_{xx}\Omega_{kk}]_{ij} = \delta_{ij} + \sum_l \Omega_{x_i x_l}
     \Omega_{k_l k_j}
     \nonumber \\
     &= \delta_{ij} - e \sum_{l,m,n} \epsilon_{i l m} B_m \epsilon_{l j n} b_n 
     \nonumber \\
     &= \delta_{ij} + e \delta_{ij} B \cdot b - e B_i b_j
     \nonumber \\
     &= (1 + e B \cdot b) [ 1 - c  ]_{ij}
\end{align}
where the matrix $c$ is defined by the last line of the equation.
Now we use the formula $\ln {\rm det} M = {\rm Tr} \ln M$ for a given matrix $M$,
one obtains 
\begin{align}
    \ln {\rm det}[ 1 + \Omega_{xx} \Omega_{kk}] = 
    3\ln( 1 + e B \cdot b) + {\rm Tr} \ln[ 1 -c]. 
\end{align}
Here 
\begin{align}
     {\rm Tr} \ln[ 1 -c] = -\sum_{n=1}^\infty \frac{1}{n}  {\rm Tr}[c^n] 
\end{align}
while one can easily confirm 
\begin{align}
     {\rm Tr} [c^n] = (1 + e B \cdot b)^{-n} e^n ( B \cdot b)^n  
\end{align}
and hence
\begin{align}
     {\rm Tr} \ln[ 1 -c] = \ln \biggl[ 1 - \frac{e B \cdot b}{ 1 + e B \cdot b}
     \biggr] = - \ln [ 1 + e B \cdot b]
\end{align}
Therefore,
    $\ln {\rm det}[ 1 + \Omega_{xx} \Omega_{kk}] = 
    2\ln( 1 + e B \cdot b)$
and 
\begin{align}
 {\cal J}=    1 + e B \cdot b 
\end{align}
With this density of states and eq.(\ref{eq:SN2}), one can construct the
Boltzmann transport theory to analyze the linear and nonlinear 
transport phenomena \cite{PhysRevB.88.104412}.

\subsection{Wannier Function and Its Localization}

Now we consider the Wannier function $| w_{n R} \rangle$ defined by 
\cite{Vanderbilt_2018}
\begin{align}
| w_{n R} \rangle  = \frac{1}{\sqrt{N}} \sum_k e^{-i k \cdot R} | \psi_{n k} \rangle.
\label{eq:Wan}
\end{align}
One obtains from eq.(\ref{eq:Wan}) 
\begin{align}
| \psi_{n k} \rangle =  \frac{1}{\sqrt{N}} \sum_R  e^{i k \cdot R} | w_{n R} \rangle. 
\label{eq:Wan2}
\end{align}
Correspondingly, we can define the Fourier transformation of energy
\begin{align}
 \varepsilon_{n R}   = \frac{1}{N} \sum_k  e^{-i k \cdot R} \varepsilon_{n k},  
\label{eq:dis}
\end{align}
and 
\begin{align}
\varepsilon_{n k} = \sum_R  e^{i k \cdot R} \varepsilon_{n R}.
\end{align}
and also
\begin{align}
 A_{n R}   = \frac{1}{N} \sum_k e^{-i k \cdot R} A_{n k},  
\label{eq:dis}
\end{align}
and 
\begin{align}
A_{n k} = \sum_R  e^{i k \cdot R} A_{n R}.
\end{align}
There are several properties of Wannier function $|w_{n R} \rangle$.
First, 
\begin{align}
    \langle x |w_{n R} \rangle = \langle x - R |w_{n 0} \rangle,  
\end{align}
and 
\begin{align}
   | \langle x |w_{n R} \rangle | \to 0,  \ \ \ \  {\rm as } 
   \ \ \   |x - R | \to \infty.  
\end{align}
Second, the inner product is given by 
\begin{align}
   &\langle w_{n R} | w_{m R'} \rangle = 
    \frac{1}{N} \sum_{k, k'}  e^{i k \cdot R}  e^{-i k' \cdot R'} 
    \langle \psi_{n k} | \psi_{m k'} \rangle  \nonumber \\
    &=  \frac{1}{N} \sum_{k, k'} \delta_{n,m} \delta_{k,k'} e^{i k \cdot (R- R')} 
     = \delta_{R,R'}\delta_{n,m}
\end{align}

In addition, 
\begin{align}
& \langle w_{n 0} | \hat{x} | w_{n R} \rangle   \nonumber \\
&= \frac{1}{N} \sum_{k, k'}  e^{- i k' \cdot R}  
 \biggl[ i  \frac{\partial \delta_{k,k'}}{\partial k} 
    + i  \langle u_{n k} |  \frac{\partial}{\partial k'}u_{n k'} \rangle \delta_{k,k'}  
 \biggr]
 \nonumber \\
&= \frac{1}{N} \sum_{k, k'}  e^{- i k \cdot R}  
 \biggl[ i  \frac{\partial \delta_{k,k'}}{\partial k} 
    + i  \langle u_{n k} |  \frac{\partial}{\partial k'}u_{n k'} \rangle \delta_{k,k'}  
 \biggr]
 \nonumber \\
 &= \frac{1}{N} \sum_{k} e^{- i k \cdot R} A_{n,k} = A_{n R}
 \label{eq:AR}
\end{align}
and
\begin{align}
& \langle w_{n 0} | \hat{H} | w_{m R} \rangle   \nonumber \\
&= \frac{1}{N} \sum_{k, k'}  e^{- i k' \cdot R}  
 \varepsilon_{n k} \delta_{k,k'} \delta_{n,m}  
 \nonumber \\
 &= \frac{1}{N} \sum_{k} e^{- i k \cdot R} \varepsilon_{nk} \delta_{n,m} 
 = \varepsilon_{n R} \delta_{n,m}.
 \label{eq:ER}
\end{align}

From eq.(\ref{eq:AR})
\begin{align}
\bar{x}_n = 
\langle w_{n 0} | \hat{x} | w_{n 0} \rangle = \frac{1}{N} \sum_{k} A_{nk} = A_{n R=0}
 \label{eq:AR2}
\end{align}
Here some remarks are in order on the gauge dependence of $\bar{x}$.
When $| u_{n k} \rangle $ is transformed to  
 $| \tilde{u}_{n k} \rangle = e^{i \beta_k} | u_{n k} \rangle  $,
 the corresponding $A_{n k}$ is transformed to 
 $\tilde{A}_{n k} = A_{n k} + \nabla_k \beta_k $.
Because the Bloch wavefunction must be periodic 
for $k$ with period $G$ ($G$: reciprocal lattice vector),
$e^{i \beta_k} = e^{i \beta_{k+G}} $, and 
$\beta_k = \beta^p_k + 2 \pi k \cdot R$ where $\beta^p_k= \beta^p_{k+G}$ 
is a periodic function. It is seen that the part $2 \pi k \cdot R$ corresponds to 
relabeling $w_{n R'}$ to $w_{n R'+R}$, and does not change the set of
Wannier function. Furthermore, the periodic part 
$\beta^p_k$ does not modify $\bar{x}_n$ because
\begin{align}
\bar{\tilde{x}}_n = 
 \frac{1}{N} \sum_{k} (A_{nk} + \nabla_k \beta^p_k ) =  \frac{1}{N} \sum_{k} A_{n,k}
 = \bar{x}_n
 \label{eq:AR3}
\end{align}
Therefore, one can regard $\bar{x}_n$ as gauge invariant.
Now let us proceed to $\langle w_{n 0} | \hat{x}^2 | w_{n 0} \rangle$.
\begin{align}
& \langle w_{n 0} | \hat{x}^2 | w_{n 0} \rangle   \nonumber \\
&= \frac{1}{N} \sum_{k, k'}  \int d^d x  
x^2 u_{nk}^*(x) u_{nk'}(x)  e^{- i k \cdot x}   e^{ i k' \cdot x}  
 \nonumber \\
 &= \frac{1}{N} \sum_{k,k'} \int d^d x   u_{nk}^*(x) u_{nk'}(x) 
 \frac{\partial e^{- i k \cdot x}}{\partial k} 
 \frac{\partial e^{ i k' \cdot x}}{\partial k'} 
 \nonumber \\
 &= \frac{1}{N} \sum_{k,k'} \int d^d x   
 \frac{\partial u_{nk}^*(x)}{\partial k} 
 \frac{\partial u_{nk'}(x)}{\partial k'} e^{i (k'-k) \cdot x}
 \nonumber \\
 &= \frac{1}{N} \sum_{k} \int d^d x  
 \frac{\partial u_{nk}^*(x)}{\partial k} 
 \frac{\partial u_{nk}(x)}{\partial k} 
\nonumber \\
&= \frac{1}{N} \sum_{k} 
  \biggl\langle \frac{\partial u_{nk}}{\partial k} \bigg| 
 \frac{\partial u_{nk}}{\partial k} \biggr\rangle 
\nonumber \\
 &= \frac{1}{N} \sum_{k} \biggl[ 
  \biggl\langle \frac{\partial u_{nk}}{\partial k} \bigg| (1 - P) \bigg|
 \frac{\partial u_{nk}}{\partial k} \biggr\rangle
  \nonumber \\
  &+ \biggl\langle \frac{\partial u_{nk}}{\partial k} \bigg| u_{nk} \biggr\rangle
 \biggl\langle u_{nk} \bigg|  \frac{\partial u_{nk}}{\partial k} \biggr\rangle \biggr]
 \nonumber \\
  &= \frac{1}{N} \sum_{k} \sum_{i=1}^d [ g_{n, ii}(k) + (A^i_{nk})^2 ]
  \label{eq:ER2}
\end{align}

 Therefore, the quantity $\Omega_n$ characterizing the extent of the Wannier function
is given by \cite{PhysRevB.56.12847,RevModPhys.84.1419}
\begin{align}
&\Sigma_n = \langle w_{n0} | \Delta\hat{x}^2 | w_{n 0} \rangle = 
\langle w_{n0} | \hat{x}^2 | w_{n 0} \rangle -
\langle w_{n0} | \hat{x}| w_{n 0} \rangle^2 
\nonumber \\
&= \frac{1}{N} \sum_{k} \sum_{i=1}^d[g_{n, ii}(k) + (A^i_{nk}- \bar{A^i}_n)^2 ]
\nonumber \\
 &= \tilde{\Sigma}_n + \frac{1}{N} \sum_{k} \sum_{i=1}^d (A^i_{nk}- \bar{A^i}_n)^2 
  \label{eq:ER}
\end{align}
where the first term $\tilde{\Omega}_n = {\rm Tr}[g_{n}]$ is 
gauge invariant and is given by the quantum metric tensor, while
the second term is gauge dependent, i.e., depends on the phase choice of the
Bloch function to construct the Wannier function. 
It is noted that the lower bound for the square of the real 
space extend of the Wannier function $\Omega_n$, i.e.,
$\tilde{\Omega}_n$, is given by quantum metric tensor which  
characterizes the uncertainty principle and also quantum fluctuation.

Here we consider an application of the non-negative nature of 
eq.(\ref{eq:positive})
to the Bloch wavefunctions in two-dimensions \cite{PhysRevB.104.045103}.
Since $G_{ij}(k)$ is a hermitian matrix, it has 
real eigenvalues $\lambda_i$ ($i = 1,2$, which satisfies the equation 
\begin{align}
 (\lambda - g_{11}(k) )(\lambda - g_{22}(k) ) - g_{12}^2 - b(k)^2/4 =0    
\end{align}
where $b(k)= F_{12}$ is the Berry curvature.
The condition that both $\lambda_1$ and $\lambda_2$ are nonnegative 
is given by 
\begin{align}
 &g_{11}(k) + g_{22}(k) \ge 0 
 \nonumber \\
 &g_{11}(k)g_{22}(k)- g_{12}(k)^2 \ge b(k)^2/4.  
\end{align}
One can also integrate eq.(\ref{eq:positive}) with respect to 
$k$ in the 1st BZ.
Then 
\begin{align}
 {\rm det} M^R \ge C^2/4.  
\end{align}
where 
\begin{align}
 C = \int_{\rm 1st \  BZ} \frac{d^2 k}{2 \pi} b(k)   
\end{align}
is the Chern number, while
\begin{align}
 M^R_{ij} = \int_{\rm 1st \  BZ} \frac{d^2 k}{2 \pi} g_{ij}(k)   
\end{align}
The extent of the Wannier function is limited by 
${\rm tr} M^R = M^R_{11}+ M^R_{22}$ from below, which satisfies 
\begin{align}
 ({\rm tr}M^R)^2 \ge 4 M^R_{11} M^R_{22} \ge 4 (M^R_{12})^2 + C^2   
\end{align}

\subsection{Linear Response and Quantum Geometry}

Let us start with the linear response theory 
\cite{doi:10.1143/JPSJ.12.570,komissarov2024quantum,
verma2025instantaneousresponsequantumgeometry,verma2025quantumgeometryrevisitingelectronic}
for the conductivity
of the non-interacting electrons, i.e., Bloch waves, 
which is given by
\begin{align}
    &\sigma_{\mu \nu} (\omega) = \frac{ine^2}{m\omega_+}\delta_{\mu \nu}
    + \frac{1}{\omega_+} \int_0^t d t e^{i \omega_+ t} \langle 
    [ \hat{J}_\mu (t),  \hat{J}_\nu (0) ] \rangle
    \nonumber \\
    &=  \frac{ine^2}{m\omega_+}\delta_{\mu \nu} 
     + \frac{i}{\omega_+} \sum_{n,m} \int (dk) 
       \frac{f_{nm} (J_\mu)_{nm} (J_\nu)_{mn} }{\omega_+ + \omega_{nm}}     
       \label{eq:Kubo}
       \end{align}
where we put $\hbar=1$, $\omega_+ = \omega + i \delta$ ($\delta$: 
infinitesimal positive number),
and $n,m$ represent both the band index and the crystal wavevector $k$.
$(dk)= d^d k/(2\pi)^d$ is related to $[ dk]$ introduced before
by $[dk] = V_{\rm cell} (dk)$.

$f_{nm}= f(\varepsilon_n)-f(\varepsilon_m)$ and $\omega_{nm} = \varepsilon_n -\varepsilon_m$.
The matrix element of the current $\hat{J}$ is related to that of the position operator
$\hat{x}$ as
$(J_\mu)_{nm} = - i e \omega_{nm} (x)_{nm}$. 
The first term in the right hand side of eq.(\ref{eq:Kubo}) is 
the diamagnetic response while the second one is the paramagnetic response.
The gauge invariance is an important issue relating the diamagnetic and paramagnetic 
terms.
To show that, let us calculate the following integral:
\begin{align}
     & \sum_{n,m} \int (dk) 
       \frac{f_{nm} (J_\mu)_{nm} (J_\nu)_{mn} }{\omega_{nm}}     
       \nonumber \\
       &= e^2 \sum_{n,m} \int (dk) 
       \frac{f_{nm} (v_\mu)_{nm} (v_\nu)_{mn} }{\omega_{nm}}     
       \nonumber \\
       &= i e^2 \sum_{n,m} \int (dk) 
       f_{nm} (x_\mu)_{nm} (v_\nu)_{mn}      
       \nonumber \\
       &= i e^2 \sum_{n,m} \int (dk) 
       f_{n}[  (x_\mu)_{nm} (v_\nu)_{mn} - (x_\mu)_{mn} (v_\nu)_{nm} ]        
       \nonumber \\
        &= i \frac{e^2}{m} \sum_{n} \int (dk) 
       f_{n}  ([x_\mu, p_\nu ] )_{nn} 
       \nonumber \\
         &= -\frac{e^2}{m} \sum_{n} \int (dk)  f_{n}  \delta_{\mu \nu} = - \frac{n e^2}{m}
       \label{eq:sum}
       \end{align}
Therefore, eq.(\ref{eq:Kubo}) can be rewritten as 
\begin{align}
    &\sigma_{\mu \nu} (\omega) = i \sum_{n,m} \int (dk) 
       \frac{f_{nm}}{\omega_{nm}} (J_\mu)_{nm} (J_\nu)_{mn} 
       \nonumber \\
       &\times \biggl( \frac{1}{\omega_+ + \omega_{nm}} - \frac{1}{\omega_{nm}} \biggr)      
       \nonumber \\
      &= - i \sum_{n,m} \int (dk) 
       \frac{f_{nm}}{\omega_{nm}} \frac{(J_\mu)_{nm} (J_\nu)_{mn}}{\omega_+ + \omega_{nm}} 
       \nonumber \\
       &= \sigma_{\mu \nu}^{{\rm intra}} (\omega) +
       \sigma_{\mu \nu}^{\rm inter} (\omega)
       \label{eq:Kubo2}
       \end{align}

In eq.(\ref{eq:Kubo2}), the intraband contribution 
$\sigma_{\mu \nu}^{{\rm intra}} (\omega)$ from $n=m$ and 
the interband contribution $\sigma_{\mu \nu}^{\rm inter} (\omega)$
from $n \ne m$ are separated.
In the former, the factor  
$\frac{f_{nn}}{\omega_{nn}}$ is $0/0$ and needs some care.
Considering the small wavevetor difference 
between the two states, it is replaced by $d f_n/d \varepsilon_n =
- \delta(\varepsilon_n - \mu)$ where $\mu$ is the chemical potential. 

\begin{align}
    & \sigma_{\mu \nu}^{{\rm intra}} (\omega)
     = \frac{i}{\omega_+} \sum_{n} \int (dk) 
       \delta(\varepsilon_n - \mu) (J_\mu)_{nn} (J_\nu)_{nn} 
       \label{eq:Kubo3}
       \end{align}
which vanishes in insulators.
The dc limit of 
$\sigma_{\mu \nu}^{\rm inter} (\omega \to 0)$ can be 
related to the imaginary part of the QGT as
follows.
\begin{align}
    & \sigma_{\mu \nu}^{{\rm inter}} (0)
     = - i \sum_{n \ne m} \int (dk) f_{nm} 
       \frac{(J_\mu)_{nm} (J_\nu)_{mn}}{\omega_{nm}^2} 
   \nonumber \\
   &= - i e^2  \sum_{n \ne m} \int (dk) f_{nm}
       (x_\mu)_{nm} (x_\nu)_{mn}
    \nonumber \\
  &= - i e^2  \sum_{n \ne m} \int (dk)f_{n}
     (  (x_\mu)_{nm} (x_\nu)_{mn} - (x_\nu)_{nm} (x_\mu)_{mn} )
     \nonumber \\
  &= -i e^2  \sum_{n, m} \int (dk) f_{n}
     (  (x_\mu)_{nm} (x_\nu)_{mn} - (x_\nu)_{nm} (x_\mu)_{mn} )
     \nonumber \\
  &= - i e^2  \sum_{n} \int (dk) f_{n}
      ([ x_\mu, x_\nu ])_{nn} 
     \nonumber \\
      &= - i e^2  \sum_{n} \int (dk) f_{n}
       (\langle \partial_\mu n | \partial_\nu n \rangle -
       \langle \partial_\nu n | \partial_\mu n \rangle )
 \nonumber \\
  &= - e^2  \sum_{n} \int (dk) f_{n}
       F_{n, \mu \nu}        
       \label{eq:Kubo4}
       \end{align} 
where $F_{n, \mu \nu}$ is the Berry curvature of $n$-th band, 
which is the celebrated TKNN formula. 

 Now we discuss the sum rule for the conductivity.
 Let us introduce $\sigma_{\mu \nu}^{{\rm abs.}}$
 which corresponds to the absorption of energy and dissipation 
satisfying the energy conservation described by the $\delta$-function 
 $\delta(\omega + \omega_{nm})$.

\begin{align}
    &\sigma^{{\rm abs.}}_{\mu \nu} (\omega) = -\pi \sum_{n,m} \int (dk) 
       \frac{f_{nm}}{\omega_{nm}} (J_\mu)_{nm} (J_\nu)_{mn} 
       \delta(\omega+ \omega_{nm})
       \nonumber \\
      &= - i \frac{\pi e^2}{m} \sum_{n,m} \int (dk) 
       f_{nm}(x_\mu)_{nm} (p_\nu)_{mn}  \delta(\omega+ \omega_{nm}) 
       \label{eq:Kubo3}
       \end{align}
Here we used the fact 
$\frac{1}{\omega_+ + \omega_{nm}}  = 
P\frac{1}{\omega + \omega_{nm}} - i \pi \delta(\omega+ \omega_{nm})$, 
and $J_\nu = - e v_\nu = - ep_\nu/m$.

From eq.(\ref{eq:Kubo3}), one immediately obtains 
\begin{align}
    & \int_{- \infty}^{\infty} \frac{ d \omega}{2 \pi} \sigma^{{\rm abs.}}_{\mu \nu} (\omega) 
    \nonumber \\
    &= - \frac{ie^2}{2m} \sum_{n,m} \int (dk) f_{nm}(x_\mu)_{nm} (p_\nu)_{mn} 
             \nonumber \\
      &= - \frac{ ie^2}{2m} \sum_{n,m} \int (dk) 
       f_{n}[(x_\mu)_{nm} (p_\nu)_{mn} - (x_\mu)_{mn} (p_\nu)_{nm} ]   
        \nonumber \\
      &= - \frac{ ie^2}{2m} \sum_{n} \int (dk) 
       f_{n}[x_\mu, p_\nu]_{nn}
 \nonumber \\
      &= \frac{e^2}{2m} \delta_{\mu \nu} \sum_{n} \int (dk) f_{n}
\nonumber \\
      &= \frac{e^2n }{2m} \delta_{\mu \nu}, 
      \label{eq:Kubo4}
       \end{align}
which is the optical sum rule.

Furthermore, one can derive another sum rule as

 \begin{align}
    & \int_{- \infty}^{\infty} \frac{ d \omega}{2 \pi} 
    \frac{\sigma^{{\rm abs.}}_{\mu \nu} (\omega)}{\omega} 
    \nonumber \\
    &= \frac{e^2}{2m} \sum_{n,m} \int (dk) f_{nm}(x_\mu)_{nm} (x_\nu)_{mn} 
             \nonumber \\
      &=  \frac{e^2}{2m} \sum_{n \ne m} \int (dk) 
       f_{n}[(x_\mu)_{nm} (x_\nu)_{mn} - (x_\mu)_{mn} (x_\nu)_{nm} ]   
        \nonumber \\
      &=  \frac{e^2}{2m} \sum_{n} \int (dk) 
       f_{n} (\langle \partial_\mu n | \partial_\nu n\rangle - 
              \langle \partial_\mu n | n\rangle 
              \langle n | \partial_\nu n\rangle )
 \nonumber \\
      &=  \frac{e^2}{2m} G_{\mu \nu}, 
      \label{eq:sum4}
       \end{align}
which is called Souza-Wilkens-Martin (SWM) sum rule
\cite{PhysRevB.62.1666}.

The many-body formulation of the sum rule has been also discussed 
originally by Kohn \cite{PhysRev.133.A171}.
He considered the many-body Hamiltonian
given by 
\begin{align}
    H(k) = \sum_j \biggl[ \frac{1}{2} ( p_j + k)^2 + V_j \biggr]
     + U
     \label{eq:KHam}
\end{align}
where $U$ represent the many-body interaction.
The mass $m$ of the particle and $\hbar$ are set to be 1, and 
$V_j$ is the single-particle interaction.
Suppose the 
system in the shape of a ring with length $L$.
Then the many-body wavefunction satisfies the 
periodic boundary condition.
\begin{align}
    &\Phi(x_1, y_1, z_1; x_2, y_2, z_2; \cdots  ;x_j+L , y_j, z_j; \cdots  ;
    x_N, y_N, z_N )
    \nonumber \\
    &= \Phi(x_1, y_1, z_1; x_2, y_2, z_2; \cdots  ;x_j , y_j, z_j; \cdots  ;
    x_N, y_N, z_N )
    \label{eq:pbc}
\end{align}
The parameter $k$ is not the wavenumber but represents the
constant vector potential corresponding to the magnetic flux $\phi$
penetrating the center of the ring, i.e.,
\begin{align}
    k = \frac{\phi}{e L}
\end{align} 
Let $\Phi_\alpha(k)$ and $E_\alpha(k)$ be the eigenfunction and energy
eigenvalue of the Hamiltonian eq.(\ref{eq:KHam}),
Feynman-Hellmann theorem gives 
\begin{align}
   \frac{d E_\alpha(k)}{dk} &= \biggl\langle \Phi_\alpha(k) \biggl| 
   \frac{\partial H(k)}{\partial k} \biggr| \Phi_\alpha(k) \biggr\rangle
    \nonumber \\
    &= \langle \Phi_\alpha(k) | \sum_j (p_j + k) | \Phi_\alpha(k) \rangle
    \nonumber \\
    &= \langle \Phi_\alpha(k) | P  | \Phi_\alpha(k) \rangle + Nk
    \nonumber \\
    &= - V J_\alpha
\end{align} 
where $P = \sum_j p_j$ and $J_\alpha$ are the total momentum 
and current density along the $x$-direction.
From the Feynman-Hellmann theorem, one can derive
\begin{align}
   \frac{d | \alpha \rangle }{dk} &= | \alpha \rangle \langle \alpha | 
   \frac{d}{dk} | \alpha \rangle +
   \sum_{\alpha' (\ne \alpha)} 
   \frac{| \alpha' \rangle \langle \alpha' | P | \alpha \rangle  }
   { E_\alpha - E_{\alpha'} }  
\end{align} 
and hence
\begin{align}
   \frac{d^2 E_\alpha(k)}{dk^2} &= N + 
   \biggl[ \frac{d}{d k} \langle \alpha | \biggr] P | \alpha \rangle
   + \langle \alpha | P \biggl[ \frac{d}{d k} |\alpha \rangle \biggr] 
    \nonumber \\
    &= N - 2 \sum_{\alpha' (\ne \alpha)} \frac{|\langle \alpha' | P | \alpha \rangle|^2  }
   { E_{\alpha'} - E_{\alpha} }.  
\end{align} 
From the Lehman representation of Kubo formula 
for conductivity $\sigma(\omega)$,
one can see that \cite{PhysRev.133.A171}
($\sigma^{''}_\alpha(\omega)= {\rm Im}\sigma_\alpha(\omega)$)
\begin{align}
   D= {\rm lim}_{ \omega \to 0} \sigma^{''}_\alpha(\omega) 
   = - \frac{1}{ V \omega} \frac{ d^2 E_\alpha(k)}{d k^2} \biggr|_{k=0}.
   \label{eq:Kohn}
\end{align} 
When the real part ${\rm Re} \sigma(\omega)$ has the Drude part 
$D\delta(\omega)$ with $D$ being the Drude weight,
the corresponding complex $\sigma(\omega)$ contains 
\begin{align}
\sigma_{\rm Drude}(\omega)= \frac{1}{\pi} \frac{D}{- i\omega + \delta}      
\end{align}
with infinitesimal positive number $\delta$,
and corresponding ${\rm Im}\sigma_{\rm Drude}(\omega)= D/\omega$.
For ordinary metals, on the other hand, $\delta$ is replaced
by the inverse of the transport lifetime $\tau$, and
the Dude weight $D$ defined by eq.(\ref{eq:Kohn}) is 
$D=0$. 
Kohn argued that the insulator can be characterized by the
exponentially small dependence of $E_\alpha(k)$ on the length $L$,
and hence $D=0$.
SWM \cite{PhysRevB.62.1666} also discussed the 
many-body formulation of their sum-rule and
$G_{\mu \nu}$ is finite for insulators, while is it diverges for 
ideal and dirty metals. Therefore, $D$ and $G_{\mu \nu}$ give the
classification of insulators, ordinary metals, and ideal metals.

An interesting application of sum-rule is the bound of the gap
$E_g$ of the optical absorption \cite{PhysRevB.62.1666,PhysRevX.14.011052}.
From eq.(\ref{eq:sum4}),
one can derive 
\begin{align}
    & \frac{e^2}{2m} G_{\mu \mu} =
    \int_{- \infty}^{\infty} \frac{ d \omega}{2 \pi} 
    \frac{\sigma^{{\rm abs.}}_{\mu \mu} (\omega)}{\omega} 
    \nonumber \\
    & \leq \frac{1}{E_g} \int_{- \infty}^{\infty} \frac{ d \omega}{2 \pi} 
    \sigma^{{\rm abs.}}_{\mu \mu} (\omega) = \frac{e^2 n }{ 2 m E_g}
  \label{eq:ineq}
       \end{align}
Therefore,
\begin{align}
    E_g \leq \frac{n}{G_{\mu \mu}}
  \label{eq:ineq2}
\end{align}

There are other related quantities to the conductivity.
One is the noise spectrum \cite{PhysRevB.87.245103}
which is given by 
 \begin{align}
     S_{\mu \nu}(\omega) = 
     \int_{- \infty}^\infty d t e^{i \omega t} 
     \langle 0 | \hat{J}_\mu (t) \hat{J}_\nu (0) | 0 \rangle 
 \end{align}
 where $|0 \rangle$ is the ground state and $\hat{J}_\mu$ is the 
 $\mu$-component of the current.
 Following the similar procedure, one can obtain the following expression:
\begin{align}
     S_{\mu \nu}(\omega) = 2 \pi \omega^2 &\int (dk) \int (d k')
     \sum_{n,m}f_n (1 - f_m)  
     \nonumber \\
     &\times \int_{- \infty}^\infty d t e^{i \omega t} 
     \langle n | \hat{J}_\mu (t) |m \rangle
     \langle m | \hat{J}_\nu (0) | n \rangle
\nonumber \\
        = \frac{2 \pi e^2}{m^2} \sum_{n,m}f_n (1 &- f_m)  
      \omega_{nm}^2 \delta(\omega + \omega_{nm} )
      \langle n | \hat{x}_\mu |m \rangle
     \langle m | \hat{x}_\nu (0) | n \rangle.
     \nonumber \\
 \end{align}
 Therefore, 
 \begin{align}
     & \int_{- \infty}^{\infty} 
     \frac{d \omega}{2 \pi} \frac{S_{\mu \nu}(\omega)}{\omega^2}
     = \frac{e^2}{m^2} {\rm Tr} [ P \hat{x}_\mu ( 1 - P) \hat{x}_\nu P]
     = \frac{e^2}{m^2} G_{\mu \nu} 
 \end{align}

 A concept related to the linear response theory is 
 the ``quantum weight", which is given by the
 equal-time density-density correlation function $S_q$ \cite{PhysRevLett.133.206602,onishi2024quantumweight}.
 The Fourier component of the density operator $\rho_q$ is given by
 \begin{align}
 \rho_q &= \int d^d x e^{-i q \cdot x} \psi^\dagger (x) \psi(x)
 \nonumber \\
 &= \sum_{n,m} \langle u_{nk} | u_{mk+q} \rangle c^\dagger_{nk} c_{mk+q}
 \end{align}
and 
 \begin{align}
 S_q &= \frac{1}{V} \langle \rho_q \rho_{-q} \rangle 
 \nonumber \\
 &= \frac{1}{V} \sum_{n,m,k} \sum_{n',m',k'}\langle u_{nk} | u_{m k+q} \rangle 
 \langle u_{n'k'+q} | u_{m'k'} \rangle 
 \nonumber \\
 &\ \ \ \ \ \ \ \ \ \ \ \ \   \times \langle c^\dagger_{nk} c_{mk+q}
c^\dagger_{n'k'+q} c_{m'k'} \rangle
 \end{align}
For $q \ne 0$, 
\begin{align}
    &\langle c^\dagger_{nk} c_{mk+q}
c^\dagger_{n'k'+q} c_{m'k'} \rangle 
\nonumber \\
&= \delta_{n,m'} \delta_{m,n'} \delta_{k,k'}
 \langle c^\dagger_{nk} c_{nk} \rangle 
  \langle c_{mk+q} c^\dagger_{m k+q} \rangle
\nonumber \\
&= \delta_{n,m'} \delta_{m,n'} \delta_{k,k'}
 f_n(k) ( 1 - f_m(k+q)).
\end{align} 
where $f_n(k)$ is the Fermi distribution function at $T=0$ K.
Therefore, 
\begin{align}
 S_q &= \frac{1}{V} \sum_k {\rm Tr} [ P(k) ( 1 - P(k+q))] 
 \end{align}
where the projector $P(k) = \sum_n |u_{nk} \rangle f_n(k) \langle u_{nk} |$
Expanding $S_q$ up to the second order in $q$,
one obtain 
\begin{align}
 S_q &= - \frac{1}{2V} \sum_k {\rm Tr} 
 [ P(k) \partial_i \partial_j P(k)] q_i q_j
  \nonumber \\
 &= \frac{1}{2V} \sum_k {\rm Tr}[ \partial_i P(k) \partial_j P(k)] q_i q_j
 \nonumber \\
 &= \frac{1}{2V} \sum_k g_{ij}(k) q_i  q_j 
 \end{align}
where $\partial_i = \partial / \partial k_i$.
Here $P(k) ( 1 - P(k))=0$, and 
$\sum_k P(k) \partial_{i} P(k) = (1/2) \sum_k \partial_i P(k)^2 
= 0$ are used.

\subsection{Landau Levels and Fractional Quantum Hall Effect}

Let us start with the basic facts about the Landau levels.
The Hamiltonian of two-dimensional electrons under a magnetic field 
$\mathbf{B}= (0,0,B)$ $(B>0)$ is 
 \begin{align}
 H = \frac{1}{2m}(\mathbf{p} + e \mathbf{A} )^2 = 
 \frac{1}{2m}\mathbf{\pi}^2     
 \end{align}
with $\mathbf{A}$ being the vector potential satisfying
$\nabla \times \mathbf{A}= \mathbf{B}$.
From the dimensional analysis, the relevant length scale $\ell_B$ is given by
 \begin{align}
 \frac{\hbar}{\ell_B}  = e B \ell_B      
 \end{align}
where $A \sim B \ell_B$ is used.
This gives the expression for the magnetic length 
 \begin{align}
 \ell_B =  \sqrt{\frac{\hbar}{e B}}.   
 \end{align}
Correspondingly, the relevant energy scale $\hbar \omega_c$
($\omega_c$: the cyclotron frequency ) is
given by 
 \begin{align}
 \hbar \omega_c = \frac{\hbar^2}{m\ell_B^2} = \frac{\hbar e B}{m}.   
 \end{align}
 
${\bf \pi} = (\mathbf{p} + e \mathbf{A}) $
is the gauge invariant momentum, and its component 
satisfies
 \begin{align}
 [ \pi_i, \pi_j ] = - i \epsilon_{ij}
 \hbar e (\nabla \times \mathbf{A})_z =- i \epsilon_{ij}\hbar e B,     
 \end{align}
 where $\epsilon_{ij}$ is the antisymmetric tensor with 
 $\epsilon_{12}= - \epsilon_{21} =1$.
 Defining  $\pi_{\pm} = \pi_1 \pm i \pi_2$, and $a^\dagger$, $a$
 by $\pi_+ = \sqrt{2 \hbar e B} a^\dagger$, and $\pi_- = \sqrt{2 \hbar e B} a$,
 one can see that $a\dagger$ and $a$ are creation and annihilation operators
 of bosons satisfying 
  \begin{align}
 [ a, a^\dagger] = 1.     
 \end{align}
The Hamiltonian $H$ is expressed as
\begin{align}
 H = \hbar \omega_c \biggl( a^\dagger a + \frac{1}{2} \biggr),
 \end{align}
 and the energy eigenstate $| n \rangle$ satisfies
 $a^\dagger a | n \rangle = n |n \rangle$.
 However, the degeneracy of the landau levels should be considered, which 
 is taken care by the operator $\tilde{\mathbf{\pi}} = (\mathbf{p} - e \mathbf{A})$.
 In the gauge where 
 \begin{align}
 \mathbf{A}= \frac{1}{2}(\mathbf{B} \times \mathbf{x} ),     
 \end{align}
 one can easily confirm 
 \begin{align}
 [ \pi_i, \tilde{\pi}_j ] = 0,     
 \end{align}
 and 
 \begin{align}
 [ \tilde{\pi}_i, \tilde{\pi}_j ] = i \epsilon_{ij} \hbar e B,     
 \end{align}
Therefore,  
$\tilde{\mathbf{\pi}}$ does not appear in the Hamiltonian and describe
the degeneracy within each Landau level. 
Defining, $\tilde{\pi}_{\pm}$ is a similar way, and also
$\tilde{a}^\dagger$, $\tilde{a}$
by $\tilde{\pi}_+ = \sqrt{2 \hbar e B} a$, and $\tilde{\pi}_- = \sqrt{2 \hbar e B} 
a^\dagger$, these form the boson creation and annihilation operators with 
$[\tilde{a}, \tilde{a}^\dagger ] = 1$.
Therefore, the electronic states are denoted by two quantum numbers, i.e.,
the Landau level index $n$, and that for $\tilde{n} = \tilde{a}^\dagger\tilde{a}$.

In order to define the QGT for Landau levels in momentum space $k$,
we introduce the unit cell and corresponding Brillouin zone
\cite{PhysRevB.104.045103}.
Let us define the square unit cell with the size of 
$a= \sqrt{ 2 \pi} \ell_B$ with the area $a^2$.
The lattice vector $R$ is given by the 
linear combination of $a_1= (a,0)$ and $a_2= (0,a)$, 
and the translation symmetry operation is defined by 
\begin{align}
 M(R) = e^{-i \xi_R(x) } \cal{T}(R)   
 \label{eq:translation}
 \end{align}
where $\cal{T}(R) f(x) = f(x+R)$ is the translation operator and
the phase $\xi_R(x)$ satisfies the condition 
\begin{align}
 eA(x+R) = eA(x) -  \nabla \xi_R(x).   
 \label{eq:translation2}
 \end{align}
 where we put $\hbar=1$.
One can show that $M(R_1) M(R_2) = M(R_1+R_2)$ as
follows.
\begin{align}
 M(R_1) M(R_2) f(x) &= M(R_1) e^{- i \xi_{R_2} (x)} f(x+ R_2)
 \nonumber \\
 &= e^{- i \xi_{R_1} (x)} e^{- i \xi_{R_2} (x+R_1)} f(x+ R_1+ R_2)    
 \label{eq:translation3}
 \end{align}
On the other hand, 
\begin{align}
 M(R_1+R_2) f(x) = e^{- i \xi_{R_1+R_2} (x)} f(x+ R_1+ R_2).    
 \label{eq:translation4}
 \end{align}
Now we take the derivative of eq.(\ref{eq:translation2}) with respect to
$x$ 
\begin{align}
 & \nabla_x \xi_{R_1}(x) + \nabla_x \xi_{R_2}(x+R_1) 
 \nonumber \\
 &=  \frac{1}{e} (A(x+R_1) - A(x) + A(x+R_1 + R_2) - A(x+ R_1))
 \nonumber \\
  &=  \frac{1}{e} (A(x+R_1 + R_2) - A(x))
 \nonumber \\
  &= \nabla_x \xi_{R_1+R_2}(x)
 \label{eq:translation5}
 \end{align}
Integrating this equation over $x$, we obtain $M(R_1) M(R_2) = M(R_1+R_2)$.

Another important property of $M(R)$ is that it commutes with 
$p+e A(x)$.
\begin{align}
&[ p +eA(x), M(R)] = [ p + e A(x), e^{- i \xi_R(x) } e^{i R \cdot p} ]
 \nonumber \\
 &= [ p,  e^{- i \xi_R(x) } ] e^{i R \cdot p}  + 
  e  e^{- i \xi_R(x) } [ A(x),  e^{i R \cdot p} ]   
 \nonumber \\
 &= - \nabla_x \xi_R(x) e^{- i \xi_R(x) } e^{i R \cdot p} 
  + e  e^{- i \xi_R(x) } [ A(x),  e^{i R \cdot p} ] 
 \nonumber \\
 \label{eq:translation6}
 \end{align}
Here 
\begin{align}
&[ A(x),  e^{i R \cdot p} ] f(x) 
\nonumber \\
&= A(x) f(x+R) - A(x+R)f(x+R) 
\nonumber \\
&= (A(x)- A(x+R))e^{i R \cdot p} f(x), 
 \label{eq:translation7}
 \end{align}
and we conclude $[ p +eA(x), M(R)] =0$.
Therefore, the Hamiltonian $H$ and $M(r)$ are diagonalized 
simultaneously, and the eigenstate of $H$ is 
characterized by the  
eigenvalue $e~{i k \cdot R} $ of $M(R)$ as
\begin{align}
M(R) \psi_{m,k} = e^{i k \cdot R} \psi_{m,k}, 
 \label{eq:translation8}
 \end{align}
which is the analogue of Bloch wavefunction.
However, it satisfies 
\begin{align}
\psi_{m,k}(x+R)  = e^{i k \cdot R + i \xi_R(x)} \psi_{m,k}(x), 
 \label{eq:translation9}
 \end{align}
and the orthogonality property with respect to $k$ is not 
guaranteed.
However, note that $H$ and $p + eX = m v$ ($v= \dot{x}$: velocity) 
commute with $M(R)$, and hence 
their matrix elements are diagonal in $k$.
Accordingly, the matrix elements of $x$ is also diagonal in $k$.

Consider the situation where the lowest $n$-Landau levels are completely filled,
i.e., $n=0, 1, \cdots n-1$. 
The quantum geometric tensor for this ground state is 
then given by 
\begin{align}
 G_{ij} = Tr[ P \hat{x}_i (1 - P) \hat{x}_j P]   
 \label{eq:GijLL}
 \end{align}
where $i,j = 1,2$ and $P$ is the projector to the ground state.
In terms of $a^\dagger, a$ and $\tilde{a}^\dagger, \tilde{a}$
in the symmetric gauge,
the operators $\hat{x}_1$ and $\hat{x}_2$ can be expressed as
\begin{align}
\hat{x}_1 &= - i \frac{\ell_B}{\sqrt{2}}
 ( a^\dagger - a + \tilde{a}^\dagger - \tilde{a} )
\nonumber \\
 \hat{x}_2 &=  \frac{\ell_B}{\sqrt{2}}
 ( - a^\dagger - a + \tilde{a}^\dagger +\tilde{a} )
 \label{eq:GijLL2}
\end{align}
 In eq.(\ref{eq:GijLL}), $\tilde{a}^\dagger, \tilde{a}$ do not contribute
 because they do not change the Landau index. Therefore, for example, 
\begin{align}
 G_{12} 
 &= \langle n-1 | \hat{x}_1 | n \rangle \langle n | \hat{x}_2 | n -1 \rangle
 \nonumber \\
 &= \frac{\ell_B^2}{2}  \langle n-1 | -i  (a^\dagger - a) | n \rangle \langle n | 
 (- a^\dagger - a )| n -1 \rangle  
 \nonumber \\
  &= -i \frac{\ell_B^2}{2}  \langle n-1 | aa^\dagger | n -1 \rangle 
   \nonumber \\
    &= -i \frac{\ell_B^2}{2} n.   
 \label{eq:GijLL3}
 \end{align}
Other elements can be estimated in a similar way, and one obtains
\begin{align}
 G_{ij} =  \frac{\ell_B^2}{2} n ( \delta_{ij} - i \epsilon_{ij} )   
 \label{eq:GijLL4}
 \end{align}
for each $k$ (but independent of $k$).
Therefore, one can calculate the integral of Berry curvature $F_{12}(k)$ over the
1st BZ as
\begin{align}
 \biggl( \frac{2 \pi}{a} \biggr)^2 n \times \frac{\ell_B^2}{2} = \pi n, 
 \label{eq:translation9}
 \end{align}
which gives $\pi$ times the Chern number $n$.

In the Chern insulator with the finite Chern number, the quantized Hall effect is protected by the 
band gap, and the dispersion of each band does not matter
\cite{PhysRevLett.61.2015,PhysRevLett.90.206601,
doi:10.1126/science.1234414, RevModPhys.95.011002}.
On the other hand, for the fractional quantum Hall effect 
(FQHE) in the partially filled band with Chern number, the flatness of the dispersion is needed since the electron-electron interaction is essential.
It is believed that the structure of the quantum geometric tensor, akin to that in the Landau level, is preferable to realize the FQHE in Bloch bands \cite{PhysRevLett.51.605,Parameswaran:2013pca,PhysRevB.90.165139,
PhysRevB.96.165150,PhysRevResearch.2.023237,PhysRevB.108.205144,PhysRevLett.127.246403,PhysRevX.1.021014}.
This issue is related to the quantum geometric interpretation of the 
FQHE by Haldane \cite{PhysRevLett.107.116801}, which has been extended to the flat band \cite{PhysRevLett.114.236802}.

Another interesting issue is the role of quantum geometry 
of the Landau levels under a magnetic field in flat band systems
\cite{hwang2021geometric,Rhim2020,yu2025quantumgeometryquantummaterials}

\subsection{Many-Body Phenomena and Quantum Geometric Tensor}

The strong correlation effect characterizes the flat band systems because the kinetic energy is quenched, and offers an ideal platform for the quantum many-body phenomena \cite{yu2025quantumgeometryquantummaterials}. 
Superconductivity is a representative phenomenon, and the Cooper pairs in the flat band show exotic properties.
One important issue is what determines the superfluidity density of the
flat band system because the usual contribution due to the kinetic energy is vanishing there. It turned out that the quantum metric tensor contributes to the superfluidity density as demonstrated, especially for the flat-band system of twisted bilayer graphene 
\cite{torma2022superconductivity,peotta2015superfluidity,
julku2016geometric,liang2017band,hu2019geometric,
xie2020topology,julku2020superfluid}
One way to understand this contribution is based on the derivation of the Ginzburg-Landau theory \cite{chen2024ginzburg,julku2020superfluid,hu2025anomalous} taking into account the Bloch wavefunctions 
$\phi_{nk}(x) = e^{ikx} u_{nk}(x)$.

Let us start with the attractive interaction between the
electrons given by 
\begin{align}
    H_{\rm int.} = - g \int d x \psi_\uparrow^\dagger (x) 
    \psi_\downarrow^\dagger (x) \psi_\downarrow(x) \psi_\uparrow(x),
\end{align}
where the field operator $\psi_\sigma (x)$ is expressed by
\begin{align}
    \psi_\sigma(x) = \frac{1}{\sqrt{V}} \sum_{n,k} \phi_{nk}(x) c_{nk \sigma},
 \end{align}
where $V$ is the volume of the system.
Introducing the mean field decoupling to $H_{\rm int.}$,
one obtains
\begin{align}
    H_{\rm mean \ int.} = - g \int d x \psi_\uparrow^\dagger (x) 
    \psi_\downarrow^\dagger (x) \langle \psi_\downarrow(x) \psi_\uparrow(x) 
    \rangle + \  {\rm h.c.},
    \label{eq:MF}
\end{align}
where h.c. means hermitian conjugate.
Here we introduce the order parameter with wavevector $q$ as
\begin{align}
    \Delta(x) = \langle \psi_\downarrow(x) \psi_\uparrow(x) \rangle 
    = \Delta_0 e^{i q \cdot x}
 \end{align}
Then the mean field Hamiltonian eq.(\ref{eq:MF}) becomes 
\begin{align}
    &H_{\rm mean \ int.} = - g \Delta_0 \int d x e^{i q \cdot x} 
    \psi_\uparrow^\dagger (x) \psi_\downarrow^\dagger (x) 
    + {\rm h.c.}
    \nonumber \\
    &= - g \Delta_0 \sum_{n,k} 
    c^\dagger_{n -k + q/2} c^\dagger_{n k+ q/2} 
    \langle u_{n k+q/2} |u^*_{n -k+ q/2} \rangle 
    + {\rm h.c.}
    \label{eq:MF2}
\end{align}
where 
$ \langle u_{n k+q/2} |u^*_{n -k+ q/2} \rangle  = \int d x 
    u^*_{n k+ q/2}(x)  u^*_{n k+ q/2}(x) $
    is the inner product. 
When the system has the time-reversal symmetry, 
$ |u^*_{n -k+ q/2} \rangle = |u^*_{n k- q/2} \rangle $,
and we define $\Gamma(k,q) = \langle u_{n k+q/2} |u_{n k- q/2} \rangle$,
which is related to the quantum metric tensor $g_{ab}(k)$ as
\begin{align}
    | \Gamma(k,q) |^2 = 1 - \sum_{a,b}g_{ab}(k) q_a q_b  
\end{align}
for small $q$.
Therefore, the second order term in the GL expansion is 
given by 
\begin{align}
   F_2 = \sum_q (g - g^2 \chi(q) ) \Delta(q) \Delta (-q).  
\end{align}
Here the pair susceptibility which characterizes how favorable the Cooper pair formation is for a given $q$, is expessed as
\begin{align}
   \chi(q) = \frac{1}{\beta} \sum_{\omega_n} \frac{1}{N} \sum_k 
   |\Gamma(k,q)|^2 G(k+q/2) G(-k+ q/2),
\end{align}
where $G$ is the electrons Green's function.
We separate $\chi(q)$ into two contributions as
$\chi(q)  = \chi_0(q) - \chi_2(q)$  
up to the second order in $q$ where the bare susceptibility
\begin{align}
   \chi_0(q) = \frac{1}{\beta} \sum_{\omega_n} \frac{1}{N} \sum_k 
   G(k+q/2) G(-k+ q/2) 
\end{align}
and
\begin{align}
   \chi_2(q) = \frac{1}{\beta} \sum_{\omega_n} \frac{1}{N} \sum_k 
   (1-|\Gamma(k,q)|^2) G(k) G(-k). 
   \label{eq:chi2}
\end{align}
Note that the factor $1-|\Gamma(k,q)|^2$ is already $\mathcal{O}(q^2)$
and hence $q$ is dropped in the argument of $G$ 
in eq.(\ref{eq:chi2}) because we are interested in the contribution
$\propto q^2$.
The $q$-dependence of $\chi_0(q)$ comes from the energy dispersion 
of electrons, while $\chi_2(q)$ arises from the quantum metric tensor as
\begin{align}
   \chi_2(q) = \sum_{a,b} q_a q_b \frac{1}{4N} \sum_k 
   \frac{\tanh(\beta \varepsilon_n(k)/2)}{\varepsilon_n(k)} g_{ab}(k)
\end{align}

The coefficient of $q^2$ term in the GL free energy is 
nothing but the superfluidity density $\rho_s$, which enters as $\rho_s (\nabla \theta)^2$ with $\theta$ being the phase of the 
superconducting order parameter. Therefore, $\rho_s$ consists of two contributions as
\begin{align}
   \rho_s = \rho_s^{{\rm kinetic}} + \rho_s^{{\rm geometric}}, 
\end{align}
where $\rho_s^{{\rm kinetic}} $ comes from the kinetic energy or 
energy dispersion of electrons, while $\rho_s^{{\rm geometric}}$ arises
from the quantum metric tensor.

A related phenomenon to the superconductivity is the 
superfluidity of bosons 
\cite{PhysRevLett.127.170404,PhysRevB.104.144507,PhysRevA.107.023313}. 
In the flat band system, the momentum $k_c$ of the condensed bosons
is determined by the interaction and the superfluid density or
square of the velocity of the phonon is proportional to 
the quantum metric at $k_c$ \cite{PhysRevLett.127.170404}. 
An analogous consideration has been applied to exciton 
condensation \cite{PhysRevB.105.L140506,PhysRevLett.132.236001}.

Note also that the magnetic order such as the ferromagnetism 
is influenced by the quantum geometry
\cite{PhysRevB.102.165118,kang2024quantumgeometricboundsaturated}.
When the GL free energy is expanded in the magnetic moment $M(x)$,
the gradient term, i.e., $\kappa (\nabla M(x))^2$ where $\kappa$ is the spin stiffness, arises
from the kinetic term and quantum metric because the
magnetic susceptibility is given by the similar expression 
to $\chi(q)$ for the superconductivity.

More fundamentally, it has been discussed that the geometry of the Bloch state affects the electron-phonon interaction \cite{Yu2024}
and phonon-phonon interaction \cite{behnia2025phononthermalhalllattice}.

\subsection{Nonlinear Transport and Quantum Geometry}

 Nonlinear transport phenomena have attracted recent intense interest
 \cite{jiang2025revealingquantumgeometrynonlinear}.
 Especially, the nonreciprocal transport that is second order in the applied electric field $E$ is a focus of recent studies.
 The Onsager's reciprocal theorem says that the
 linear conductivity tensor $\sigma_{ij}(q, \omega)$, 
 where $q$ and $\omega$ are the wavevector and frequency of the
 electric field, should satisfy the relation
 \begin{align}
     \sigma_{ij}( q, \omega, B) = \sigma_{ji} ( -q, \omega, -B) 
 \label{eq:Onsager}
 \end{align}
 where $B$ is the magnetic field representing the time-reversal symmetry 
 breaking. This reciprocal theorem originates from the time-reversal
 symmetry of the microscopic dynamics of the system, and is applied 
 to the linear response, rather than the nonlinear response.
 However, Rikken gave a heuristic argument extending eq.(\ref{eq:Onsager}) to the nonlinear response by replacing the momentum of stimuli $q$ by the momentum of electrons or current $I$.
 Based on this idea, he proposed the empirical expression of the
 current-dependent resistance 
 \begin{align}
    R = R_0 ( 1 + \gamma I B)
 \end{align}
 where $\gamma$ quantify the strength of the non-reciprocity
 in the unit of $A^{-1} T^{-1}$.
 From the viewpoint of theory, 
 it corresponds to the second order conductivity $\sigma^{(2)}_{ijk}$
 \begin{align}
    J_i = \sigma^{(2)}_{ijl} E_j E_l. 
 \end{align}

 When the energy dispersion $\varepsilon_n(k)$ of band $n$ is given,
 the conventional Boltzmann transport equation can be solved
 iteratively with respect to the electric field. 
 \begin{align}
    \sigma^{(2)}_{ijl} = \int (dk) e^3 \tau^2 
    \frac{\partial v_{k,i}}{\partial k_j} v_{k, l} 
    \frac{\partial f_{{\rm eq.}}}{\partial \varepsilon},  
    \label{eq:Bol2}
 \end{align}
 which depends only on the energy dispersion
 because $v_{k,i} = \frac{\partial \varepsilon(k)}{\partial k_i}$.
 (Here the band index is dropped. It is understood that the
 contribution from each band is summed independently.) 
 It is always needed to break the spatial inversion symmetry $P$ 
 to have a finite $\sigma^{(2)}$.
 In addition to $P$-breaking, also the time-reversal
 symmetry ($T$) breaking, i.e., $B$, is required in 
 the expression eq.(\ref{eq:Bol2}). This can be 
 understood intuitively that the dispersion has the 
 symmetry between $k$ and $-k$ due to the
 $T$-symmetry even when the inversion symmetry $P$ is broken.
 However, this condition is not always needed and
 nonreciprocal transport is possible without $T$-breaking in 
 some cases \cite{Tokura2018}.
 This occurs when the interband matrix elements 
are considered beyond the conventional Boltzmann theory.
Note that the Berry phase and/or Berry curvature account 
for this effect in the linear response, such as the anomalous Hall 
effect, and is also relevant in the nonlinear transport.

One representative example is the nonlinear Hall effect \cite{sodemann2015quantum}.
From eq.(\ref{eq:SN2}) with $B=0$, the current density is
given by 
\begin{align}
    J= -e \dot{x}_c = - e \frac{\partial \varepsilon_n(k)}{\partial k} 
    - e^2 E \times b(k).
    \label{eq:NLH1}
\end{align}
Assuming the symmetric energy dispersion, the first term of 
eq.(\ref{eq:NLH1}) does not contribute to the second order 
current in $E$, while the second term leads to 
\begin{align}
    J_{{\rm NLH}}= \int (dk) 
    \biggl( -e \tau E \cdot \frac{\partial \varepsilon_n(k)}{\partial k}
    \biggr)
    ( - e^2 E \times b(k) )
    \label{eq:NLH2}
\end{align}
Writing the component explicitly, we
obtain \cite{sodemann2015quantum}
\begin{align}
     (J_{{\rm NLH}})_i &= \int (dk) 
    \biggl( -e \tau E_j \cdot \frac{\partial f(k)}{\partial k_j}
    \biggr)
    ( - e^2 \epsilon_{i l m} E_l b_m(k) )
    \nonumber \\
    &= - e^3 \tau 
    \int (dk) \epsilon_{i l m}  
    f(k) \frac{\partial b_m(k)}{\partial k_j} E_j E_l
    \label{eq:NLH3}
\end{align}
Compared with eq.(\ref{eq:Bol2}), this is a sub-leading order effect, i.e.,
first order in $\tau$. The Berry curvature dipole 
$\frac{\partial b_m(k)}{\partial k_j}$ integrated over the 
occupied states  in the equilibrium distribution determines the
current perpendicular to the electric field, and is called the nonlinear
Hall effect.

When one considers the contribution which is $\tau$-independent,
the quantum metric tensor appears.
This comes from the correction of the Berry connection $a_{n \mu}$
due to the electric field $E$.
The perturbative Hamiltonian $H_E$ due to the electric field is
given by \cite{PhysRevLett.112.166601,10.21468/SciPostPhys.14.4.082,
holder2021mixedaxialgravitationalanomalyemergent}
\begin{align}
    H_E = e E_a x_a.
\end{align}
Correspondingly, the periodic part of the Bloch wavefunction 
is modified in first order in $E$ as
\begin{align}
   |\tilde{u}_{nk} \rangle =  |u_{nk} \rangle + 
  \sum_{m (\ne n)} \frac{| u_{mk} \rangle \langle u_{mk} | H' | u_{nk} \rangle }
  {\varepsilon_{nm}(k)}
   \end{align}
where $\varepsilon_{nm}(k) = \varepsilon_{n}(k)-\varepsilon_{m}(k)$.
The corresponding correction to the Berry connection $a^1_{n a}$
is given by
\begin{align}
 a^1_{n a} = - e \sum_{m (\ne n)}  
  2 {\rm Re} \biggl( \frac{ (v_a)_{nm} (v_b)_{mn} }
  {\varepsilon_{nm}^3} \biggr) E_b  = - e G^n_{ab} E_b
   \end{align}
where $v_a = \dot{x_a}$. 
This tensor $G^n_{ab}$ is not identical to the metric tensor, while
for the two-band model with $H = \vec{h}(k) \cdot \vec{\sigma}$,
one can show 
$G_{ab}$ for the ground state is 
given by \cite{PhysRevLett.112.166601,PhysRevLett.122.227402}
\begin{align}
 G_{ab} = - \frac{1}{4 h(k)} \partial_a \vec{n}(k) \cdot \partial_b \vec{n}(k)
    = - \frac{1}{h(k)} g_{ab} (k)
   \end{align}
 where $h(k) = |\vec{h}(k)|$, $\partial_a = \partial/\partial k_a$ and $g_{ab}(k)$ is the metric tensor for this 2-band model. 
 Dropping the band index $n$, the correction to the current 
 due to $b^1 (k) = \nabla_k \times a^1(k)$ is obtained as
 \begin{align}
 J^1_{i} &= e^2 \int (dk) ( E \times b^1(k) )_i f(k) 
 \nonumber \\
 &= e^2 \int (dk) \epsilon_{ijk} E_j b_k(k) f(k)
  \nonumber \\
&= e^2 \int (dk) \epsilon_{ijk} \epsilon_{k l m} (\partial_l a^1_m) f(k)  E_j 
  \nonumber \\
  &= -e^2 \int (dk) \epsilon_{ijk} \epsilon_{k l m} a^1_m(\partial_l  f(k))  E_j 
  \nonumber \\
  &= e^3 \int (dk) \epsilon_{ijk} \epsilon_{k l m} G_{mp} v_l (k) 
  \frac{\partial f(k)}{\partial \varepsilon}  E_j 
  \nonumber \\
  &= e^3 \int (dk) \epsilon_{ijl} 
   ( v_i(k) G_{j l} (k) - v_j(k) G_{i l}(k) ) 
  \frac{\partial f(k)}{\partial \varepsilon}  E_j E_l 
\end{align}
where the identity
\begin{align}
   \sum_k \epsilon_{ijk} \epsilon_{k l m} = 
  \delta_{i l} \delta_{j m} - \delta_{i m} \delta_{j l} 
\end{align}
is used.

\subsection{Riemannian Geometry in Nonlinear Optical Responses} 

\begin{figure}[htb]
\begin{center}
\includegraphics[width =3.1in]{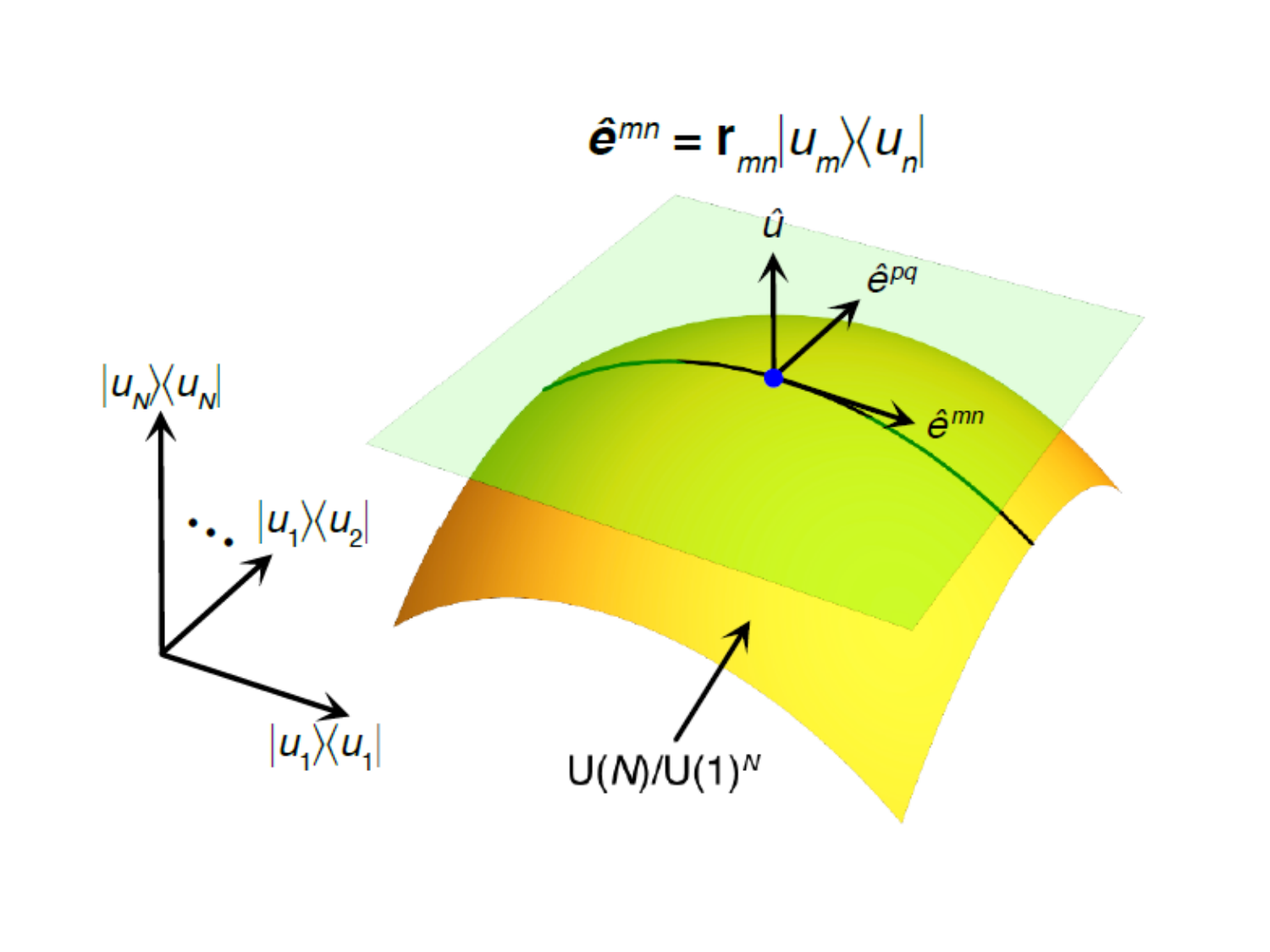} 
\caption{
{\bf Geometry of the pair of Bloch states for resonant optical transitions.}
The matrix element of dipole moment regarded as tangent basis vectors
$\hat{e}^{pq}$s and $\hat{e}^{mn}$ on the manifold 
${\rm U}(N)/{\rm U}(1)^N$ with $\hat{u}$ being the normal vectors.
Here, ${\rm U}(1)^N$ corresponds to the gauge degrees of freedom
of Bloch states. Reproduced from \cite{Ahn2022}.
}
\label{fig:geometry}
\end{center}
\end{figure}

The quantum geometric tensor discussed above is defined for each quantum state,
while the pair of states is relevant to the optical transitions, i.e.,
the initial and final states. Especially when the resonant optical 
process is considered,  the matrix element of the
transition dipole moment $r^a$ along the polarization $a$
between $m$ and $n$ states is given by 
\begin{align}
r^a_{mn}(k)=\delta_{mn}i\partial_a 
+\langle u_{m k}|i\partial_a|u_{n k} \rangle,
\end{align}
and the Riemannian geometry of quantum states
can be defined by identifying tangent vectors as (See Fig.\ref{fig:geometry})
\begin{align}
\hat{e}^{mn}_a(k)\equiv r^a_{mn}(k) |u_{m k} \rangle 
\langle u_{n k}|, 
\quad a=1,\dots,d.
\end{align}
We define the inner product of the tangent basis vectors by
\begin{align}
\label{eq:optical-metric}
Q^{mn}_{ba}
\equiv (\hat{e}^{mn}_b,\hat{e}^{mn}_a)
=r^b_{nm}r^a_{mn}.
\end{align}
The parallel component of the derivative on $\hat{e}^{mn}_a$
is given by 
\begin{align}
 \partial_c\hat{e}^{mn}_a|_{\rm parallel}=\sum_b(C^{mn})^b_{ca}\hat{e}^{mn}_b   
\end{align}
and the covariant derivative $\nabla$ is defined by
\begin{align}
    \nabla_c\hat{e}^{mn}_a=\sum_b(C^{mn})^b_{ca}e_b
\end{align}
with 
\begin{align}
C^{mn}_{bca}\equiv 
\sum_eQ^{mn}_{be}(C^{mn})^e_{ca}
=(\hat{e}^{mn}_b,\nabla_c\hat{e}^{mn}_a)
=r^b_{nm}r^a_{mn,c}.
\end{align}
Here we introduced the generalized derivative \cite{PhysRevB.52.14636} 
\begin{align}
 O_{mn,c}\equiv \partial_cO_{mn}-i[{\cal A}_c,O]_{mn}   
\end{align}
where $({\cal A}_c)_{mn}=\delta_{mn}\langle u_m|i\partial_c|u_n \rangle$.
The Riemann curvature tensor is defined by 
\begin{align}
K^{mn}_{badc}
\equiv (\hat{e}^{mn}_b,(\nabla_d\nabla_c-\nabla_c\nabla_d)\hat{e}^{mn}_a)
=-ir^{b}_{nm}[{\cal F}_{dc},r^{a}]_{mn},
\end{align}
where $({\cal F}_{dc})_{pq}=
\partial_d({\cal A}_c)_{pq}-\partial_c({\cal A}_d)_{pq}$ is the 
${\rm U}(1)^N$ Berry curvature.
In terms of these Riemannian geometry language, 
the linear optical conductivity tensor is given by
\begin{align}
\sigma^{b;a}
=\frac{\pi \omega e^2}{h}\sum_{m,n}\int_{\bf k} \delta\left(\omega-\omega_{mn}\right)f_{nm}Q^{mn}_{ba},
\end{align}
where $\int_{\bf k}=\int d^dk/(2\pi)^d$, $\hbar\omega_{mn}$ is the energy difference between $m$ and $n$ bands, and $f_{nm}=f_n-f_m$ is the difference between the Fermi-Dirac distribution of the $n$ and $m$ states.
Furthermore, the second order responses are expressed as follows.
The shift current conductivity is given by 
\begin{align}
\sigma^{c;ab}_{\rm shift}
=-\frac{\pi e^3}{2\hbar^2}\sum_{m,n}\int_{\bf k} \delta\left(\omega-\omega_{mn}\right)f_{nm}i (C^{mn}_{bca}-(C^{mn}_{acb})^*),
\end{align}
while the injection current conductivity is 
\begin{align}
\sigma^{c;ab}_{\rm inj}=-\frac{\pi e^3}{\hbar^2\Gamma}\sum_{m,n}\int_{\bf k} 
\delta\left(\omega-\omega_{mn}\right)f_{nm}Q^{mn}_{ba}(v^c_{mm}-v^c_{nn}),
\end{align}
where $v^c_{mm}$ and $v^c_{nn}$ are the group velocities of bands $m$ and $n$, 
respectively.
One can proceed with the third-order response, and the third-order injection conductivity tensor contains the Riemannian curvature as
\begin{align}
\label{eq:third-order-photoconductivity-main}
\sigma^{d;abc}_{\rm inj}
&=
\frac{\pi e^4}{6\Gamma\hbar^3}\sum_{m,n}\int_{\bf k}\delta\left(\omega-\omega_{mn}\right)f_{nm}
iK^{mn}_{cbad}
+\dots,
\end{align}

Note that the nonlinear optical responses are enhanced and sometimes
diverging in Weyl semimetals \cite{Yan:2016euz,Armitage:2017cjs,vafek2014dirac,wehling2014dirac} 
when the Fermi energy and the frequency of light are reduced 
\cite{PhysRevB.95.041104, PhysRevLett.117.216601, Ahn2020, PhysRevB.108.L161113, PhysRevB.109.104309}.

Recently, the generalization of the quantum geometric tensor 
for two states has been proposed  
\cite{avdoshkin2024multistategeometryshiftcurrent,
mitscherling2024gaugeinvariantprojectorcalculusquantum}.
The advantage of this formalism is that the U(1) gauge invariance is 
evident, and the treatment of the degeneracy can be taken into
account easily. 

The idea of Riemannian geometry has been applied also to the 
dc transport \cite{PhysRevB.106.125114}, and the
photon-drag conductivity in centrosymmetric materials where
the inversion symmetry is broken by the finite
momentum of light 
\cite{PhysRevLett.126.197402,PhysRevB.106.205423,doi:10.1073/pnas.2424294122}.

\section{Quantum Phenomena Associated with Berry Curvature and Berry Connection}
\label{sec:sec2a}

The Berry curvature characterizes the local curvature (twisting or rotation of the phase) of the wavefunction. It quips the momentum space with a local gauge structure, analogous to an effective magnetic field. Berry curvature plays a crucial role in determining the topological and transport properties of materials, leading to measurable effects in experiments. In a crystalline solid, electrons are described by Bloch wavefunctions $|\mu_n(\textbf{k})\rangle$, where $n$ is the band index and $\textbf{k}$ is the crystal momentum. The Berry curvature $\Omega_n(\textbf{k})$ is an antisymmetric tensor (like a magnetic field in $\textbf{k}$-space \cite{nagaosa2010anomalous, RevModPhys.82.1959}) defined as: 
\begin{equation}
    \Omega_n(\textbf{k})=\nabla \times A_n(\textbf{k})
\end{equation}
where $A_n(\textbf{k})= \langle{\mu_n(\textbf{k})} | i \nabla_\textbf{k} | {\mu_n(\textbf{k})}\rangle$ is the Berry connection, analogous to a vector potential. The Berry curvature arises due to the anholonomy (path dependence) of the quantum phase when the system undergoes adiabatic evolution. Unlike a real magnetic field, it is intrinsic to the band structure and depends on wavefunction geometry. 

In semiclassical electron dynamics, the Berry curvature generates an anomalous velocity $v_n(\textbf{k})$ which modifies the equations of motion for Bloch electrons \cite{RevModPhys.82.1959} under an external electric field $\bf{E}$:
\begin{equation}
    v_n(\textbf{k}) =-\frac{e}{\hbar}\textbf{E} \times \Omega_n(\textbf{k})
   \end{equation}
The anomalous velocity $v_n(\textbf{k})$ is perpendicular to the electric field $\bf{E}$, leading to a Hall current even without a magnetic field.
 
When the Berry curvature is integrated over the entire Brillouin zone (for all occupied bands), it yields a quantized topological invariant that indicates topologically non-trivial phases, such as (fractional) quantum anomalous Hall, (fractional) quantum Hall, and quantum spin Hall. When the Berry curvature is integrated for part of the Brillouin zone (the Fermi level is located in the bands), the distribution of Berry curvature will induce non-quantized quantum phenomena, such as anomalous Hall, spin Hall, valley Hall, layer Hall, and nonlinear Hall. This section will show the experimental results induced by the Berry curvature.

\subsection{Intrinsic Anomalous Hall Effect}

The intrinsic anomalous Hall effect (AHE) is observed in certain materials, particularly ferromagnets, where a transverse voltage develops across a sample when a current flows through it, without an external magnetic field. Intrinsic AHE is fundamentally tied to the Berry curvature which acts as an effective magnetic field in momentum space~\cite{nagaosa2010anomalous, RevModPhys.82.1959}. Over the past two decades, experimental advances have provided compelling evidence for the role of Berry curvature in driving the intrinsic AHE, particularly in magnetic and topological materials. The intrinsic contribution to the AHE can be expressed as ~\cite{Jungwirth:2002prl, PhysRevLett.90.206601, Yao:2004prl, nagaosa2010anomalous}:
\begin{equation}
    \sigma^{\text{int}}_{xy}=-e^2/\hbar\int_{\text{BZ}}\frac{dk}{(2\pi)^2}\Omega_z(k)f(k)
\end{equation}
where $\Omega_z(k)$ is the $z$-component of the Berry curvature in momentum space, and $f(k)$ is the Fermi-Dirac distribution. This formula shows that the intrinsic AHE is directly proportional to the integral of the Berry curvature over the occupied states in the Brillouin zone (BZ).

\begin{figure}[htp]
    \centering
    \includegraphics[width =8.5cm]{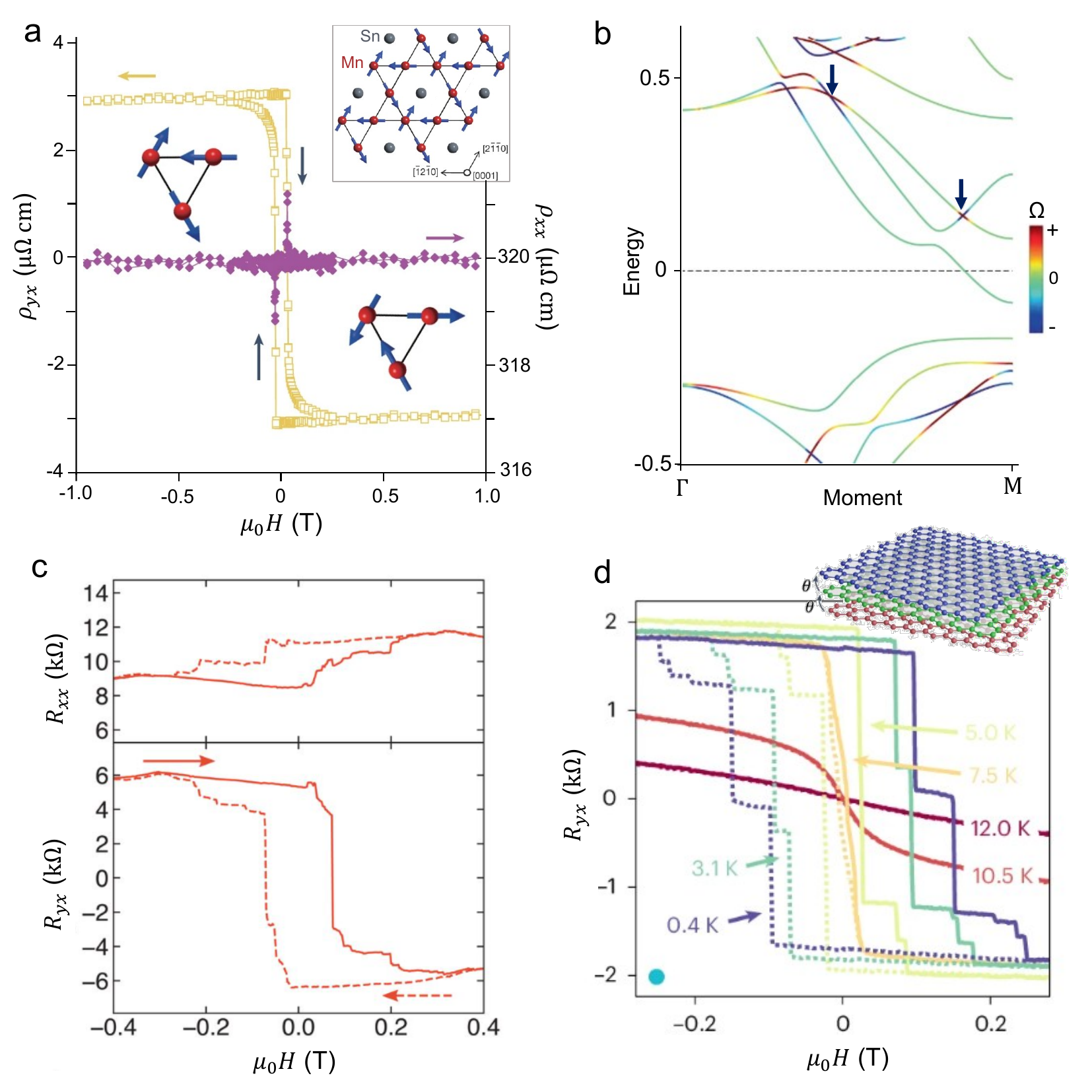} 
    \caption{\textbf{AHE in different systems.} (a) The AHE observed in the non-collinear antiferromagnet Mn$_3$Sn. Inset: an individual $a–b$ plane of Mn$_3$Sn, where the Mn spins (represented by arrows) form an inverse triangular spin structure. Each Mn spin has its local easy axis oriented parallel to the in-plane direction towards its nearest-neighbor Sn sites. From Ref.~\cite{Nakatsuji:2015Nature} (b) The band structure of Mn$_3$Sn, with the Berry curvature indicated by the color scheme. The Berry curvature hotspots at the Weyl points are highlighted by arrows. $\mathcal{B}$ denotes the Berry curvature. From~\cite{Xu:2020SA} (c) Longitudinal resistance $R_{xx}$ and Hall resistance $R_{yx}$ as a function of magnetic field for tBLG aligned with h-BN device. From~\cite{Sharpe:2019Science} (d) Inset: schematic of the helical trilayer graphene sample, consisting of three layers of graphene twisted together at the same angle. Hall resistance $R_{yx}$ as a function of magnetic field for the helical trilayer graphene device at different temperatures. From Ref.~\cite{Xia:2025NP}.}
\label{AHE}
\end{figure}

\color{blue} \underline{In Antiferromagnets:} \color{black} A substantial AHE in certain non-collinear antiferromagnets with a magnitude comparable to that observed in ferromagnets has been proposed theoretically \cite{PhysRevLett.87.116801, Chen:2014PRL}, and observed experimentally \cite{ Kübler:2014EPL, Nakatsuji:2015Nature, Ajaya:2016SA,  Vsmejkal:2022NRM}. As in Mn$_3$Sn \cite{Nakatsuji:2015Nature}, the underlying geometrical frustration results in an inverse triangular spin arrangement, which supports a tiny net magnetic moment of approximately 0.002 $\mu$B per Mn atom. However, it exhibits a large anomalous Hall conductivity, reaching the same order of magnitude as in ferromagnetic metals (Fig.~\ref{AHE}a). Although the net magnetization is negligibly small, the triangular spin configuration creates Berry-curvature hotspots in the electronic band structure, which drive the large AHE  (Fig.~\ref{AHE}b). Recently, people even observed AHE in collinear antiferromagnets known as altermagnets \cite{Libor:2022PRX}, which feature spin-split electronic bands that alternate in momentum space. In principle, the integral of the Berry curvature over the Brillouin zone should vanish. However, the AHE can be activated when the symmetry of the magnetic crystal is reduced by reorienting the Néel vector, allowing for a non-zero Berry curvature contribution~\cite{Libor:2020SA}. In addition to reorienting the Néel vector, Takagi et al. demonstrated that reducing the crystal symmetry to break $\mathcal{T}t$ (time-reversal $\mathcal{T}$ followed by translation $t$ symmetry) can also activate the AHE~\cite{Takagi:2025NM}.  

\begin{figure}[htp]
    \centering
    \includegraphics[width =8.5cm]{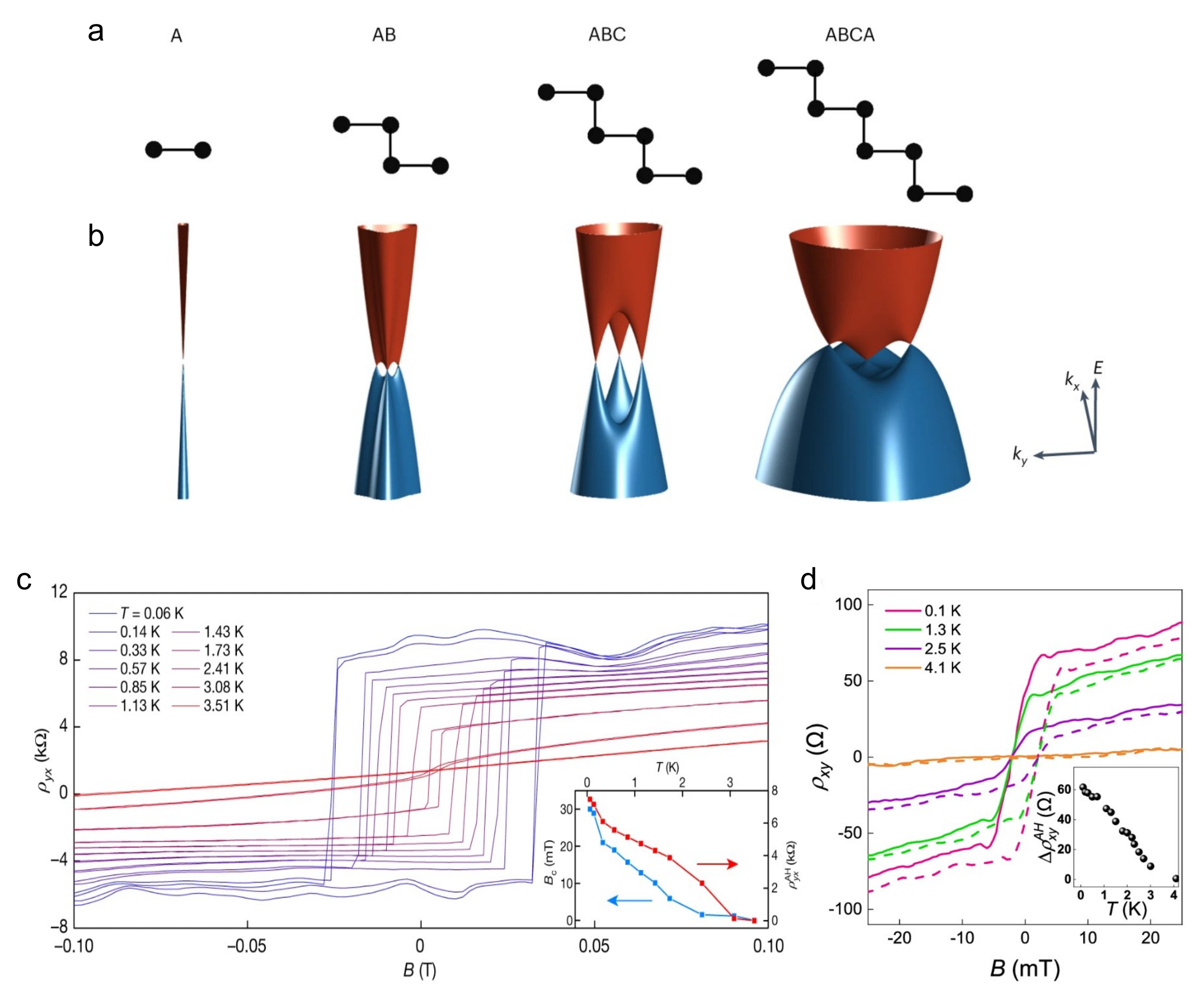} 
    \caption{\textbf{AHE in rhombohedral-stacked graphene.} (a) Side view of the rhombohedral stacking order in graphene. (b) Band structure of rhombohedral-stacked graphene. As the layer increases, the lowest energy band becomes sequentially flatter. From Ref.~\cite{Liu:2024NN} (c) Anomalous Hall resistivity $\rho_{yx}$ as a function of magnetic field for rhombohedral-stacked trilayer graphene at different temperatures. Inset: The extracted coercive field B$_c$ and anomalous Hall signal as functions of temperature. From Ref.~\cite{Chen:2020Nature} (d) Anomalous Hall resistance $\rho_{yx}$ as a function of magnetic field for the rhombohedral-stacked septuple-layer graphene device at different temperatures. From Ref.~\cite{Zhou:2024NC}.}
\label{AHE_RG}
\end{figure}

\color{blue} \underline{In Two-Dimensional Systems:} \color{black} Because the AHE originates from Berry curvature, a nonzero AHE can be observed in nonmagnetic materials with nonzero Berry curvature. It has been shown that in moiré graphene systems, the valley degeneration ($\mathcal{T}$ symmetry) can be spontaneously broken due to the strong electron-electron interaction, which induces a transition to a ferromagnetic state  (Fig.~\ref{AHE}c). This transition is associated with the emergence of ferromagnetic hysteresis and a significant AHE~\cite{Sharpe:2019Science, Nick:2020PRL, Chen:2021NP, Kuiri:2022NC, He:2021NC, Lu:2019Nature}.  By placing twist bilayer graphene (tBLG) onto WSe$_2$ to enhance the spin-orbital coupling, Lin et al. observed AHE at both quarter and half filling~\cite{Lin:2022Science}. Notably, Xia et al.~\cite{Xia:2025NP} demonstrate that supermoiré (moiré of moiré) can produce sufficiently flat bands that spontaneously break $\mathcal{T}$ symmetry, leading to the observation of the AHE at specific fillings without the requirement of alignment with h-BN  (Fig.~\ref{AHE}d). The supermoiré is formed in helical trilayer graphene, comprising three layers of graphene twisted at two independent angles. Another method to enhance the electron-electron interaction involves stacking multilayer graphene sheets in a rhombohedral configuration, where each sheet is shifted by one-third of the graphene unit cell (Fig.~\ref{AHE_RG}a). In this $n$-layer rhombohedral graphene system, the electron band dispersion follows a power law concerning the number of layers, $n$. The lowest energy bands become increasingly flat as the number of layers increases (Fig.~\ref{AHE_RG}b). The AHE has been observed in trilayer \cite{Chen:2020Nature, Zhou:2021Nature}, tetralayer \cite{Liu:2024NN}, pentalayer \cite{Han:2023Nature, Han:2024NN, morissette2025intertwined}, and septuple-layer \cite{Zhou:2024NC} rhombohedral-stacked graphene after applying an out-of-plane electric field (Fig.~\ref{AHE_RG}c,d). Rhombohedral-stacked graphene is an ideal platform for exploring many-body theory, topology, and superconductivity by manipulating symmetry, eliminating the need for Moiré superlattices.

\subsection{Quantum Anomalous Hall Effect}

The Quantum Anomalous Hall Effect (QAHE) represents one of the most remarkable discoveries in condensed matter physics over the past decade, showcasing how the geometry of electronic states can lead to surprising and profound phenomena. The story of QAHE began in 1988 when physicist Duncan Haldane~\cite{PhysRevLett.61.2015} proposed a toy model of electrons on a honeycomb lattice with a clever arrangement of magnetic fields that canceled out in real space but created a net Berry curvature in momentum space. In this model, the quantized Hall effect could occur without an external magnetic field as a result of the intrinsic Berry curvature on the band. Although Haldane's model laid the groundwork for understanding QAHE, it is hard to realize experimentally. It has been pointed out that the tight-binding model with spin-orbit interaction can show the QAHE similar to that proposed by Haldane \cite{doi:10.1143/JPSJ.71.19,PhysRevLett.90.206601}.

\begin{figure}[htp]
    \centering
    \includegraphics[width =8cm]{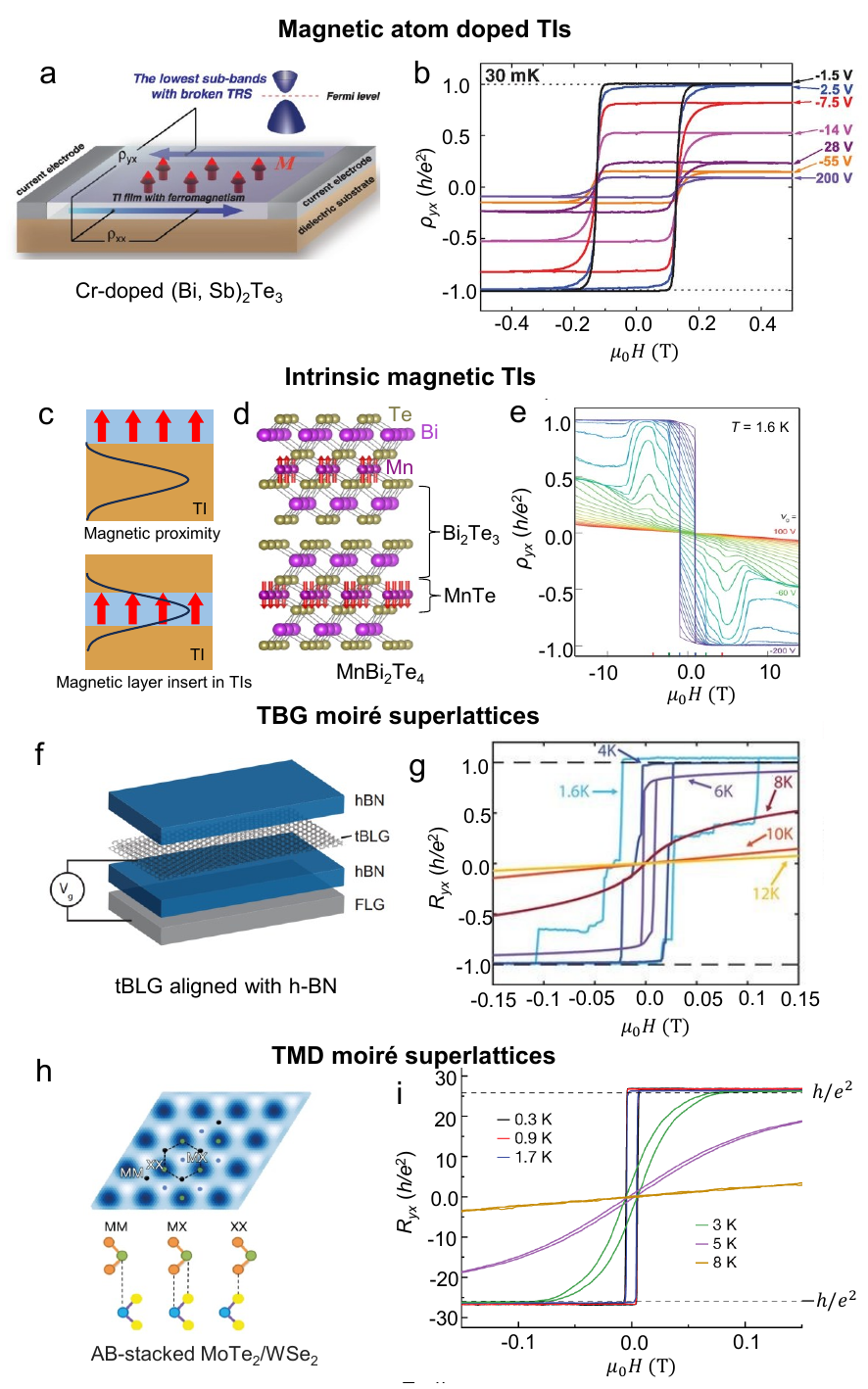} 
    \caption{\textbf{QAH effect in different systems.} (a) The schematic of Cr-doped  (Bi, Sb)$_2$Se$_3$ device. (b) Hall conductivity as a function of magnetic field for the device in (a). From Ref.~\cite{doi:10.1126/science.1234414} (c) Schematic illustration of the wavefunction distribution in magnetic TI surface states. Top panel: In artificially assembled magnet/TI heterostructures, the electron wavefunction cannot penetrate the magnetic layer. Bottom panel: In intrinsic magnetic TI, the electron wavefunction extends into the magnetic layer. (d)The chematic of the intrinsic topological antiferromagnet MnBi$_2$Te$_4$. Within each layer, the spins align uniformly along the $z$-axis, while in adjacent layers, the spins are oriented in the opposite direction. (e) The Hall resistance at different Fermi levels for 5-SL MnBi$_2$Te$_4$. From Ref.~\cite{Deng:2020Science} (f) The schematic of tBLG aligned with h-BN device. (g) Hall resistance for different temperatures.  From Ref.~\cite{Serlin:2020Science} The schematic (h) and Hall resistance (i) vs. magnetic field at different temperatures of AB-aligned TMD moiré superlattices. From Ref.~\cite{Li:2021Nature}}
\label{QAHE}
\end{figure}

The discovery of topological insulators (TIs) in the 2000s brought QAHE closer to reality~\cite{zhang：2009NP, Chen:2009Science, Neto:2009rmp}. TIs have gapless Dirac cone surface states, which can be gapped by breaking time-reversal symmetry. Therefore, it is natural to consider adding magnetization to open a gap in the surface state to realize the QAH effect. In 2013, Chang et al. observed the QAH effect in the experiment by doping (Bi, Sb)$_2$Te$_3$ with Cr \cite{doi:10.1126/science.1234414} for the first time, which confirms the theory predictions (Fig.~\ref{QAHE}a,b). Despite significant progress, the magnetic doping approach unavoidably degrades sample quality, limiting the critical temperature of the QAH effect to sub-Kelvin levels. As a result, achieving the QAH effect in intrinsic magnetic materials is highly desirable. 

Sandwiching a topological insulator, such as (Bi, Sb)$_2$Se$_3$, between intrinsic magnetic layers, like Cr$_2$Ge$_2$Te$_6$, presents a promising approach to achieving the QAHE~\cite{Mogi:2019PRL}. Ideally, the intrinsic magnetization of these magnetic layers breaks time-reversal symmetry, opening a surface gap, and then the QAHE can be realized. However, the QAHE remains unobserved due to interface potential issues and weak coupling, which hinder wavefunction penetration into the ferromagnetic layers (Fig.~\ref{QAHE}c). To address this challenge, topological magnetic insulator MnBi$_2$Te$_4$ has been proposed as a promising candidate to realize the intrinsic QAHE~\cite{Otrokov:20172D, Otrokov:2017JL, Otrokov:2019Nature}. MnBi$_2$Te$_4$ is a layered topological antiferromagnet, in each septuple layer (SL), it contains a magnetic layer MnTe inserted in between two Bi$_2$Te$_3$ layers (Fig.~\ref{QAHE}d), and therefore the topological state can significantly penetrate the magnetic layer region and direct interaction with the magnetic moments of Mn atoms (Fig.~\ref{QAHE}c). By mechanically exfoliating MnBi$_2$Te$_4$ into a few layers, Deng et al. observed the QAHE in a 5-SL sample at 1.6K (Fig.~\ref{QAHE}e)~\cite{Deng:2020Science}. Due to the space-time ($\mathcal{PT}$) symmetry, MnBi$_2$Te$_4$ is also an axion insulator for even SL~\cite{Zhang:2019PRL, Liu:2020NM}.

Another intriguing system for achieving the intrinsic QAHE is the moiré superlattices~\cite{Rafi:2011PNAS}, which does not need to involve any magnetic atoms. In moiré systems, the moiré potential can induce a topological band structure, if the system also exhibits spontaneous time-reversal symmetry breaking (e.g., due to strong correlations), the QAHE appears. By aligning tBLG with hexagonal boron nitride (h-BN), Serlin et al. observed a well-defined QAHE that persists at approximately 3K (Fig.~\ref{QAHE}f,g)~\cite{Serlin:2020Science}. This experiment has been proven difficult for other research groups to replicate, primarily due to the challenges associated with fabricating high-quality tBLG/h-BN moiré devices and weak SOC. Monolayer TMDs exhibit valley-locked Berry curvature due to their strong SOC, making them more reliable candidates for realizing the QAHE than graphene (Fig.~\ref{QAHE}h,i)~\cite{Wu:2019PRL, Li:2021Nature, Park:2023Nature, Xu:2023PRX}. Recently, twisted bilayer–trilayer graphene~\cite{Su:2025nature}, ABCA-tetralayer graphene ~\cite{Sha:2024Science, Choi:2025superconductivity}, ABC-trilayer graphene/h-BN also realized QAH ~\cite{Chen:2020Nature, Lu:2025Nature}.

\subsection{Fractional Quantum Anomalous Hall Effect}

The fractional quantum anomalous Hall effect (FQAH) is a remarkable phenomenon that combines the fractional quantization with the anomalous Hall effect in quantum systems, typically observed in two-dimensional flat-band systems in the absence of an external magnetic field. Like the fractional quantum Hall effect (FQH) observed under strong magnetic fields at low temperatures, FQAH exhibits strong correlations and topology, leading to unique transport properties characterized by fractional quantization of Hall conductance, i.e. $\sigma_{xy} = \nu\frac{e^2}{h}$, where $\nu$ is fractional. The role of quantum geometry is pivotal in understanding the FQAH, as it provides insights into the topological properties of the electron wave functions that underlie this effect.

\begin{figure*}[ht]
\begin{center}
 \includegraphics[width=18cm]{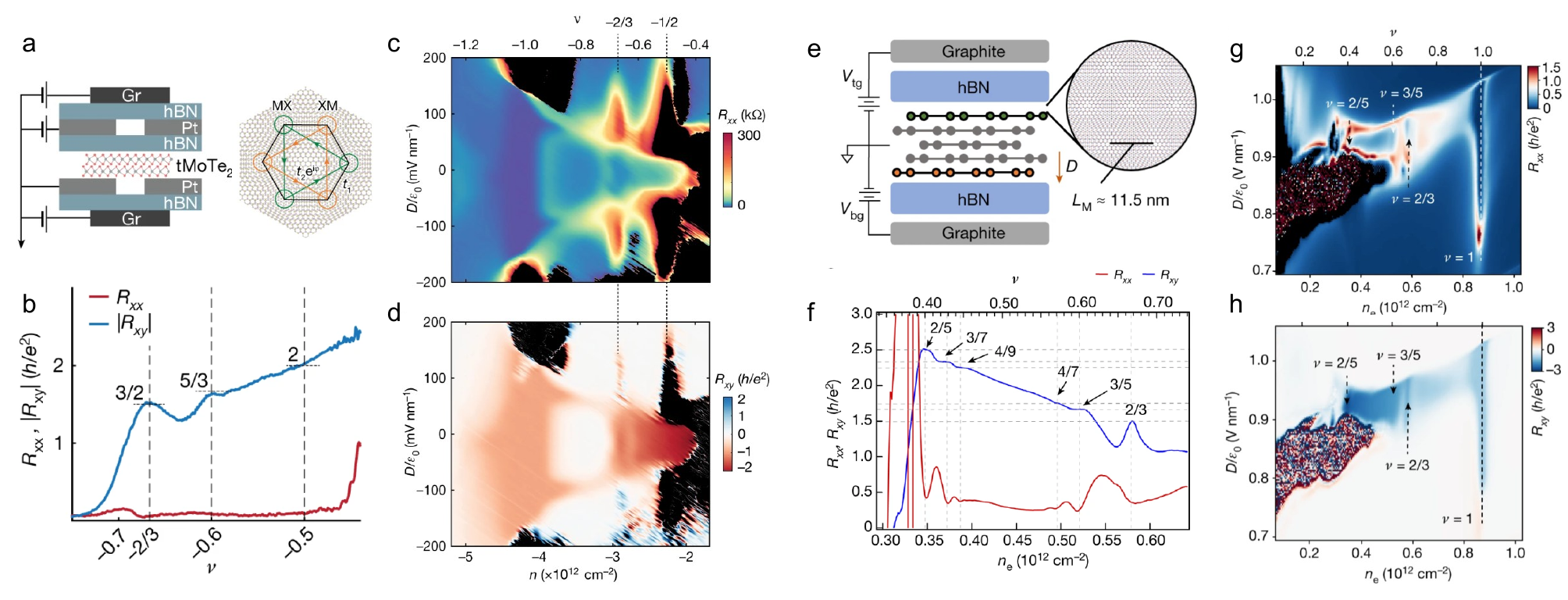}
    \caption{
    \textbf{Observation of fractional quantum anomalous Hall effect.} (a) Left panel: Schematic of the twist-bilayer MoTe$_2$ device with contact gate design. Right panel: The honeycomb superlattice formed by the two degenerate moiré orbitals localized at the MX (green) and XM (orange) sites. Nearest-neighbour hopping t$_1$ and next-nearest-neighbour complex hopping t$_2$e$^{i\varphi}$ between moiré orbitals realize the Kane–Mele model. (b) Longitudinal $R_{xx}$ and Hall $R_{xy}$ resistance as a function of filling factor $\nu$. (c,d) $R_{xx}$ (c) and $R_{xy}$ (d) as a function of electric field $D/\epsilon_0$ and carrier density $n$ at 100 mK. $R_{xx}$ and $R_{xy}$ are symmetrized and anti-symmetrized at 200 mT, respectively. From \cite{Park:2023Nature}. (e) The Schematic of the device configuration shows a moiré superlattice between the top layer of graphene and the top hBN, with a moiré period of 11.5 nm. (f)  $R_{xx}$ and $R_{xy}$ along the dashed lines in g and h, taken with a constant current measurement. Clear plateaus of $R_{xy}$ at fractional filling $\nu$= 2/5, 3/7, 4/9, 4/7, 3/5 and 2/3, as indicated by the dashed lines and arrows. $R_{xx}$ shows clear dips at the corresponding filling factors. (g,h) Phase diagrams of the device revealed by symmetrized $R_{xx}$ (b) and antisymmetrized $R_{yx}$ (c) at B = ± 0.1 T as functions of $\nu$ and $D$. The temperature is 10 mK. Large anomalous Hall signals emerge in a tilted stripe region centred at $D/\epsilon_0= 0.93$ V/nm. Clear dips of $R_{xx}$ can be seen at filling factors of the moiré superlattice $\nu=$ 1, 2/3, 3/5, and 2/5 (indicated by the dashed lines and arrows), where $R_{yx}$ shows plateaus of values. From \cite{Lu:2024fractional}.
    }
\label{fig:FQAH}
\end{center}
\end{figure*}

The moire superlattices by rhombohedral-stacked bilayer TMD generate a flat band, where dramatically enhances the electron-electron interaction. The correlation gaps can be opened by electron-electron interaction, adding the spontaneously $\mathcal{T}$-symmetry breaking, the FQAH can be realized \cite{Wu:2019PRL, Li:2021spontaneous, Devakul:2021magic, Anderson:2023programming}. The experimental signature of FQAH is in the twist bilayer MoTe$_2$ system near 3.7$^\circ$ (Fig. \ref{fig:FQAH}a).  Cai et al \cite{Cai:2023signatures} and Zeng et al \cite{Zeng:2023thermodynamic} first observed the signature of FQAH by photoluminescent and thermodynamic measurements. After improving the electric contact, the direct observations of FQAH have been realized by electrical transport measurement in twist bilayer MoTe$_2$ \cite{Park:2023Nature, Xu:2023PRX} (Fig. \ref{fig:FQAH}b-d). 
Another system for the observation of FQAH is rhombohedral graphene sitting on top of hBN with a very small twist angle (Fig. \ref{fig:FQAH}e). Lu et al measured the Hall resistance and longitudinal resistance of rhombohedral pentalayer graphene/hBN moiré superlattice and observed the FQAH at filling factors $\nu=$ 2/3, 3/5, 4/7, 4/9, 3/7, and 2/5 of the moiré superlattice (Fig. \ref{fig:FQAH}f-h) \cite{Lu:2024fractional}.  

\subsection{Spin Hall, Valley Hall and Layer Hall Effects}

Electrons possess multiple degrees of freedom beyond charge, such as spin, valley, and layer. Encoding Berry curvature with these degrees of freedom can lead to novel types of Hall effects. These effects arise due to the interplay between the electron's intrinsic geometric effect of band structure. Below, we’ll discuss how encoding Berry curvature with these degrees of freedom leads to new Hall effects.

\begin{figure*}[ht]
\begin{center}
 \includegraphics[width=16cm]{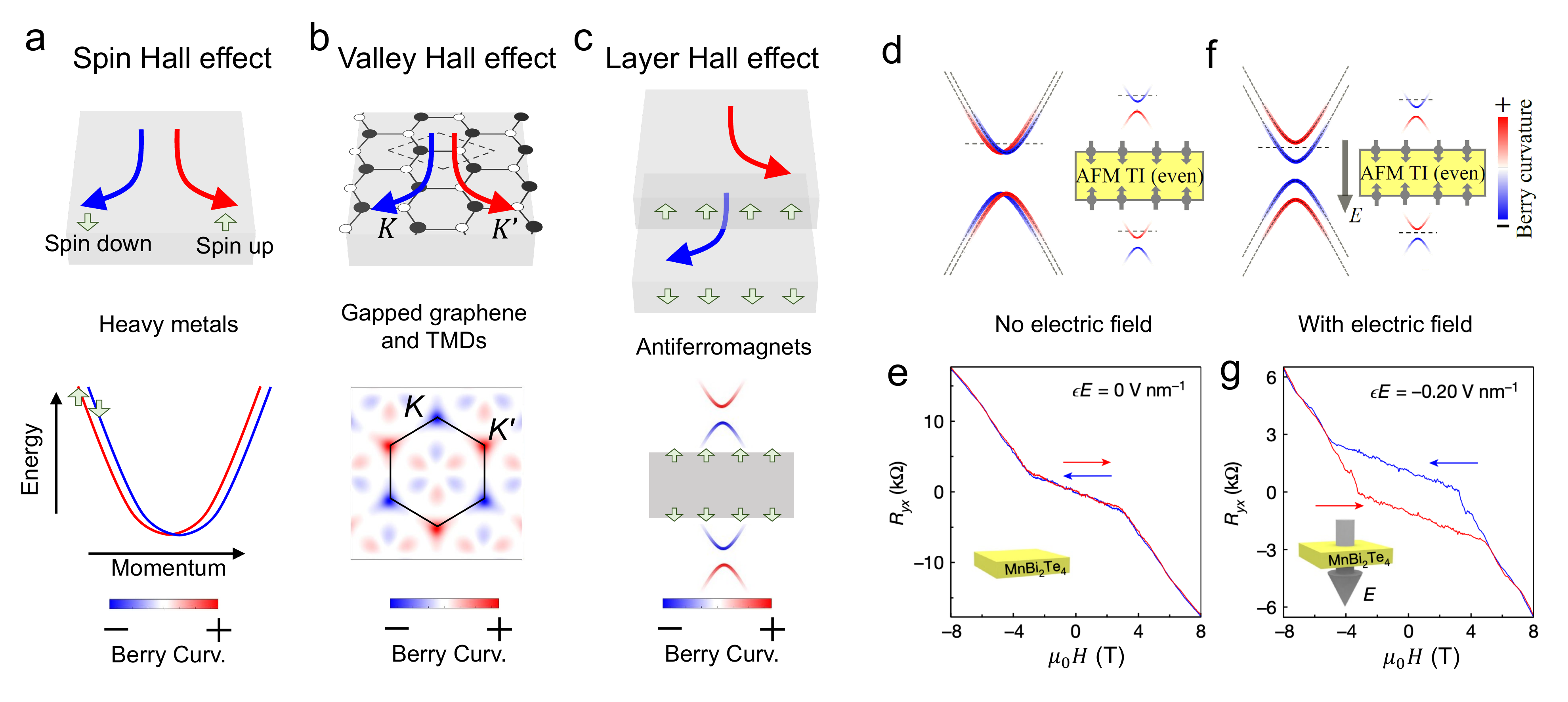} 
    \caption{
    \textbf{Unconventional Hall effects associated with the distribution of Berry curvature.}  (a) The spin Hall effect in heavy metals induced by spin-locked Berry curvature. (b) The valley Hall effect in gapped graphene and transition metal dichalcogenides induced by the valley-locked Berry curvature. (c) The layer Hall effect in antiferromagnets induced by the layer-locked Berry curvature. (d) The inclusion of the A-type antiferromagnetic order gaps the Dirac fermions. The resulting Berry curvature of the top Dirac fermion exactly cancels that of the bottom Dirac fermion. (e) Applying an electric field can break the degeneracy between the top and bottom Dirac fermions, leading to a large, layer-polarized AHE, as long as the Fermi level is away from the bandgap. (f,g) Hall resistance as a function of magnetic field without (f) and with (g) vertical electrical field. From Ref.~\cite{Gao:2021Nature}.}
\label{Spin_Valley_Layer_Hall}
\end{center}
\end{figure*}

\color{blue} \underline{Spin Hall Effect:} \color{black} In materials with strong SOC, the spin and momentum of electrons are coupled \cite{sinova:2015spin}. This coupling splits the bands into spin-polarized subbands. Each spin-polarized subband can have a different Berry curvature, leading to spin-locked Berry curvature. When an electric field $\bf E$ is applied to the material, electrons with opposite spins experience a transverse force due to spin-locked Berry curvature. This force causes electrons with opposite spins to move in opposite transverse directions, leading to a separation of spins across the material, named intrinsic spin Hall effect (SHE) (Fig.~\ref{Spin_Valley_Layer_Hall}a) as proposed theoretically \cite{doi:10.1126/science.1087128, Sinova:2004PRL} and observed experimentally \cite{kato:2004Science, Wunder:2005PRL}. Materials with large Berry curvature and strong spin-orbit coupling exhibit a significant SHE. A straightforward method to measure the SHE involves using a magneto-optical Kerr microscope to scan the spin polarization across the sample~\cite{kato:2004Science, Wunder:2005PRL}. Although the SHE was first proposed in a time-reversal-reserved system, it has also been observed in the magnetic systems due to the momentum-dependent spin splitting~\cite{Kimata:2019nature, Dai:2024NC, Chen:2021NM, Bai:2023PRL}. Recently, the SHE has also been observed in AB-stacked MoTe$_2$/WSe$_2$ heterostructure~\cite{Tschirhart:2023NP, Tao:2024NN}, which demonstrates long-range spin Hall transport and efficient non-local spin accumulation.

\color{blue} \underline{Valley Hall Effect:} \color{black} Two-dimensional TMDs exhibit strong spin-orbit coupling, and monolayer TMDs break $\mathcal{P}$ symmetry while preserving $\mathcal{T}$ symmetry. As a result, in monolayer TMDs, electrons in different valleys possess opposite Berry curvatures due to $\mathcal{T}$ symmetry, meaning the Berry curvature is locked to the valley. However, the two valleys are energetically degenerate, no net AHE is generated. By breaking the valley degeneracy through methods such as circularly polarized light, an anomalous Hall signal can be observed, known as the valley Hall effect (VHE) \cite{Xiao:2007PRL, Xiao:2012PRL, Mak:2014Science}. In the monolayer MoS$_2$, the Berry curvature at the K and K' valleys is equal in magnitude but opposite in sign (Fig. \ref{Spin_Valley_Layer_Hall}b). Without breaking the valley degeneracy, the anomalous Hall signals generated by the opposing Berry curvatures cancel each other out. Mak et al. observed the AHE by exposing a monolayer MoS$_2$ sample to circularly polarized light, which created an imbalance in the population of the two valleys~\cite{Mak:2014Science}. Lee et al. observed the Valley Hall in bilayer MoS$_2$ with an out-of-plane electric field breaking inversion symmetry $\mathcal{P}$ \cite{Lee:2016Electrical}. Notably, thanks to the high mobility, long-range nonlocal VHE has been demonstrated in graphene/h-BN superlattices \cite{Gorbachev:2014Science} and gapped bilayer graphene systems \cite{Yin:2022Science} after breaking $\mathcal{P}$ symmetry. Moreover, in ultrahigh-mobility graphene, the ballistic transport properties can contribute to the quantum VHE \cite{Komatsu:2018SA}, showcasing its potential for valley-Hall transistors~\cite{Gorbachev:2014Science, Li:2020NN}.

\color{blue} \underline{Layer Hall Effect:} \color{black} Although the external manifestations are different, AHE, SHE, VHE, and Berry-curvature dipole (BCD) nonlinear Hall are all due to the momentum-space distributed Berry curvature. This naturally raises the question: Does the real-space distributed Berry curvature lead to new quantum phenomena? Because the Berry curvature changes sign under $\mathcal{T}$ operations, which is similar to the behavior of electron spins. This naturally leads us to consider antiferromagnets, particularly the antiferromagnetic topological insulator MnBi$_2$Te$_4$~\cite{Otrokov:2019Nature, Deng:2020Science, Liu:2020NM}. In MnBi$_2$Te$_4$, the spins within the same layer are aligned parallel and oriented out of the plane, while spins in adjacent layers are aligned antiparallel. This configuration results in the Berry curvature being spatially locked within the layers; specifically, the Berry curvature exhibits the same sign within a given layer but opposite signs between adjacent layers. Consequently, the Berry curvature of adjacent layers forms a BCD in real space (Fig.~\ref{Spin_Valley_Layer_Hall}c). 

In 2021, Gao et al. first observed the real-space-locked Berry curvature induced AHE in even-layer MnBi$_2$Te$_4$ samples~\cite{Gao:2021Nature}. They found that without the application of an external electric field, there was no observable anomalous Hall signal, as the anomalous Hall contributions resulting from the layer-locked Berry curvature in adjacent layers effectively cancel each other out (the AHE in the spin-up layer is equal in magnitude but opposite in sign to that in the spin-down layer) (Fig.~\ref{Spin_Valley_Layer_Hall}d and e). However, when an electric field is applied, the degeneracy of the Berry curvature across the layers is lifted, leading to a net anomalous Hall signal, where the AHE in the spin-up layer becomes larger than that in the spin-down layer (Fig.~\ref{Spin_Valley_Layer_Hall}f and g). Gao et al. refer to this AHE induced by the layer-locked Berry curvature as the layer Hall effect (LHE). This effect offers an effective method for electrically probing the fully compensated antiferromagnetic state. Recently, the quantized version of LHE has also been proposed theoretically~\cite{Dai:2022PRB, Liu:2024NPJ, Tian:2024NL}. 

\subsection{Berry Curvature Dipole Nonlinear Hall Effect}
\begin{figure}[htp]
    \centering
    \includegraphics[width =8.5cm]{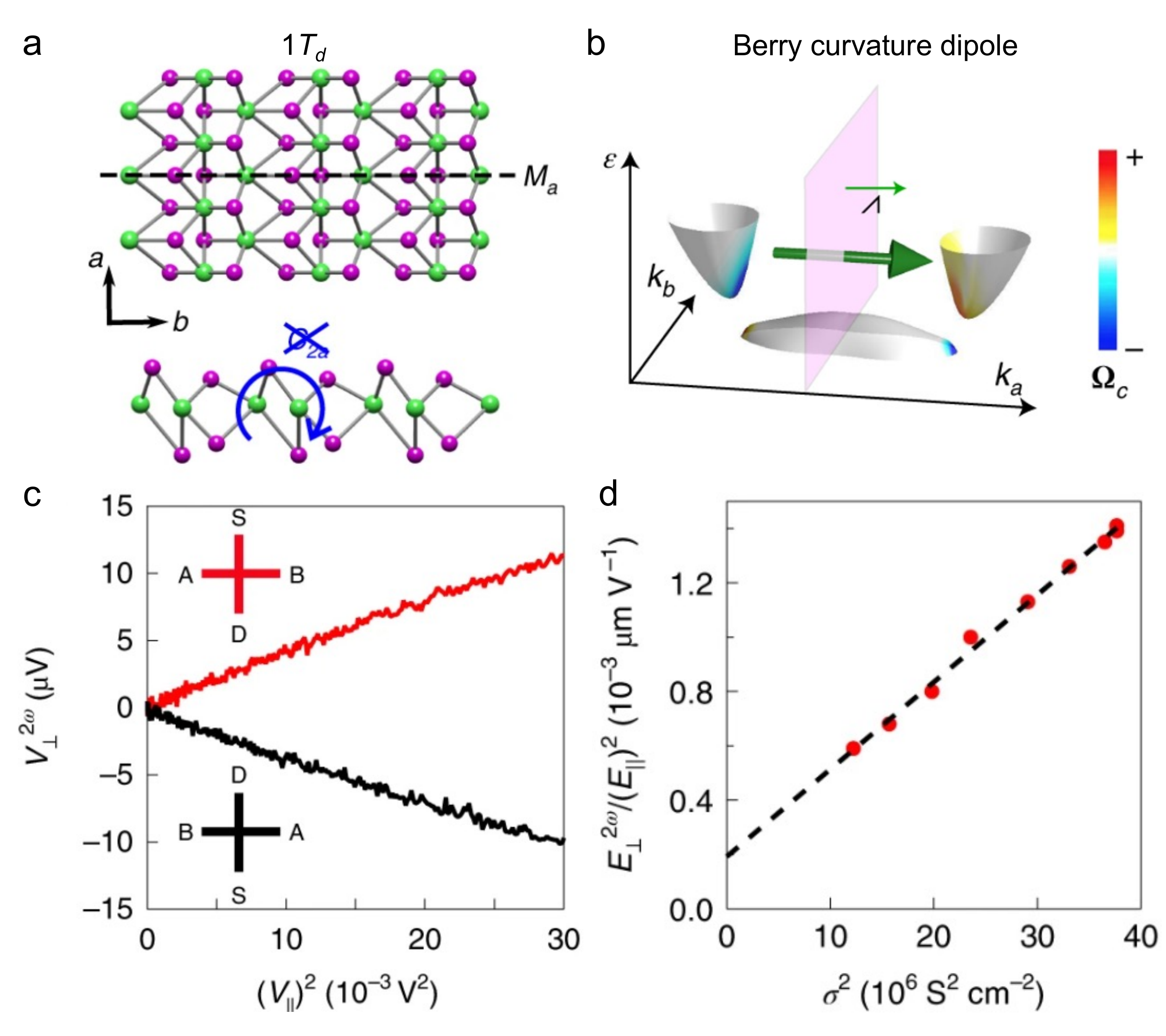} 
    \caption{\textbf{Berry Curvature Dipole Nonlinear Hall Effect.} (a) The lattice of 1T$_d$-phase of monolayer WTe$_2$. It has only the $\mathcal{M}_a$ without 2-fold rotational symmetry along $a$, $C_{2a}$. (b) Band structure of WTe$_2$ color-coded with Berry curvature $\Omega_c$, showing the Berry dipole moment. The positive and negative Berry curvatures are related by the mirror-symmetry $\mathcal{M}_a$. From Ref.~\cite{Xu:2018NP} (c) The second-order NHE as a function of the square of longitudinal voltage. (d) Square of the nonlinear Hall electrical field over the square of the longitudinal electrical field as a function of the square of the longitudinal conductivity. From Ref.~\cite{Kang2019nonlinear}}
\label{BCD_NHE}
\end{figure}

In a quantum sample lacking $\mathcal{P}$ symmetry but possessing $\mathcal{T}$ symmetry, the further breaking of rotational symmetry $C_z$ leads to the emergence of a BCD \cite{sodemann2015quantum, Xu:2018NP}, analogous to an electric dipole moment (Fig.~\ref{BCD_NHE}a,b). In equilibrium, the Berry curvatures at opposite $k$-points in momentum space cancel each other out, which prevents the observation of an anomalous Hall signal. However, when an electric field is applied in the direction of the BCD, the Fermi surface tilts, resulting in a net non-zero total Berry curvature for the system. Subsequently, under the effect of a second electric field, an anomalous Hall signal can be generated. This anomalous Hall signal ($J^{2\omega}$) is proportional to the square of the electric field ($\sigma\propto E^{2}$), thus, it is referred to as the second-order nonlinear Hall effect (NHE) of the BCD.
\begin{equation}
    J^{2\omega}_\alpha=\sigma^{(\tau)}_{\alpha\beta\gamma} E^{\omega}_{\beta} E^{\omega}_{\gamma}
\label{equ_NLH}
\end{equation}
where $\sigma^{(\tau)}_{\alpha\beta\gamma}$ represents the nonlinear Hall conductivity, which is proportional to the scattering time $\tau$. The electric fields $E^{\omega}_{\beta}$ and $E^{\omega}_{\gamma}$ are applied along the $\beta$ and $\gamma$ directions, respectively (see Eq. \ref{eq:NLH3}). 
\begin{equation}
    \sigma^{(\tau)}_{\alpha\beta\gamma}=\frac{e^3\tau}{\hbar^2}\int\frac{d^nk}{2\pi^n}\epsilon_{\alpha\beta\delta}(\partial_{\gamma}{\Omega_{\delta}})f_k^0
\label{equ_NLH_Con}
\end{equation}
where $\tau$ is the scattering time, $n$ is the dimensionality, $\epsilon_{\alpha\beta\delta}$ is Levi-Civita symbol, $\partial_{\gamma}=\frac{\partial}{\partial{k_\gamma}}$, $\Omega_{\delta}$ is Berry curvature and $f_k^0$ is the Fermi 
distribution. Non-zero $\partial_{\gamma}{\Omega_{\delta}}$ will induce the BCD in momentum space. 

In Weyl semimetals, Weyl points exhibit Berry curvature hot spots, making them promising candidates for investigating the NHE due to the BCD (Fig. \ref{BCD_NHE}b). The experimental observation of NHE was first reported in bilayer\cite{Ma:2019Nature} and few-layer\cite{Kang2019nonlinear} WTe$_2$. They observed a second-order nonlinear signal characterized by quadratic $I–V$ relationships, which is perpendicular to the current direction (i.e., the Hall response), while such signal was zero along the longitudinal direction. Additionally, these measurements demonstrated that the second-order nonlinear voltages remained consistent after reversing the current direction (Fig. \ref{BCD_NHE}c). By analyzing the temperature dependence of both the linear longitudinal conductivity and the nonlinear Hall conductivity, it was revealed that the NHE in WTe$_2$ can be induced by both the BCD (proportional to the scattering time $\tau$) and skew scattering (proportional to the scattering time $\tau^3$ (in magnetic materials, the skew scattering contribution has been shown to scale as $\tau^2$ \cite{Tian:2009proper, nagaosa2010anomalous, Ye:2018massive}) (Fig.~\ref{BCD_NHE}d). After breaking the three-fold rotational symmetry of strain, BCD can appear in TMD, which supports the NHE~\cite{Qin:2021CPL, Huang:2023NSR}. Following the initial observations, new experimental studies of the NHE have generated a rapidly growing interest in this phenomenon. The NHE has also been detected in various other materials and systems. Ho et al. demonstrated that BCD NHE can be produced through periodic strain~\cite{Ho:2021NE}.

In addition to the ability of BCD-induced NHE to directly visualize the band geometry of quantum devices, it also holds significant potential for applications. Firstly, the intrinsic BCD-induced NHE can be utilized to probe the symmetry-breaking in quantum devices due to the stringent symmetry requirements for the observation of NHE. According to eq.(\ref{equ_NLH}), the $\mathcal{P}$ will cause the BCD NHE to become zero. Thus, it serves as a powerful tool for probing ferroelectric~\cite{Xiao:2020NP} and topological phase transitions~\cite{Hinha:2022NP}. Secondly, the NHE can be employed for efficient energy harvesting and next-generation wireless technologies~\cite{Hemour:2014IEEE, kumar2021room}. The quadratic relationship between the transverse current and input voltage allows for the rectification of incident microwave or terahertz electric fields into direct current. Since the NHE arises from the intrinsic properties of the material, it does not face the limitations of small-voltage or high-frequency inputs that are typical of semiconductor diodes~\cite{Hemour:2014IEEE, Isobe:2020SA}. 
 
\subsection{Magnetoelectric Coupling}

Magnetoelectric (ME) coupling is the phenomenon in which magnetic polarization can be induced by an electric field and vice versa. The ME effect is typically represented by the equation $M_i = \alpha_{ij} E_j$ or $P_i = \beta_{ij} H_j$, where $M_i$ denotes the magnetization induced by the external electric field $E_j$,  $P_i$ denotes the electrical polarization induced by the magnetic field $H_j$, and $\alpha_{ij}$, $\beta_{ij}$ represents the ME coupling coefficients. This equation indicates that ME coupling occurs in systems that break both inversion symmetry $\mathcal{P}$ and time-reversal symmetry $\mathcal{T}$. This characteristic underscores the interaction between electric and magnetic order parameters, rendering ME materials highly appealing for potential applications in spintronics. Recent theoretical and experimental advancements have revealed a significant relationship between ME coupling and Berry curvature, which arises from the geometric properties of band structures in materials. 

\begin{figure}[htp]
    \centering
    \includegraphics[width =8cm]{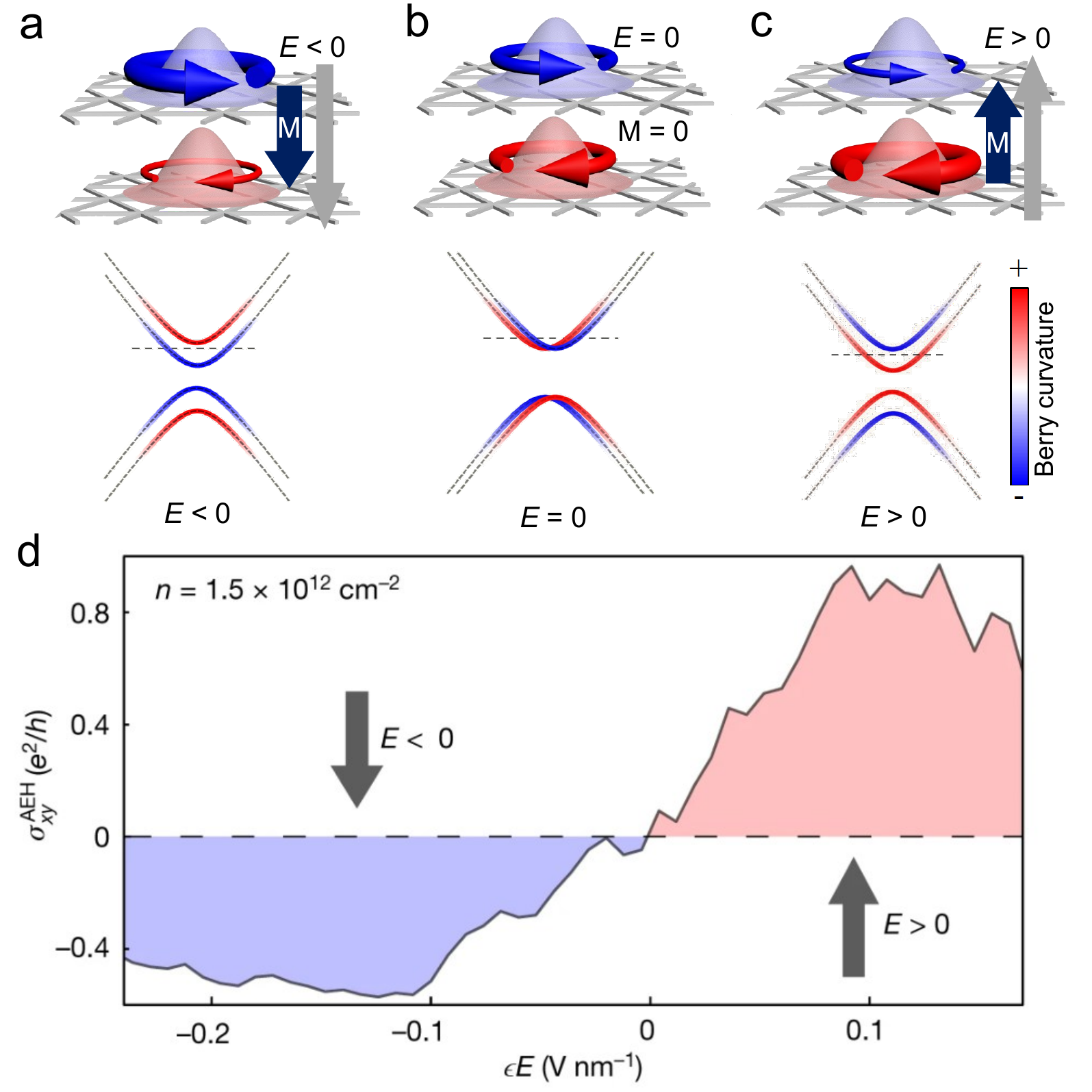} 
    \caption{\textbf{Orbital magnetoelectic effect.} (a-c) Illustration of the layer-locked Berry curvature that generates orbital magnetization under different electric field conditions. In the absence of an electric field (b), the Berry curvature in adjacent layers cancels each other out, resulting in no net orbital magnetization. However, when an electric field is applied (a,c), breaking the layer degeneracy, orbital magnetization emerges. (d) Anomalous Hall conductivity as a function of the electric field. Anomalous Hall conductivity is proportional to the orbital magnetization. From Ref.~\cite{Gao:2021Nature}}
\label{ME}
\end{figure}

\begin{figure*}[ht]
   \begin{center}
 \includegraphics[width=17cm]{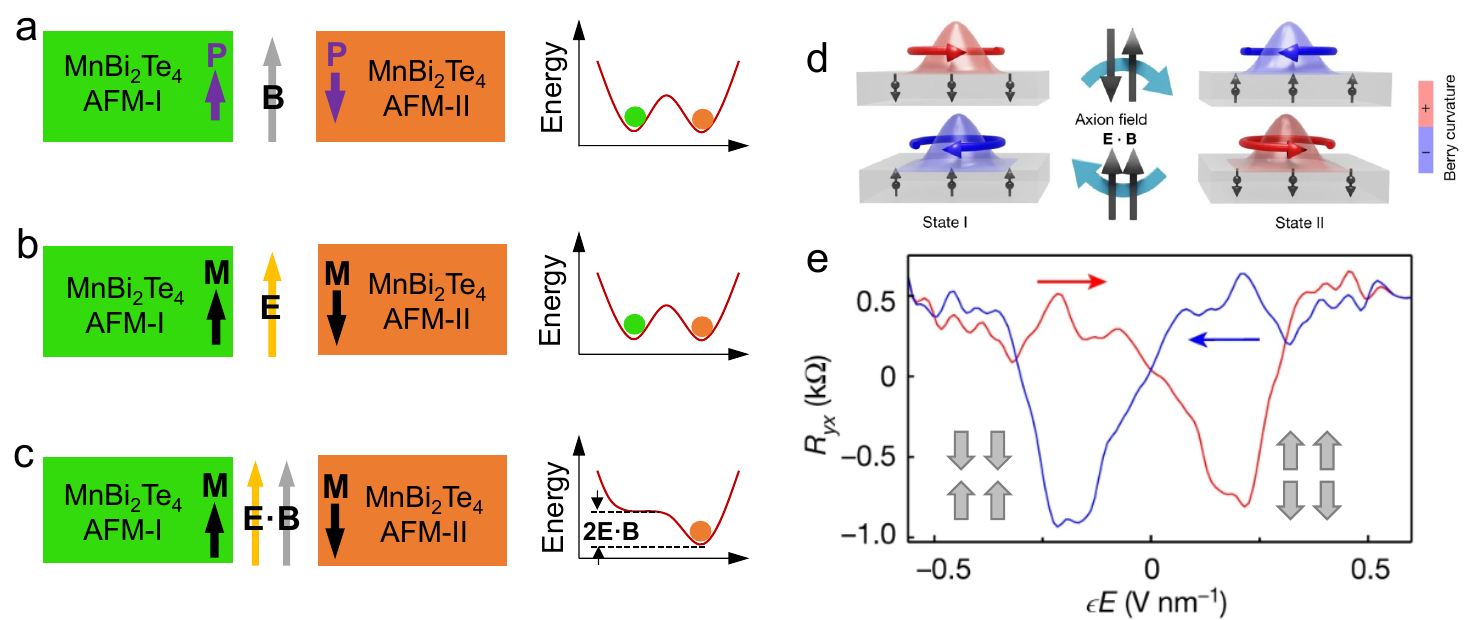} 
    \caption{\textbf{E $\cdot$ B manipulate the AFM state through ME.} (a-b) The application of an electric field $\bf E$ (magnetic field $\bf B$) induces electrical polarization $\bf P$ (magnetization $\bf M$) due to the orbital ME. At this stage, the two AFM states remain energetically degenerate. (c) When an $\bf E$ is applied, it generates the opposite $\bf M$ in the two AFM states. The subsequent application of a $\bf B$ couples to the magnetization and breaks the energy degeneracy between the two states. If the product $\bf E \cdot B$ is sufficiently large to shift the energy of one AFM state over the energy barrier, switching from one AFM state to another can occur. (d) The schematic of $\bf E \cdot B$ switching mechanism for the AFM state, where $\bf B$ remains fixed while $\bf E$ is varied. (e) Hall resistance as a function of the electric field at a magnetic field of $B = 1$ T, with the two AFM states indicated in the inset. From Ref.~\cite{Gao:2021Nature}}
\label{ME_switch}
\end{center}
\end{figure*}

In A-type antiferromagnets, such as MnBi$_2$Te$_4$, the material breaks both $\mathcal{T}$ and $\mathcal{P}$ symmetries, thereby enabling the ME effect. Unlike Cr$_2$O$_3$ \cite{Dzyaloshinski:1960OnTM, Astrov:1960magnetoelectric} and CrI$_3$ \cite{Huang:2018electrical, Jiang:2018electric}, where the lowest energy bands are associated with spins arising from localized magnetic ions, the lowest energy bands of MnBi$_2$Te$_4$ exhibit Berry curvature that is attributed to orbital contributions rather than localized spins \cite{Li:2019Intrinsic, Otrokov:2019Nature}. Consequently, the ME effect in MnBi$_2$Te$_4$ is characterized as orbital magnetization, directly related to the Berry curvature of the occupied bands \cite{Essin:2010PRB, Lu:2024orbital}.

In MnBi$_2$Te$_4$, the layer-locked Berry curvature, resulting from spatially resolved spins, generates a Berry curvature dipole in real space. With layer degeneracy, there is no net orbital magnetization (Fig.~\ref{ME}b). However, when an external electric field breaks this layer degeneracy, the positive Berry curvature becomes greater than the negative Berry curvature, leading to a net orbital magnetization (Fig.~\ref{ME}a and c). This real-space Berry curvature dipole induced orbital ME has been observed by Gao et al. in even-layered MnBi$_2$Te$_4$ (Fig.~\ref{ME}d) \cite{Gao:2021Nature}. In particular, the topological contribution of ME from low-energy electrons can be visualized as an orbital magnetization induced by edge states~\cite{Zhang:2019PRL, essin09, armitage2019matter}. Theoretical calculations showed that this topological contribution is expected to be quantized in the fundamental unit of $e^2/h$, which is one or two orders of magnitude larger than the conventional localized spin contribution~\cite{armitage2019matter}.

The orbital magnetoelectric effect can be utilized to switch fully compensated AFM states. To achieve this switching, it is essential to break the energy degeneracy between the two AFM states. As such, these fully compensated AFM states cannot be switched solely by an electric field or magnetization, as neither can effectively tune the free energy of the AFM state (see Fig.~\ref{ME_switch}a and b). The switching of the two AFM states can be accomplished through the application of $\bf E \cdot B$, representing an energy term resulting from the interaction of the electric field with the magnetic field (see Fig.~\ref{ME_switch}c). The underlying phenomenological understanding is that the electric field induces orbital magnetization due to the ME (see Fig.~\ref{ME_switch}b). This induced magnetization, when multiplied by the applied magnetic field, creates an energy term that alters the free energy difference between the two states (see Fig.~\ref{ME_switch}c), thus facilitating the switching process. The $\bf E \cdot B$ modifying the free energy of the two degenerated states to choose and switch the state has been demonstrated by Gao et al. in even-layered MnBi$_2$Te$_4$ \cite{Gao:2021Nature} and Han et al. in pentalayer rhombohedral graphene \cite{ Han:2023Nature}.

\subsection{Edelstein Effect and Natural Optical Activity}

\color{blue} \underline{Edelstein Effect:} \color{black} The Edelstein effect describes the generation of magnetization $M$ in a material when an electric current $J$ is applied, $M=\alpha J$. The Edelstein effect has two origins, spin and orbital \cite{yoda2018orbital, yoda2015current}. The spin part can be straightforwardly visualized in any strongly SOC-coupled noncentrosymmetric systems \cite{edelstein:1990spin, johansson:2018edelstein, ganichev:2002spin, el:2023observation, bihlmayer:2022rashba}. On the other hand, the orbital part is essentially the orbital magnetization dipole in momentum space, which arises from the Fermi surface. This orbital magnetization is closely tied to Berry curvature — in fact, in two-band systems, it is directly proportional to it. Berry curvature, which can be viewed as an effective magnetic field in momentum space, can also lock with momentum in TMDs, Weyl semimetals, and gapped graphene. Consequently, the Berry curvature-induced Edelstein effect can be observed in these materials.

\begin{figure}[htp]
    \centering
    \includegraphics[width =8cm]{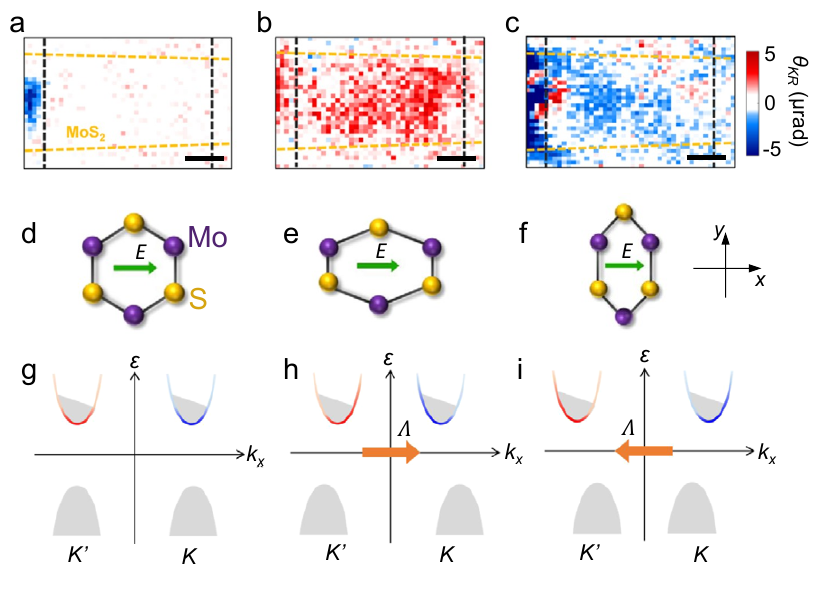} 
    \caption{\textbf{Edelstein effect in the strained monolayer MoS$_2$.} (a-c) The Kerr rotation maps are presented for different strain conditions: (a) under zero strain, (b) with strain applied along the zigzag direction (the $x$ direction), and (c) with strain applied along the armchair direction (the $y$ direction) while subjected to a finite bias in the $x$ direction. The scale bar is 5 $\mu$m. (d-f) illustrate the monolayer MoS$_2$ sample under varying strain conditions: (d) at zero strain, (e) under strain along the $x$ direction, and (f) under strain along the $y$ direction. The green arrows indicate the direction of the biased electric field. (g-i) Schematics of the modified band structure and Berry curvature distribution under variable strain corresponding to panels (d-f), with the orange arrows denoting the direction of the Berry curvature dipole. The Fermi level is tilted as a result of the applied electric field. From Ref.~\cite{Son:2019Strain}.
   }
\label{E_Effect}
\end{figure}

In monolayer TMDs, such as MoS$_2$, the Berry curvature exhibits valley locking. When a uniaxial strain is applied, breaking the $C_{3z}$ symmetry, a Berry curvature dipole is introduced in monolayer MoS$_2$. In the presence of an electric field, this leads to the formation of a current, causing the Fermi surface to tilt in the direction of the bias. Although the two valleys retain equal populations, the altered band structure along with the Berry curvature distribution, combined with the non-equilibrium carrier distribution, results in a net orbital magnetization (Fig. \ref{E_Effect}d-i). The expression for the Edelstein effect in strained monolayer TMDs can be written as \cite{Lee:2017Valley}: 

\begin{equation}
    M_\gamma^{or}=\chi_{\gamma \alpha \beta} E_{\alpha}^p \times J_{\beta}
\end{equation}
In this equation, $M_\gamma^{or}$ represents the orbital magnetization in the $\gamma$ direction, $\chi_{\gamma \alpha \beta}$ is the coefficient associated with the Edelstein effect, $E_{\alpha}^p$ denotes the strain-induced piezoelectric field in the $\alpha$ direction, and $J_{\beta}$ is the current density in the $\beta$ direction. Lee et al. and Son et al. have measured the orbital magnetization through Kerr rotation in strained monolayer MoS$_2$ \cite{Lee:2017Valley, Son:2019Strain} (Fig. \ref{E_Effect}a-c). 

In recent years, numerous new noncentrosymmetric superconductors have been discovered, showcasing a diverse range of structural properties \cite{yip2014noncentrosymmetric,  smidman2017superconductivity}. These include superconductors with chiral lattice symmetry, which exhibit unique electronic behaviors due to their lack of inversion symmetry, as well as transition metal dichalcogenides with various lattice configurations that influence their superconducting characteristics \cite{lu2015evidence, hossain2025tunable, wakatsuki2017nonreciprocal, PhysRevLett.118.087001, xia2025superconductivity, guo2025superconductivity, devarakonda2020clean, xu2025signatures, xu2025chiral}. From a theoretical perspective, Edelstein effect—where spin polarization occurs as a response to supercurrents—can also be observed in these newly discovered materials, particularly those belonging to gyrotropic point groups \cite{he2020magnetoelectric}. These developments open up new avenues for understanding unconventional superconductivity and potential applications in quantum computing.

\begin{figure}[htp]
    \centering
    \includegraphics[width =8cm]{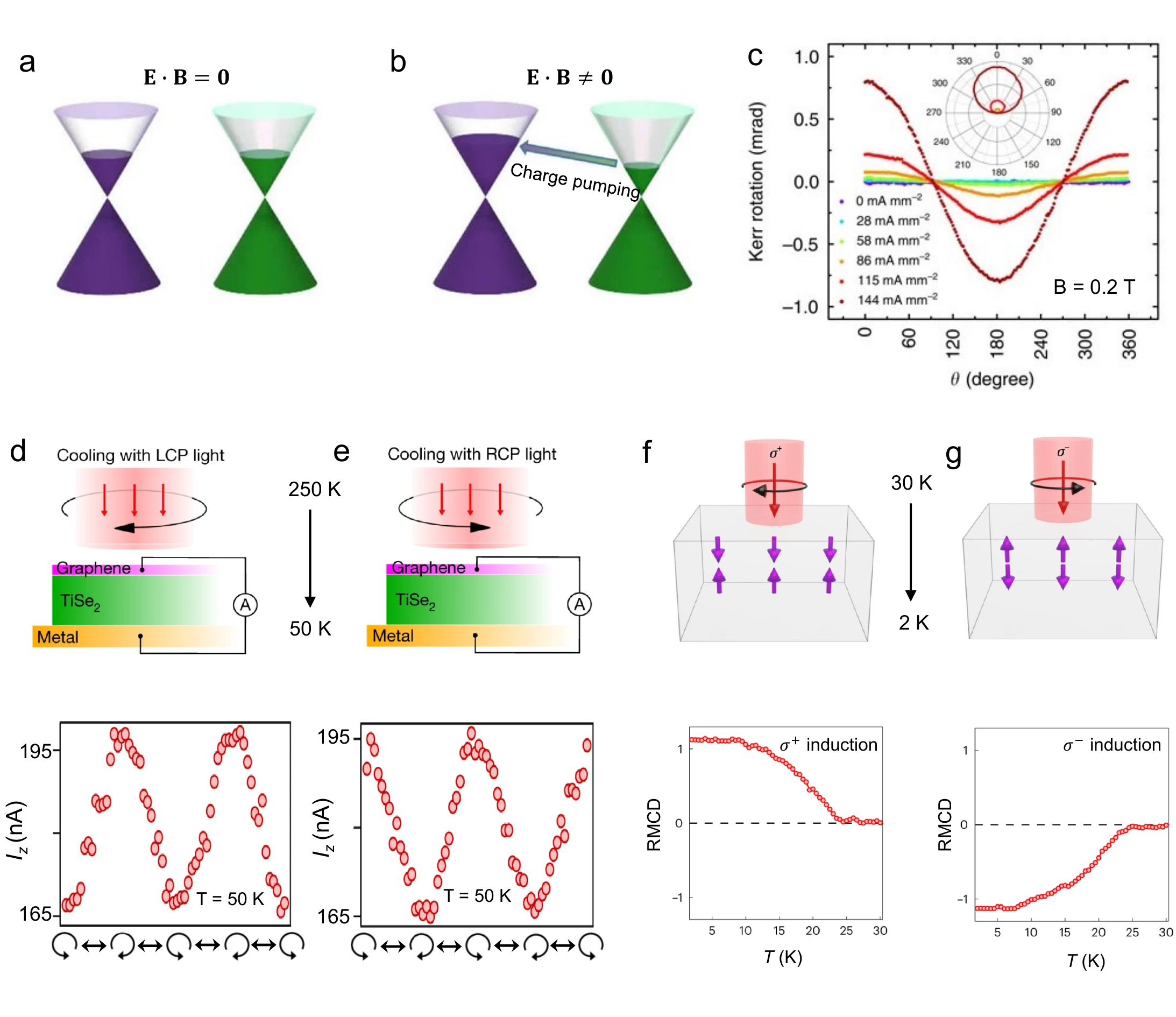} 
    \caption{\textbf{Natural optical activity and induction.} (a) Illustration of two degenerate Weyl nodes with different chiralities. (b) Charge pumped from one Weyl node to the other in the presence of an $\bf E \cdot B$ field due to chiral anomaly. (c) MOKE data for Cd$_3$As$_2$ as a function of the angle $\theta$ between the bias field $E$ and $B$ at varying current densities. Inset: Data plotted in polar coordinates. Adapted from Ref.\cite{Zhang:2017Room}. (d, e) Schematic of light induction, where left-circularly polarized (LCP) and right-circularly polarized (RCP) light are applied while cooling from 250 K to 30 K. Bottom panel: Out-of-plane photocurrent $I_z$ at $T = 50$ K as a function of light polarization. The incident wavelength is 10.6 $\mu$m, and the thickness of TiSe$_2$ is approximately 200 nm. Adapted from Ref.\cite{Xu:2020Spontaneous}. (f, g) Application of LCP (f) and RCP (g) light on eight-septuple-layered MnBi$_2$Te$_4$ during cooling from 30 K to 2 K. Bottom panel: Recorded reflection circular dichroism (RCD) signals as a function of temperature with different chiral light inductions, where RCD = $\frac{\sigma^+-\sigma^-}{\sigma^++\sigma^-}$. The light wavelength is 840 nm, and the induction power is about 1 mW. Adapted from Ref.~\cite{Qiu:2023Axion}.
    }
\label{Optical_activity}
\end{figure}

\color{blue} \underline{Natural Optical Activity:} \color{black} Natural optical activity refers to the ability to rotate the plane of polarization of linearly polarized light as it passes through some materials without mirror symmetry. This phenomenon results from the interaction of light with chiral molecules or crystals. The direction of rotation — whether clockwise or counterclockwise — depends on the chirality of the structural components. In chiral crystals, the inherent chirality of the lattice selectively couples with specific chiral light, leading to effects such as circular dichroism (CD) and optical rotation. This optical activity induced by chiral structures has been well understood \cite{Barron:2009Molecular, Kizel:1975Experimental}. Microscopically, the optical activity arises from the spatially dispersive part of the optical activity, where one can expand the optical activity $ \sigma_{ab}(\bf{q},\omega)$ as a function of the light wavevector $\bf q$ \cite{Malashevich:2010Band}:

\begin{equation}
    \sigma_{ab}(\bf{q},\omega)=\sigma_{ab}^{(0)}(\omega)+\sigma_{abc}(\omega)q_c+\dots
\end{equation}
where $\sigma_{abc} (\omega)$ is the spatially dispersive parts of the optical activity, and $q_c$ is the light wavevector in the $c$ direction. In time-reversal invariant but noncentrosymmetric crystals, $\sigma_{abc} (\omega)$ is nonzero and describes the natural optical activity. In addition to chiral structures, the quantum geometric effects of wavefunctions, specifically Berry curvature, can also give rise to natural optical activity. Theoretically, the antisymmetic part of $\sigma_{abc} (\omega)$ can be expressed as \cite{Malashevich:2010Band}:
\begin{equation}
    \sigma^{\text {A}}_{abc}(\omega)=ic(\epsilon_{bcd}\widetilde{\beta}_{ad}-\epsilon_{acd}\widetilde{\beta}_{bd})
\end{equation}
where
\begin{equation}
    \widetilde{\beta}_{ab}=\beta_{ab}+\frac{\omega}{4ic}\epsilon_{bcd}(2\xi_{acd}-\xi_{cda})
\end{equation}

Hence, natural optical activity, characterized by the response tensor $\sigma_{abc}(\omega)$, consists of two contributions: the current-induced magnetization ($\beta$) and the quadrupole term ($\xi$). The current-induced magnetization is directly proportional to the Berry curvature dipole, demonstrating the deep connection between $\sigma_{abc}(\omega)$ and Berry curvature.

Physically, this Berry-curvature-driven optical activity can be understood as follows: The Berry curvature imparts a momentum-space circular motion to electrons, and this orbital motion selectively couples with chiral light, thereby generating optical activity.

 Weyl semimetals are characterized by Berry curvature monopoles positioned at the Weyl points, each associated with a topological quantum number known as chirality. These Weyl points always appear in pairs, with each point exhibiting opposite chiralities (Fig. \ref{Optical_activity}a). By applying an $\bf E \cdot B$ field to transfer electron charge from one Weyl point to the other, due to chiral anomaly, one can observe natural optical activity through CD \cite{Hosur:2015Tunable, Ma:2015Chiral, Zhong:2016Gyrotropic} (Fig. \ref{Optical_activity}b). While a Dirac node consists of two degenerate Weyl nodes with opposite chirality, this degeneracy can be lifted upon applying a magnetic field \cite{Wang:2013Three}. Consequently, natural optical activity can also be observed in Dirac semimetals. Zhang et al. observed the magneto-optic Kerr effect (MOKE) in the three-dimensional Dirac semimetal Cd$_3$As$_2$ under an $\bf E \cdot B$ field, which serves to polarize one Weyl point \cite{Zhang:2017Room} (Fig. \ref{Optical_activity}c).

Chiral light not only provides a means to probe the chirality of quantum devices but also enables the manipulation of order through light-matter interactions. 
Xu et al. \cite{Xu:2020Spontaneous} demonstrated the induction of chiral light-driven gyrotropic electronic order by illuminating 1T-TiSe$_2$ with mid-infrared chiral light while cooling it below the critical temperature (Fig. \ref{Optical_activity}d,e). In the topological antiferromagnet MnBi$_2$Te$_4$, chiral light with opposite helicities can couple differently to the distinct AFM domains, thus altering the free energy between the two AFM orders \cite{Qiu:2023Axion} (Fig. \ref{Optical_activity}f,g).

\subsection{Photocurrent Driven by Quantum Geometry and Topology \label{sec: OPT}}

\begin{figure*}[ht]
\begin{center}
 \includegraphics[width=18cm]{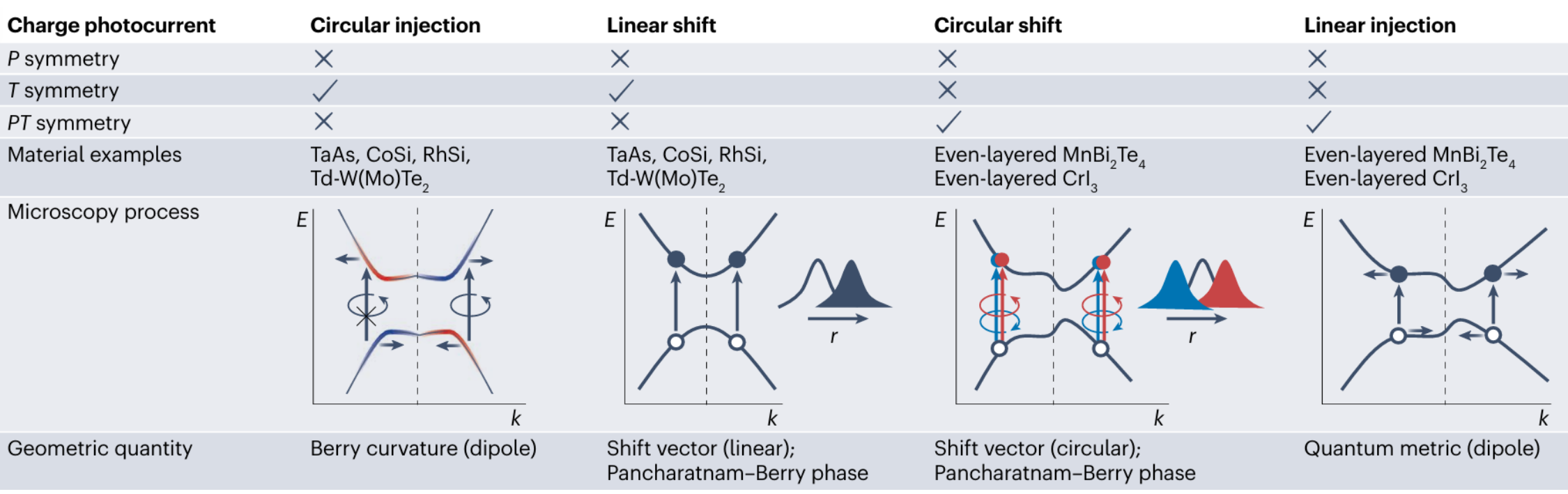}
 \caption{
{\bf Various mechanisms of photocurrent due to quantum geometry.} Time-reversal ($\cal{T}$) and spatial inversion ($\cal{P}$) are two important symmetries to determine the various types of second-order photocurrent with respect to the strength of the electric field of light. At the bottom, the corresponding geometric quantity relevant to each mechanism is shown. From Ref.\cite{Ma2023}.
}
\label{fig:geometry2}
\end{center}
\end{figure*}

Nonlinear optics is an important subject, both in fundamental physics and applications developed by advanced laser techniques. Especially, it has been recently recognized that quantum geometry and topology play an essential role in the nonlinear optical processes where the photocurrent or electric polarization are induced as the higher-order responses to the electric field of light \cite{Ma2021, Ma2023, doi:10.7566/JPSJ.92.072001,
annurev:/content/journals/10.1146/annurev-conmatphys-032822-033734}. Usually, the photocurrent is produced in two steps. The first step is the creation of photo-carriers by the interband transitions induced by the photo-radiation. The second step is the acceleration of the photo-carriers by the external dc electric field or the potential gradient due to e.g. $pn$-junction. When the electron-hole Coulomb attraction forms the exciton, the separation into electrons and holes is necessary to obtain the dc photocurrent. This conventional photocurrent can be described by the Boltzmann transport theory, where the group velocity of the photocarriers gives the photocurrent, i.e., the intraband matrix element of the current.
On the other hand, recent studies have revealed that the interband matrix elements of the current operator also play a crucial role in the photocurrent. The interband matrix elements of the current operator are closely related to the geometry and topology of the Bloch states, such as the Berry connection/curvature. This is a new aspect because the geometric nature originates mostly from the adiabatic processes and low-energy phenomena discussed thus far.  

A representative phenomenon of this nontrivial and geometric optical effect is the bulk photovoltaic effect (BPVE), where the light irradiation to bulk crystals leads to a dc current generation without the external bias in noncentrosymmetric crystals \cite{Boyd, Bloembergen, Sturman}. Figure \ref{fig:geometry2} shows the summary of the quantum geometric photovoltaic effect classified according to the symmetry of the system \cite{Ma2023}.
The time-reversal $\cal{T}$, spatial inversion $\cal{P}$, and the product of both $\cal{PT}$ are the important symmetries. It is the second-order optical response with respect to the electric field of light as described by
\begin{align}
    J_\gamma=\sigma^{(2)}_{\gamma\alpha\beta}(\omega) E_\alpha(\omega)E_\beta(-\omega)
    \label{eq: sigma (2)}.
\end{align}
Here the dc current density $J_\gamma$ along the $\gamma$ direction is induced by the combination of $E_\alpha(\omega)$ and $E_\beta(-\omega)$ corresponding to the $\alpha$ and $\beta$ directions with the frequency $\omega$ and $- \omega$, respectively.
Note that the wavevector of light is zero here.

There are two main mechanisms of BPVEs; (i) shift current, which comes from the difference of the Berry connection, i.e., the intracell coordinates, between the two Bloch states before and after the optical transition 
\cite{Baltz, Belinicher82, Sipe, Young-Rappe, Young-Zheng-Rappe, Cook17,
Morimoto-Nagaosa16, Tan16, Sotome19, Burger19, Hatada20}. (ii) injection and/or ballistic current which comes from a ballistic motion of photocarriers with their group velocities 
\cite{Belinicher-Sturman80, Belinicher-Sturman88, Sipe, deJuan17, Orenstein21}. In the presence of time-reversal symmetry, the injection current is possible only with the 
circularly polarized light, i.e.,  circular photogalvanic effect (CPGE) \cite{Orenstein21}, while the shift current is possible even for the linearly polarized light. Below, we describe each of the mechanisms in more detail.

\subsubsection{Shift Current} As discussed in Sec. IIIA, the electric polarization $P$ of Bloch electrons is described by the Berry phase as the intracell coordinates. When electron-hole pairs are created, the change in the polarization occurs, i.e., $\dot{P}$, occurs in the steady nonequilibrium state, and the dc current is produced. This is an intuitive picture of the shift current. Microscopic theory for the shift current has been developed by several authors \cite{Baltz, Sipe, Young-Rappe, Parker19}. 
The diagrammatic approach gives the second-order nonlinear conductivity 
$ \sigma^{(2)}_{\mu\alpha\alpha}(\omega)$ as \cite{Parker19,PhysRevResearch.2.033100}

\begin{align}
    &\sigma^{(2)}_{\mu\alpha\alpha}(\omega) =
    - \frac{e^3}{\hbar^3\omega^2} \int \frac{dk}{(2\pi)^d} \Bigg[ 
    \sum_{a,b}
    \frac{f_{ab}(\partial_{k_\mu}v_\alpha)_{ab}v_{\alpha,ba} }{\hbar \omega - \epsilon_{ba}+i\delta}
    \nonumber \\
    &
    + \sum_{a,b,c\neq a}
    \frac{v_{\mu,ac}v_{\alpha,cb}v_{\alpha,ba}}{\epsilon_{ac}}
    \left( \frac{f_{ab}}{\hbar\omega-\epsilon_{ba}+i\delta}-\frac{f_{cb}}{\hbar\omega-\epsilon_{bc}-i\delta} \right) 
    \bigg] 
    \nonumber \\
    &+ (\omega \leftrightarrow -\omega),
    \label{eq:sigma2perturbation}
\end{align}
where $a,b,c$ are band indices, $f_{ab}=f_a-f_b$ ($\epsilon_{ab}=\epsilon_{a}-\epsilon_{b}$) is the difference in the Fermi distribution function (energy dispersion), and $\delta$ is the relaxation rate. 
The velocity operator $v_\mu=\partial_{k_\mu} H(k)$ along the $\mu$ direction
has the matrix element $v_{\mu,ab}=\langle u_a | v_\mu | u_b \rangle$. This generic expression is reduced to the more comprehensive one when the resonant contribution in a time-reversal symmetric system is picked up as \cite{Sipe, Young-Rappe}:
\begin{align}
J_{{\rm shift}}^{\mu\alpha\alpha}
&= \frac{\pi e^3}{\hbar^3 \omega^2} |E_\alpha(\omega)|^2
\sum_{ab}\int[dk] f_{ba}|v_{\alpha,ab}|^2 R_{\mu\alpha,ba} \delta(\hbar\omega-\epsilon_{ba}),
\label{eq:shiftcurrent}
\end{align}
where 
\begin{align}
    R_{\mu\alpha,ba} &=
\frac{\partial}{\partial k_\mu}\mathrm{Im}(\log v_{\alpha,ab})
+ a_{\mu,b}- a_{\mu,a}
\label{eq:Rk}
\end{align}
is called the shift vector, which is the gauge invariant "difference" of the Berry connection between the band $a$ and band $b$. Therefore, eq.(\ref{eq:shiftcurrent}) can be read as "every time the optical transition occurs with the probability proportional to $|v_{ab}|^2$ satisfying the energy conservation $\hbar\omega=\epsilon_{ba}$, the shift $R_{\mu\alpha,ba}$ of the wavepacket occurs to result in the dc current." \cite{nakamura2017shift, nakamura2018impact}

Note that the shift current is not the transport current of the photocarriers. In eq.(\ref{eq:shiftcurrent}), the photocarriers are shifted when they are created, i.e., they are not driven by the external electric field. This means that the shift current is the geometric current similar to the polarization current, although it is dc current, and hence shows several highly nontrivial physical properties. One is that it is insensitive to the transport lifetime of the photocarriers. Even the strong localization does not suppress the shift current \cite{Hatada20,doi:10.1073/pnas.2023642118} It is also discussed that there is no shot noise of the shift current \cite{Morimoto-IV18}. An even more surprising fact is that the shift current does not require the free photocarriers at all. This was pointed out for the exciton shift current, where the continuous creation of polarization change is enough for the dc current \cite{Morimoto-exciton16, Chan21}. The shift current of excitons is observed in a semiconductor CdS \cite{Sotome21}, in CuI \cite{Nakamura2024}, and WSe$_2$ and black phosphorus \cite{Akamatsu21}. This idea without the photocarriers is extended to lower energy excitations such as electromagnons in multiferroics
\cite{Pimenov06,Aguilar09,Takahashi12, kubacka2014large,Morimoto-magnon19,
Morimoto-magnon21}, and phonons in ferroelectric BaTiO$_3$ \cite{Okamura22}.

%

\subsubsection{Circular Photogalvanic Effect}

The circular photogalvanic effect (CPGE), an example of injection current, is an intriguing phenomenon whereby circularly polarized light induces a dc electric current in non-centrosymmetric materials~\cite{Sipe}. This effect results in a reversal of the generated dc signal when the chirality of the light is inverted. The general formula for the CPGE can be expressed as:

\begin{equation}
    J^{\text{CPGE}}_\alpha = \eta_{\alpha\beta\gamma} E_{\beta}(\omega) E_{\gamma}(-\omega)
    \label{equ_CPGE}
\end{equation}
where, $E_{\beta}(\omega)$ represents the incident electric field with frequency $\omega$ along the $\beta$ direction. The indices $\alpha$, $\beta$, and $\gamma$ denote the directions of the optical electric fields. The tensor $\eta_{\alpha\beta\gamma}$ is a third-rank tensor that is non-zero only when the system lacks inversion symmetry.

In materials that do not possess inversion symmetry, the degeneracy of spin or Berry curvature may be broken. When subjected to circularly polarized light, an uneven distribution of charge carriers occurs across the energy bands due to optical selection rules, resulting in a net electric current. Next, we will examine how Berry curvature generates CPGE in various quantum materials~\cite{Hosur:2011PRB}.

\begin{figure}[htp]
    \centering
    \includegraphics[width =8cm]{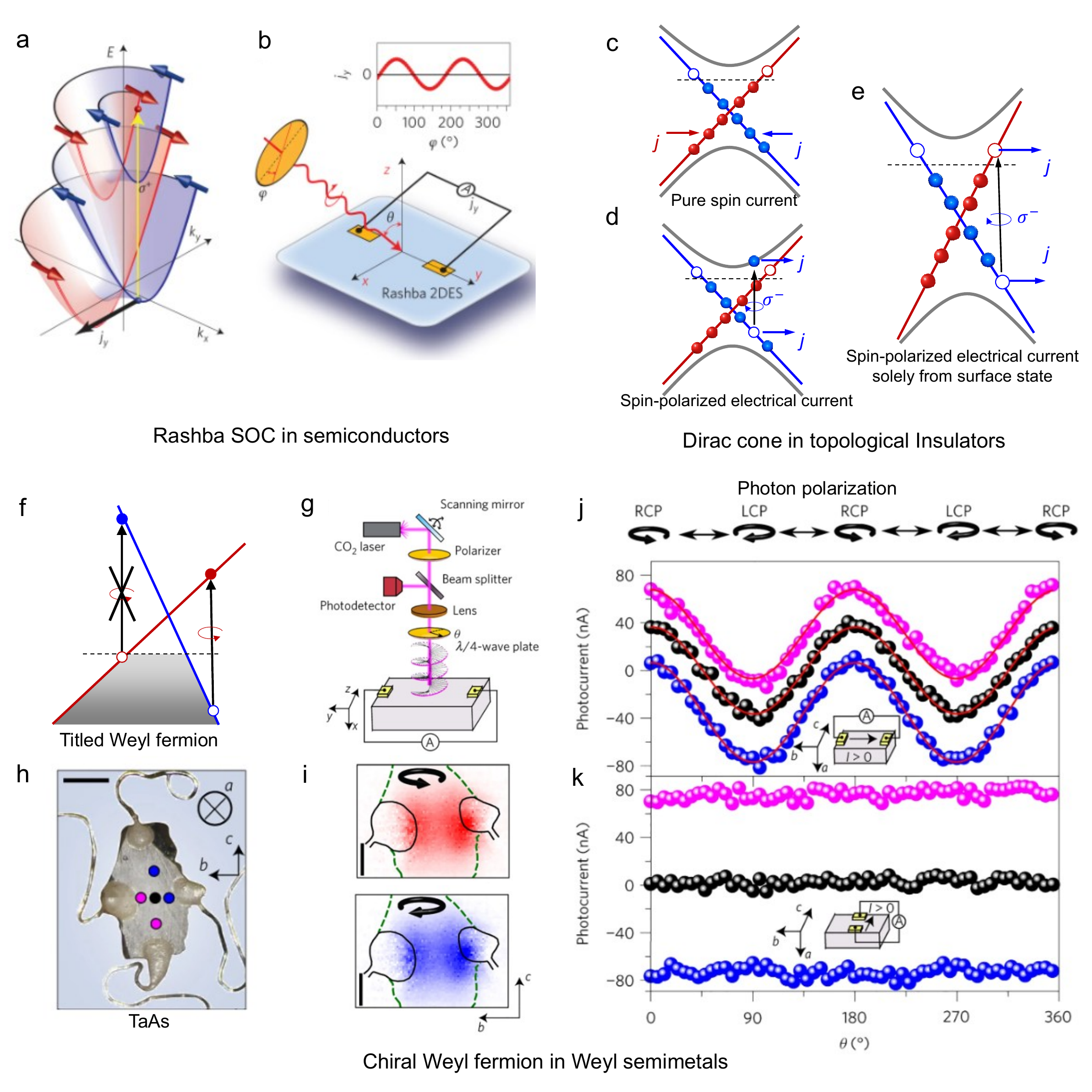} 
    \caption{\textbf{Circular photogalvanic effect in nonmagnetic materials.} (a) The schematic illustrates the SOC-induced CPGE within a Rashba spin-splitting band diagram. The yellow arrow denotes the inter-subband transition. (b) The schematic of the CPGE measurement setup, where $\theta$ indicates the angle between the incident light and the $z$ direction. The helicity of the laser light is modulated by a rotatable quarter-wave plate angle $\phi$. Inset: the generated CPGE current as a function of $\phi$. From Ref.~\cite{Yuan:2014generation}. (c) A schematic of pure spin currents emanating from the Dirac cone in an equilibrium state, where the currents from opposite spins cancel each other out. (d) A schematic of the spin-polarized electrical current induced by optically exciting the Dirac cone. (e) A schematic of the spin-polarized electrical current generated solely from the topological surface state. (f) A schematic of the CPGE generated in a tilted Weyl fermion system. Due to the Pauli blockade, excitation occurs only on one side of the Weyl fermion, resulting in a net current. (g) A schematic of the mid-infrared photocurrent microscope setup with the laser energy $\sim$120 meV. (h) A photograph of the measured TaAs sample, with a scale bar 300 $\mu$m. The photocurrent mappings with right-handed polarized light (top panel) and left-handed polarized light (bottom panel). (j, k) The photocurrents in the $c$ and $d$ directions of the sample as a function of the photon polarization angle at 10 K. The curves in different colors correspond to measurements taken in various areas of the sample in (h). From Ref.~\cite{Ma:2017direct}.}
\label{CPGE}
\end{figure}

\color{blue} \underline{Semiconductor with Spin-Momentum Locking:} \color{black} In materials with significant SOC, the spin degeneracy of sub-bands can be removed. When circularly polarized light is incident on these materials, the chirality of the light interacts with the electron spin through conservation of angular momentum. This interaction leads to spin and momentum-locked energy bands with distinct spin momenta (or Berry curvature) that exhibit different absorption characteristics for the two types of circularly polarized light. As depicted in Fig.~\ref{CPGE}a, right-handed circularly polarized light selectively excites electrons from one spin-direction band to the band of the opposite spin direction. This results in a net drift of charge carriers within the conduction band along the momentum direction dictated by the spin locking (Fig.~\ref{CPGE}b). The SOC-induced CPGE has been observed in various systems, including 2D electron gas systems \cite{ganichev:2002spin}, TMDs \cite{Yuan:2014generation}, silicon nanowires \cite{Dhara:2015Science}, chiral tellurium crystals \cite{Niu:2023NL}, and chiral lead halide perovskites \cite{Niesner:2018PNAS}.

\color{blue} \underline{Topological Insulators:} \color{black} Three-dimensional TIs demonstrate non-trivial topology characterized by helical Dirac fermions, which are linked to substantial Berry curvature. It is well established that a pure spin current can be generated when the Fermi level intersects the topological surface state (Fig.~\ref{CPGE}c). When TIs are driven out of equilibrium by circularly polarized light, this pure spin current can be transformed into a spin-polarized net electrical current (Fig.~\ref{CPGE}d) \cite{Hosur:2011PRB, Mciver:2012control}. The mechanism of the CPGE in TIs is akin to the SOC-induced CPGE. However, in TIs, the CPGE current originates solely from the surface states and is predominantly influenced by contributions from Berry curvature. Due to the inversion symmetry present in the bulk of TIs, the contribution from the bulk is zero, since CPGE—being a second-order nonlinear effect—is forbidden in inversion symmetric systems. Consequently, the CPGE observed in topological materials originates from the isolated surface states, underscoring their significant potential applications in spintronics \cite{Wunderlich:2010spin}. There is still a need for experimental demonstration of the CPGE arising exclusively from surface interband transitions in TIs (Fig.~\ref{CPGE}e). Such validation would further affirm the theoretical predictions \cite{Hosur:2011PRB} and enhance the understanding of surface-related phenomena in spintronics applications. 

\color{blue} \underline{Weyl Semimetals:} \color{black} Weyl semimetals, as a new class of topological materials, host chair Weyl node, which is a monopole of Berry curvature. Weyl cones possess specific chiralities and always exist in pairs, functioning as topological monopoles or antimonopoles of Berry curvature. Consequently, the overall sum of the photocurrents generated from a pair of Weyl nodes must necessarily vanish, resulting in no net photocurrent being observed. However, the Weyl cone can be tilted after breaking $\mathcal{P}$. The tilt introduces asymmetries in the excitation processes, allowing for the generation of a net photocurrent (Fig.~\ref{CPGE}f) \cite{PhysRevB.95.041104, Ma:2017direct, Xu:2018NP, Ma:2019nonlinear, ji2019spatially}. Therefore, CPGE can be observed in a titled Weyl node. Recent theoretical advancements indicate that when the inter-band transition is limited to just the lowest two bands, the CPGE in three-dimensional bulk materials can be attributed to a clear microscopic origin related to the nontrivial Berry curvature \cite{PhysRevB.95.041104}. This notion was supported by the work of Ma et al., who observed the CPGE in the bulk Weyl semimetal TaAs using mid-infrared circularly polarized light (Fig.~\ref{CPGE}f-k) \cite{Ma:2017direct}. Their findings further elucidate the CPGE can be used to determine the chirality of Weyl fermion. Xu et al. observed notable CPGE in monolayer type-II Weyl semimetal WTe$_2$ which directly connected to the BCD \cite{Xu:2018NP}.  Ma et al. observed CPGE in the bulk type-II Weyl semimetal TaIr$_2$Te$_4$, utilizing a built-in electric field to tilt the Fermi levels between oppositely-chirality Weyl nodes \cite{Ma:2019nonlinear}.

Furthermore, Juan et al. proposed that the CPGE can be quantized in Weyl semimetals \cite{deJuan17}. When a class of Weyl semimetals has opposite chirality Weyl nodes situated at different energy levels, the magnitude of the photocurrent generated by circularly polarized light depends uniquely on the Chern number of a single Weyl node. In this scenario, the CPGE can indeed be fully quantized. However, this phenomenon is still awaiting experimental observation.

\section{Quantum Phenomena Associated with Quantum Metric}
\label{sec:sec2b}

The studies of Berry curvature have led to many breakthroughs, including the intrinsic anomalous Hall effect in ferromagnetic materials, the optical activity in chiral media, and various topological phases of matter. In sharp contrast to Berry curvature, for a long time, there have been virtually no studies on the quantum metric. However, the situation has changed drastically in recent years: quantum metric is receiving extensive interest, many new theoretical proposals are emerging, and exciting experimental discoveries are being made. In this section, we review the research progress on the quantum metric.

\textbf{Studies of Quantum Metric in Qubits and Photonic Microcavities}

Initial studies of the quantum metric have occurred in qubit and photonic microcavity (polariton) systems. The quantum metric is the distance between adjacent wavefunctions in a band structure. In these systems, the quantum wavefunction can be directly measured, thus allowing for a straightforward measurement of the quantum metric.

\begin{figure}[h]
\centering
\includegraphics[width=8.5cm]{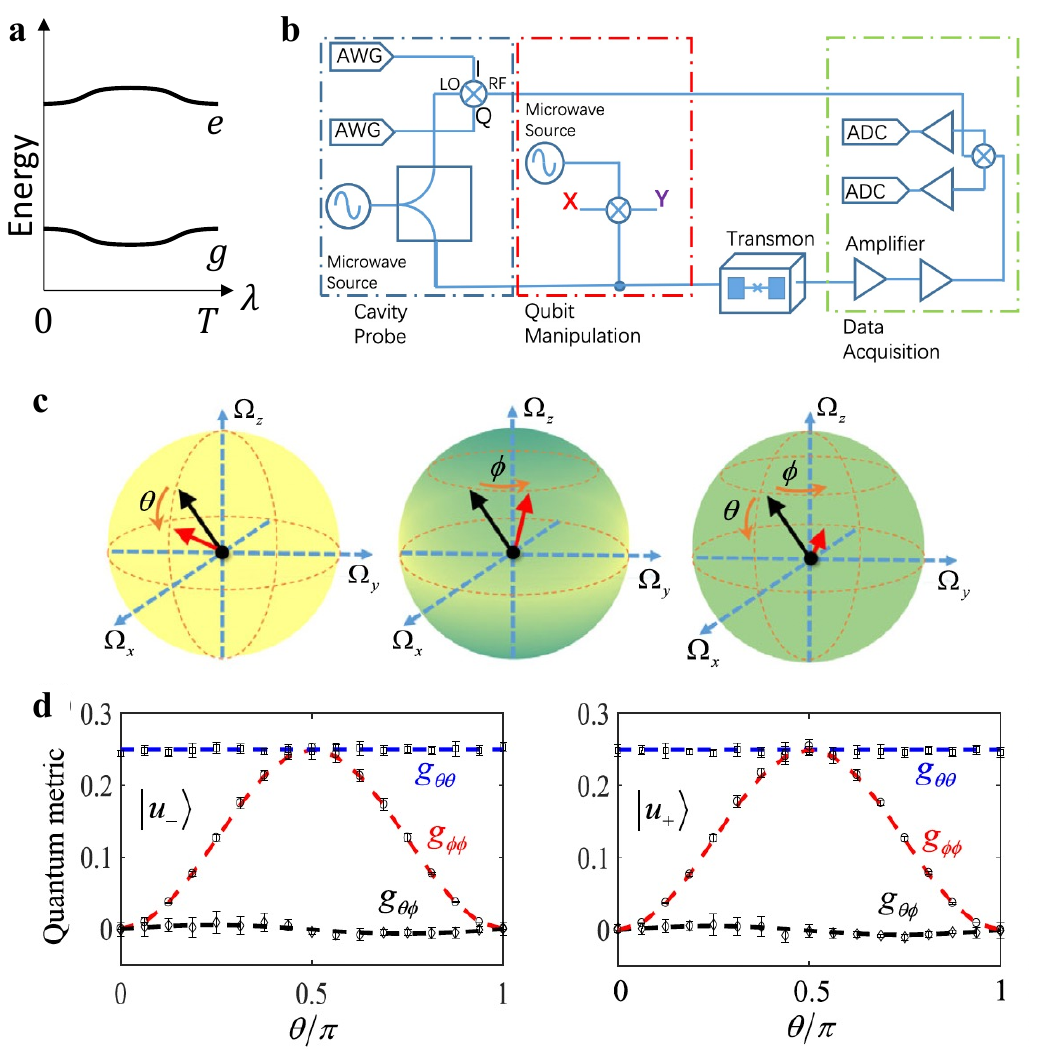}
\caption{\textbf{Simulation of the quantum metric in qubit systems.} (a) A qubit consists of two quantum levels. Under a periodic driving field $\lambda$, they simulate two electronic bands.  (b) Schematic of the transmon superconducting qubit. (c) Illustration of the periodic driving field. (d) Measured quantum metric of the ground and excited states under the period driving. From Ref. \cite{Tan2019}}
\label{Qubit}
\end{figure}

\begin{figure*}
\centering
\includegraphics[width=17cm]{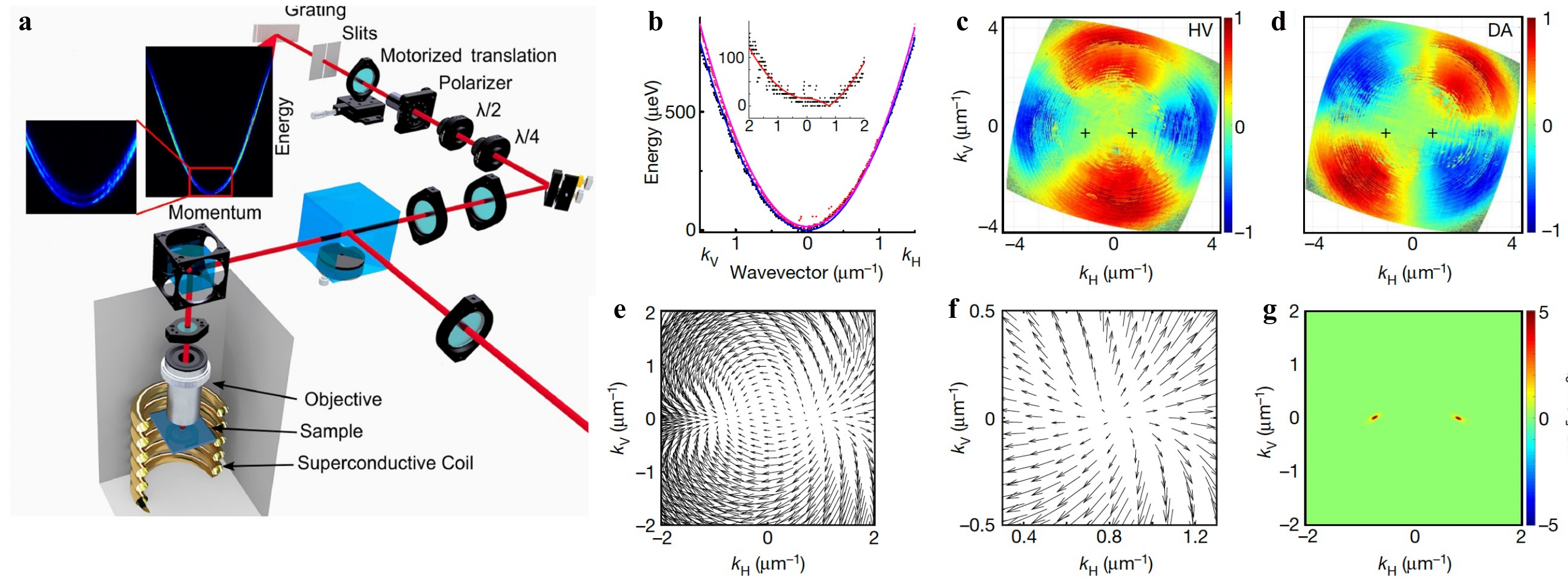}
\caption{\textbf{Experimental measurements of the quantum geometrical tensor of the exciton polariton bands in microcavities.} (a) Experimental setup. (b) Measured dispersion of the exiton polariton. (c,d) Degree of polarization of the lower band in $k$ space for HV and DA. (e,f) $k$-space in-plane pseudospin ($S_1$, $S_2$) texture. And its magnification near one of the crossing points that demonstrates a monopole texture. g, Quantum metric tensor trace ($g_{\textrm{HH}}$ + $g_{\textrm{VV}}$). From Ref. \cite{Gianfrate2020measurement}.}\label{Microcavity}
\end{figure*}

A qubit consists of two discrete quantum levels, but coupling the qubit with an external drive that is periodic in time can simulate an electronic band structure. Essentially, let us consider a periodic drive $\lambda$ that controls the quantum wavefunction of the qubit. The qubit wavefunction as a function of $\lambda$ is therefore directly analogous to the Bloch wavefunction as a function of $k$ (also periodic). Therefore, measuring the quantum distance between $\lambda$ and $\lambda+\delta\lambda$ in a qubit is equivalent to that between $k$ and $k+\delta k$ in a solid, which enables the simulation of the quantum metric. Using a superconducting transmon qubit, the authors in Ref. \cite{Tan2019} simulated the quantum metric. The transmon qubit realizes a highly tunable two-level system, where the wavefunctions of both the ground and the excited states can be fully controlled using microwave driving fields. Figure~\ref{Qubit} shows the experimental results using the sudden quench protocol: The system is initially prepared at an eigenstate of the Hamiltonian $\hat{H}(\lambda_0)$. Then the Hamiltonian is rapidly swept to $\hat{H}(\lambda_0+\delta\lambda)$, followed by a state tomography to obtain the transition probability. From state tomography, the quantum metric at $\lambda_0$ was extracted from the measured transition probability: $g\simeq P^{+}/\delta \lambda^2$, where $ P^{+}$ is the transition probability of the quantum state being excited to the excited state. Similar experiments have also been performed in other qubit systems such as the nitrogen vacancy center in a diamond \cite{yu2020experimental} and in the non-Abelian systems \cite{zheng2022measuring}.

\begin{figure*}
\centering
\includegraphics[width=17.2cm]{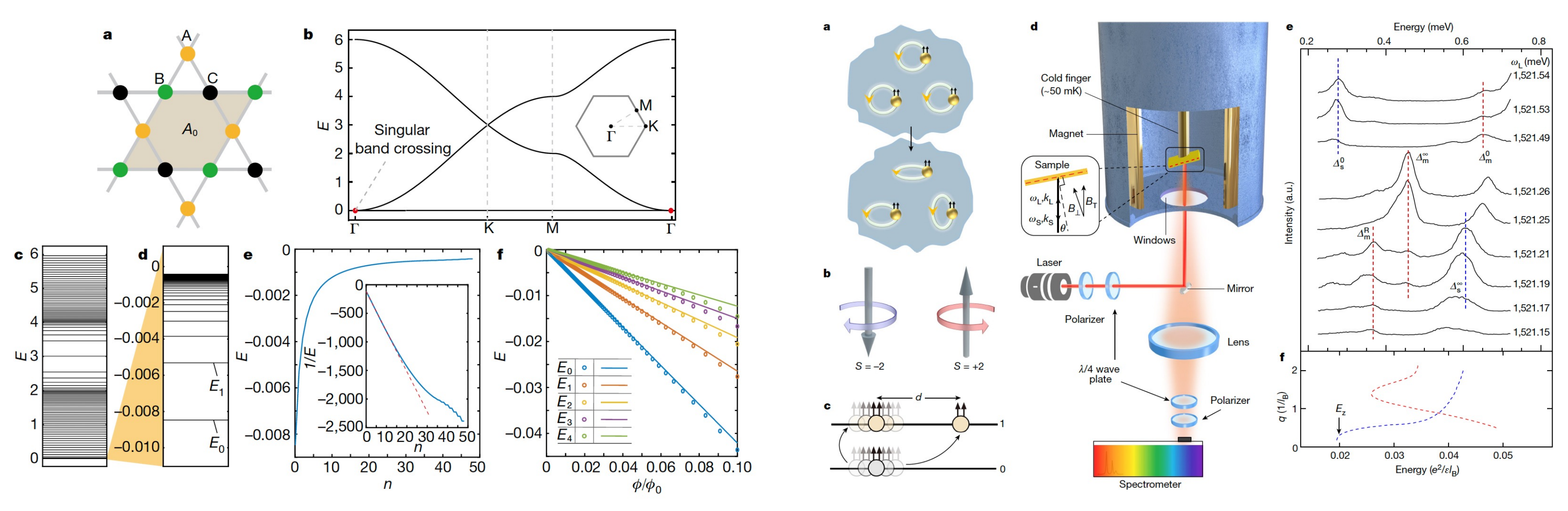}
\caption{\textbf{Quantum metric in Landau levels and quantum Hall systems.} (a) Theoretical modeling of the singular flat band in a model Kagome lattice and its the novel Landau level spectrum. From Ref. \cite{Rhim2020}. (b) Experimental detection of the chiral graviton modes in fractional quantum Hall states of a GaAs quantum well. The chiral graviton modes are proposed as quanta of fluctuations of an internal quantum metric under a quantum geometry description. From Ref. \cite{liang2024evidence}}\label{LL}
\end{figure*}

A microcavity hosts strongly coupled exciton-photon modes (exciton polaritons). The exciton polariton dispersion ($\omega-k$) is directly analogous to the electronic band structure. The polariton quantum wavefunction is a superposition between the transverse electric (TE) and transverse magnetic (TM) basis. In Ref. \cite{Gianfrate2020measurement}, the authors experimentally measured the quantum geometric tensor and the anomalous Hall drift in such an exciton polariton system. The microcavity adapted was a high quality-factor $3/2\lambda$ GaAs/AlGaAs planar cavity. Upon optical excitation, the microcavity shows  photoluminescence, which goes through waveplates, polarizers, and a monochromator with a grating and is captured by a CCD camera. This allows one to map out both the energy (i.e., $\omega$) as well as the wavefunction (i.e., the polarization of the photoluminescence) as a function of $k$. Figure~\ref{Microcavity} shows the measured dispersion of the two exciton polariton bands. Along $k_{\textrm{H}}$, the two polariton bands cross at around $\pm 0.8 \mu$m$^{-1}$, therefore realizing a pair of Dirac points. Figure~\ref{Microcavity}c,d show the measured polarization of the lower polariton band, from which the wavefunction of the lower band at every $k$ is extracted, as shown in Fig.~\ref{Microcavity}e,f. Clearly, the two Dirac points at $\pm 0.8 \mu$m$^{-1}$ serve as monopole and antimonopole of the wavefunction. Figure ~\ref{Microcavity}g shows the extracted quantum metric, which diverges at the Dirac points. In Ref.~\cite{Liao2021}, the authors studied the organic microcavity consisting of the DPAVBi molecule, which exhibits a polarization-dependent strong coupling. By tuning the gain and loss of the photonic cavity, exceptional points were achieved and the authors demonstrated a diverging quantum metric around the exceptional points.

\vspace{1cm}
\textbf{Studies of Quantum Metric in Quantum Materials}

Around the similar time when the quantum metric was measured in AMO systems, the studies of quantum metric in condensed matter started to accelerate. A crucial question is that: How does quantum metric influence the electronic motion in solids, giving rise to novel electronic transport and optical phenomena? Theoretical studies of this question have recently led to the prediction of a wide range of exotic quantum metric responses, with the initial focus on the flat band systems and nonlinear responses. Because a rigorous theoretical derivation has been given in Section III, below, we hope to provide an intuitive physical picture, which aims to illustrate (1) how quantum metric can lead to electronic motion and (2) why the effect is particularly strong in flat band systems and nonlinear responses.

\vspace{5mm}
\subsection{An Intuitive Physical Picture of Quantum-Metric-Driven Electron Motion}

Within the framework of the intuitive picture shown in Fig. 1, we can visualize how the quantum metric can lead to electronic motion and why the effect is strong in flat band systems and in the nonlinear responses. This is because the quantum metric is directly related to the location of the Bloch wave within the unit cell (more precisely, the location of the Wannier charge center). In particular, let us consider the application of an electric field that accelerates the wavevector from $k_1$ to $k_2$. In the case of a trivial atomic insulator with zero quantum metric shown in the top left panel of Fig.~\ref{QM_BC}, the Bloch wavefunction of both $k_1$ and $k_2$ are located at the same site; therefore, the change of wavevector does not lead to a shift of the Wannier charge center. By contrast, in the case of a band-inversion insulator with a large quantum metric shown in the bottom left panel of Fig.~\ref{QM_BC}, the Bloch wavefunction with $k_1$ locates away from that with $k_2$. Therefore, the change of wavevector directly generates a shift of the Wannier charge center. This is a longitudinal motion. In any metal, the longitudinal conductivity already has the regular Drude conductivity. Therefore, in order to make the quantum metric effect manifest strongly, one needs to reduce the Drude conductivity contribution; this may be achieved by having a flat band system (such as Landau levels, moiré materials, or Kagome lattices) or by going to the nonlinear order.

\vspace{5mm}
\subsection{Quantum Metric in Landau Levels and Quantum Hall Systems}

Quantum metric can manifest in a wide range of ways in Landau level systems. 

First, for an isolated flat band, theory \cite{hwang2021geometric} has shown that the Landau level spreading of isolated flat bands is a manifestation of the non-trivial wave function geometry of the flat band arising from inter-band couplings. 

Second, for a flat band that has a band crossing with other dispersive bands, theory \cite{Rhim2020} proposed that Landau levels develop an anomalous structure, which directly manifests the quantum geometry of the wave function associated with the singularity at the band crossing point. In particular, the Landau levels of a singular flat band develop in the empty region, in which no electronic states exist in the absence of a magnetic field, and exhibit an unusual $1/n$ dependence on the Landau level index $n$. The authors constructed a singular flat band in a model Kagome lattice and demonstrated the novel Landau level spectrum in their model calculations.

Third, in fractional quantum Hall system, novel collective excitations called chiral graviton modes are proposed as quanta of fluctuations of an internal quantum metric under a quantum geometry description. Such modes are condensed-matter analogues of gravitons that are hypothetical spin-$2$ bosons. They are characterized by polarized states with chirality of $+2$ or $-2$, and energy gaps coinciding with the fundamental neutral collective excitations (namely, magnetorotons) in the long-wavelength limit. In Ref. \cite{liang2024evidence}, the authors observed the chiral spin-$2$ long-wavelength magnetorotons in the fractional quantum Hall states of a GaAs quantum well using inelastic scattering of circularly polarized light.

\vspace{5mm}
\subsection{Flat Band Superconductivity Enabled by Quantum Metric}

\begin{figure}[h]
\centering
\includegraphics[width=8.5cm]{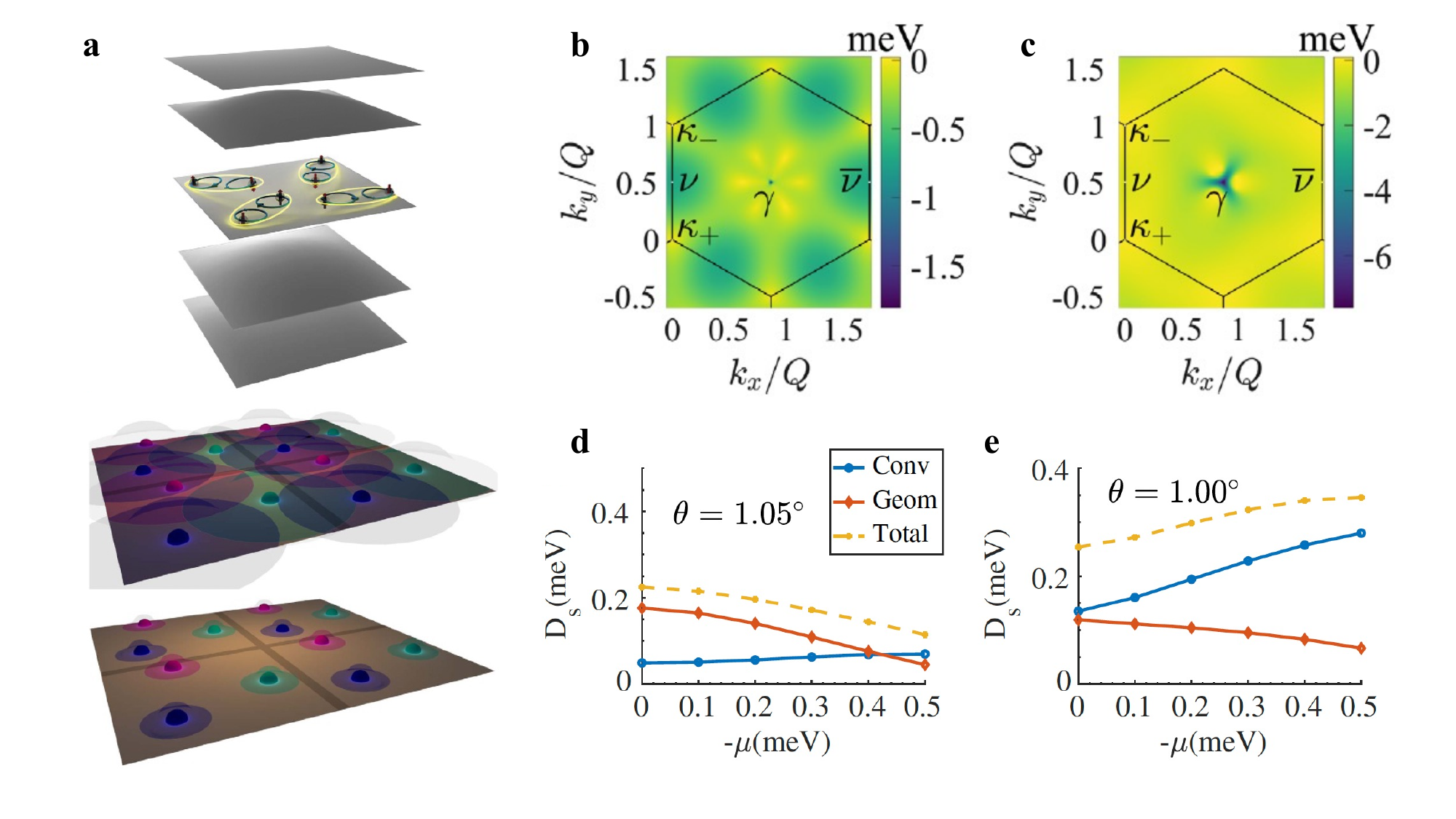}
\caption{\textbf{Theoretical studies of the superconductivity enabled by quantum metric.} (a) Schematic illustration of the extended Wannier orbital with nontrivial quantum metric and experimentally localized Wannier orbital with trivial quantum metric. From Ref.~\cite{peotta2015superfluidity}. (b,c) Calculated band structure of twisted bilayer graphene with twist angle $\theta=1.05^{\circ}$ and $\theta=1.00^{\circ}$. (d,e) Calculated superfluid stiffness arising from the conventional contribution and the quantum metric contribution. From Ref.~\cite{hu2019geometric}. }
\label{Flat_band_SC_THY}
\end{figure}

In a flat band system, charge carriers are nearly immobile. As such, it becomes difficult to understand superconductivity in a flat band system. In fact, simply adapting the established expressions from the BCS theory may lead to a paradox. For example, the superconducting coherence length is given by $\xi\sim\frac{\hbar v_{\textrm{F}} }{k_{\textrm{B}} T_{\textrm{c}}}$, which becomes zero in a flat band because $v_{\textrm{F}}\rightarrow 0$. Following that, the superconducting gap, which can be expressed as $\Delta \sim \hbar v_{\textrm{F}} \xi$, also approaches zero. The superfluid stiffness $D_s\sim\frac{e^2 n_s}{m}$ would also approach zero because $m\rightarrow \infty$. However, in a flat band with a nontrivial quantum metric, the quantum metric can generate a new contribution, therefore enabling superconductivity in a flatband system. 

\vspace{3mm}
\color{blue}\underline{Theory:} \color{black} The theory of this quantum metric induced superconductivity was initially proposed in Ref.~\cite{peotta2015superfluidity}. However, at the time, there wasn’t a material realization for such flat band superconductivity. The discovery of magic angle twisted bilayer graphene superconductivity (TBG) \cite{cao2018unconventional} in 2018 changed the situation. Indeed, quickly after this discovery, theory \cite{hu2019geometric,xie2020topology,julku2020superfluid} evaluated the quantum metric contribution and the conventional contribution to the superconductivity in TBG. Indeed, it was found that the quantum metric contribution is significant and may be dominating under certain parameter conditions.

\begin{figure*}
\centering
\includegraphics[width=17.2cm]{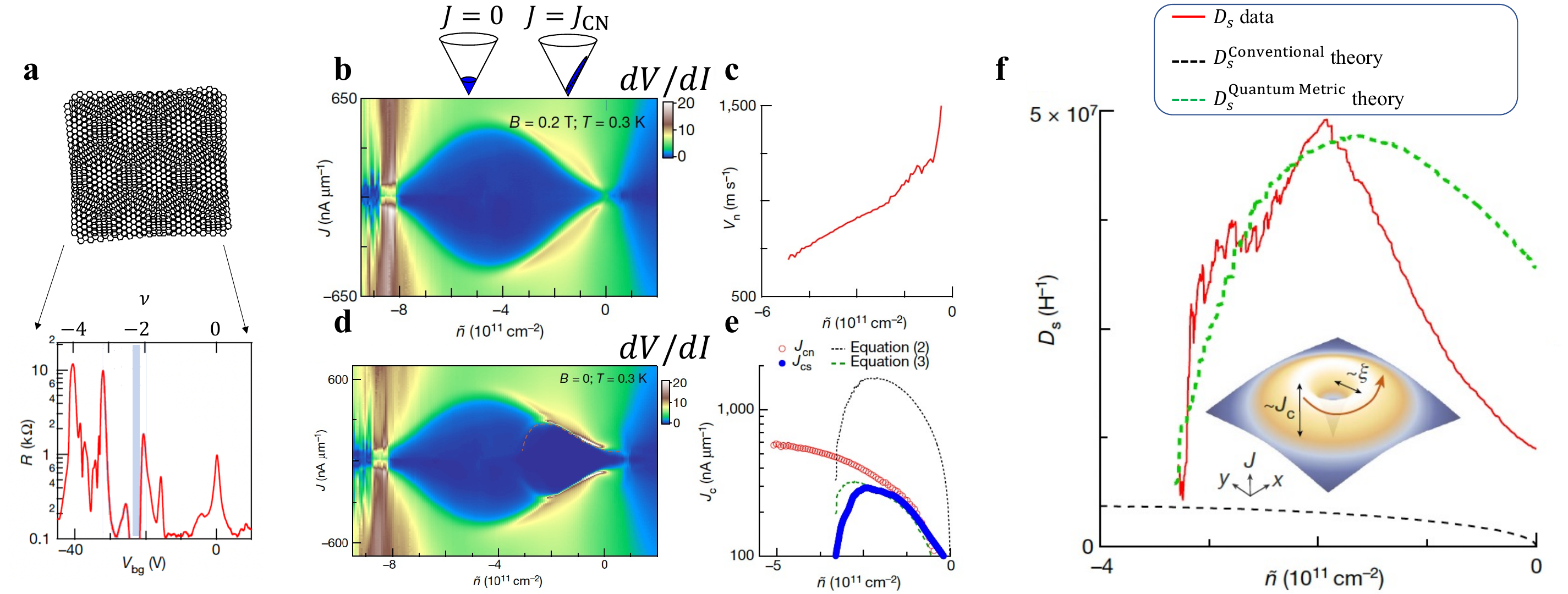}
\caption{\textbf{Transport measurements for the quantum metric induced flat band superconductivity in TBG.} (a) Twisted bilayer graphene. The sample is a $1.08^{\circ}$ TBG device. It hosts a superconducting domain near half-filling at $\nu=2$. (b) Measured differential resistance $dV/dI$ as a function of the applied current density $J$ and carrier density $n$ for the normal state at $B=0.2$ T and $T=0.3$ K. The $dV/dI$ shows a highly nonlinear dependence with a critical current density $J_\textrm{CN}$. This reveals the ``Schwinger mechanism'', where TBG's Fermi surface is entirely tilted over toward one side of the Dirac point by current (schematics above this panel). (c) the band velocity is derived from the measured $J_\textrm{CN}$ by $v=\frac{J_\textrm{CN}}{ne}$. (d) Same $dV/dI$ measurements but in the superconducting state by removing the magnetic field $B=0$. A superconducting dome is observed, which is the dark blue regime. (e) The superconducting critical current density $ J_\textrm{CS}$ coincides with the normal state critical current $J_\textrm{CN}$ for the carrier density range of $0<n<3\times10^{11}$ cm$^{-2}$. (f) Red curve: the superfluid stiffness can be obtained from the data following $D_s=(2\pi J_\textrm{CS} \xi)/\Phi_0$. Green and black curves: Theoretically calculated superfluid stiffness arising from the conventional mechanism and the quantum metric mechanism. From Ref. \cite{tian2023evidence}}\label{Flat_band_SC_EXP}
\end{figure*}

\vspace{3mm}
\color{blue}\underline{Experiment:} \color{black} Ref.~\cite{tian2023evidence} carried out electrical transport of the superconductivity in twisted bilayer graphene (TBG) sample, showing signature of the quantum metric contribution. The sample they studied is a $1.08^{\circ}$ TBG device. It hosts a superconducting domain near half-filling at $\nu=2$ (see Fig.~\ref{Flat_band_SC_EXP}\textbf{a}) with a transition temperature $T_{\textrm{c}}$ around $2$ K.  First, the authors aimed to demonstrate the normal state bands that form superconductivity are indeed flat bands. They probed the normal state by applying a magnetic field $B=0.2$ T while keeping the temperature at $T=0.3$ K (superconductivity is destroyed by $B$ field). Specifically, they measure the differential resistance $dV/dI$ as a function of the applied current density $J$, shown Fig.~\ref{Flat_band_SC_EXP}\textbf{b}. For a normal metal, the differential resistance $dV/dI$ is expected to be a constant, independent of the applied $J$. By contrast, their data in Fig.~\ref{Flat_band_SC_EXP}\textbf{b} shows a highly nonlinear dependence, where the $dV/dI$ is roughly constant at small $J$ but increases abruptly at a critical current density $J_\textrm{CN}$. Such a highly nonlinear dependence arises from the so-called ``Schwinger mechanism''. Essentially, flowing a current in the normal metal state means tilting the chemical potential. In a regular metal, this tilting is a small perturbation with respect to the original Fermi surface. However, because TBG’s flat bands and tiny Fermi surface, the Fermi surface can be entirely tilted over toward one side of the Dirac point (see inset of Fig.~\ref{Flat_band_SC_EXP}\textbf{b}), which corresponds to the critical current density $J_\textrm{CN}$. Therefore, from $J_\textrm{CN}$, one can derive the band velocity by $v=\frac{J_\textrm{CN}}{ne}$. As shown in Fig.~\ref{Flat_band_SC_EXP}\textbf{c}, the velocity of the TBG is only on the order of $10^3$ m/s, which is about a 1000 times lower than that of the monolayer graphene. As a side note, this nonlinear differential resistance due to the Schwinger mechanism has already been observed in graphene moir\'e superlattice in Ref.~\cite{berdyugin2022out}. Nevertheless, it is important for the authors of Ref.~\cite{tian2023evidence} to repeat such measurements because they will compare with the superconducting state next. Second, the authors performed similar differential resistance $dV/dI$ measurements in the superconducting state by removing the magnetic field $B=0$, based on which they show that the superconductivity is also derived from the flat band. The data in Fig.~\ref{Flat_band_SC_EXP}\textbf{d} shows a superconducting dome, which is the dark blue regime. Interestingly, the superconducting critical current density $ J_\textrm{CS}$ coincides with the normal state critical current $J_\textrm{CN}$ for the carrier density range of $0<n<3\times10^{11}$ cm$^{-2}$. In fact, for superconductivity arising from a regular metal, BCS theory predicts a much larger superconducting critical current as shown by the black dashed line in Fig.~\ref{Flat_band_SC_EXP}\textbf{e}, which is dictated by the superconducting gap given by $J_\textrm{CS}^{\textrm{BCS}}=n_se\frac{\Delta}{\hbar k_{\textrm{F}}}= n_se\frac{\alpha k_\textrm{B} T_\textrm{c}}{\hbar k_{\textrm{F}}}$. This is clearly inconsistent with the data. Rather, the coincidence of the critical current data of the normal and superconducting suggests that $J_\textrm{CS}=J_\textrm{CN}=nev$; in other words, the supercurrent, just like the normal current, is limited by the velocity, i.e., the flat band. Third, the authors aim to derive the superfluid stiffness $D_s$ from the measured superconducting critical current $J_\textrm{CS}$, because previous theory has calculated the normal contribution and quantum metric contribution to the superfluid stiffness, therefore allowing for a comparison between data and calculations. Specifically, the superfluid stiffness is given by $D_s=(2\pi J_\textrm{CS} \xi)/\Phi_0$. Figure~\ref{Flat_band_SC_EXP}\textbf{f} shows the obtained $D_s$ data and its comparison with the theoretical contributions from the conventional band dispersion and from the quantum metric. The conventional contribution is too small to account for the data. By contrast, a reasonable agreement is found between the data and the quantum metric contribution, therefore providing evidence for quantum-metric-induced superconductivity.

\begin{figure*}
\centering
\includegraphics[width=17.2cm]{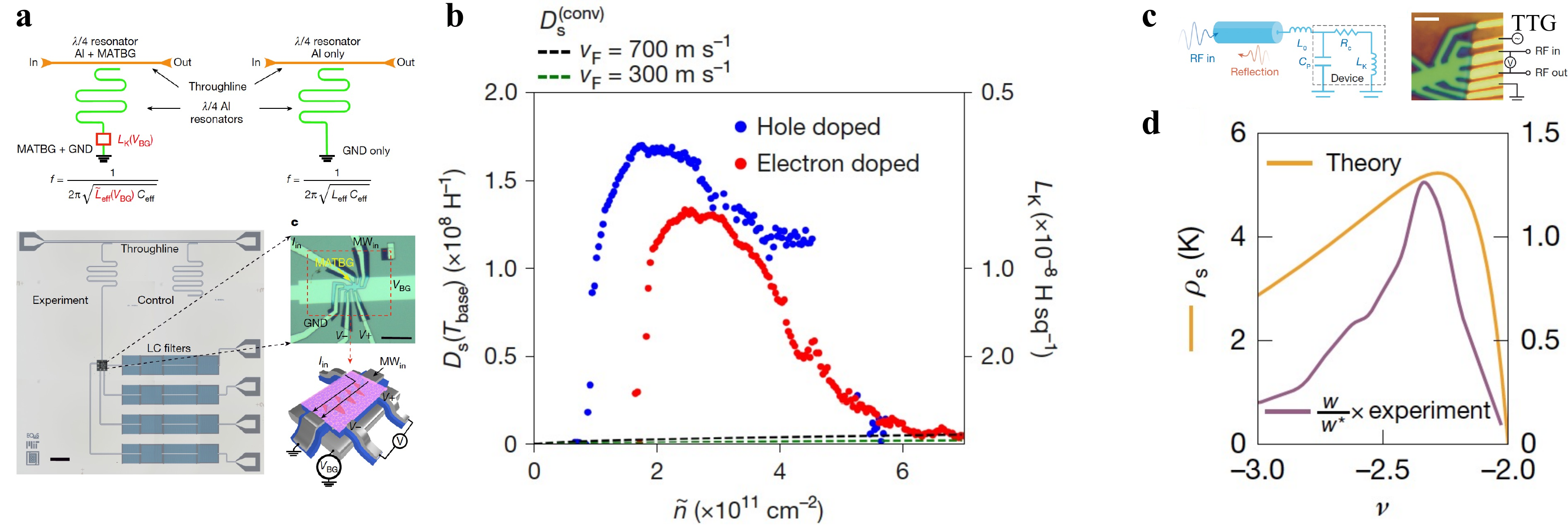}
\caption{\textbf{Microwave measurements for the quantum metric induced flat band superconductivity in graphene moir\'e samples.} (a) Schematic illustration of the superconducting resonator device used to measure the microwave transmission of the TBG. The kinetic inductance $L_{\textrm{K}}$ can be derived from the measurements and the superfluid stiffness is the inverse of the kinetic inductance, $D_s=1/ L_{\textrm{K}}$. (b) Experimentally measured superfluid stiffness $D_s$ and its comparison with the theoretical conventional contribution. From Ref. \cite{tanaka2025superfluid}. (c) Schematic of the microwave circuit used to measure the microwave transmission of the twisted trilayer graphene (TTG). (d) Experimentally measured superfluid density $\rho_s$, which is proportional to the $D_s$. From Ref. \cite{banerjee2025superfluid}.}\label{Flat_band_SC_RF}
\end{figure*}

More recently, using RF resonators, Refs. \cite{tanaka2025superfluid, banerjee2025superfluid} have directly measured the superfluid stiffness $D_s$ in the graphene moiré superconductivity. In particular, one can measure the microwave transmission coefficient through the graphene superconductor, which reveals the kinetic inductance of the superconductivity $L_{\textrm{K}}$. The superfluid stiffness is the inverse of the kinetic inductance, $D_s=1/ L_{\textrm{K}}$. Ref. \cite{tanaka2025superfluid} performed such microwave experiments on the magic-angle TBG (Fig. ~\ref{Flat_band_SC_RF}\textbf{a}). The measured superfluid stiffness $D_s$ (Fig. ~\ref{Flat_band_SC_RF}\textbf{b}), which is consistent with that derived from DC transport measurements \cite{tian2023evidence}, is much higher than the conventional contribution. Therefore, the authors suggest that the quantum metric contribution is significant. On the other hand, Ref. \cite{banerjee2025superfluid} studied the twisted trilayer graphene (TTG) and measured a similar superfluid stiffness $D_s$ (Fig. ~\ref{Flat_band_SC_RF}\textbf{c,d}).

\vspace{3mm}
\color{blue} \underline{Beyond Superconductivity:} \color{black}  Quantum metric can contribute to other Fermi surface instabilities \cite{PhysRevB.102.165118, PhysRevB.105.L140506, ying2024flat, hofmann2023superconductivity, chen2025effect} apart from superconductivity in flat band systems. Specifically, theory has considered the quantum metric induced charge density waves \cite{hofmann2023superconductivity}, ferromagnetic order \cite{PhysRevB.102.165118}, exciton condensation \cite{PhysRevB.105.L140506, ying2024flat}, etc. These exciting possibilities await experimental exploration.

\vspace{5mm}
\subsection{The Intrinsic Nonlinear Hall Effect}

The studies of the Hall effect have led to important breakthroughs, including the discoveries of Berry curvature and topological Chern invariants. Historically, the Hall effect has been studied in the linear regime, which means the generation of a transverse current that is linearly proportional to the applied electric field.  However, in principle, the Hall effect is not limited to the linear order. Indeed, recently, the second-order nonlinear Hall effects have been discovered, attracting great interest. In this case, the Hall current is quadratic to the applied electric field. Interestingly, such a nonlinear Hall effect is allowed in a wide range of quantum materials which do not support the linear Hall effect. As such, since its discovery, the nonlinear Hall effects have quickly emerged as a powerful probe of the quantum geometry of the conduction electrons in various classes of quantum materials. At the same time, their nonlinear nature suggests novel applications such as rectification and frequency mixing. The first kind of nonlinear Hall effect that has been experimentally studied is the Berry curvature dipole induced nonlinear Hall effect, which has been discussed above in Section IV.B. $D^{\textrm{BC}}$ is the Berry curvature dipole, given by  $D^{\textrm{BC}}=\int_k \frac{\partial \varepsilon}{\partial k_x} \Omega$. This Berry curvature dipole nonlinear Hall effect requires the breaking of space-inversion symmetry but can respect time-reversal symmetry. Since the theoretical prediction in 2015 \cite{sodemann2015quantum}, the Berry curvature dipole nonlinear Hall effect was first observed in the 2D WTe$_2$ systems \cite{Ma:2019Nature, Kang2019nonlinear} and then explored in a wide range of noncentrosymmetric systems \cite{kumar2021room, Lai2021third,he2022graphene, hu2022nonlinear}. In this section, we focus on the intrinsic nonlinear Hall effect that arises from the quantum metric dipole. 

\vspace{3mm}

\begin{figure*}
\centering
\includegraphics[width=15cm]{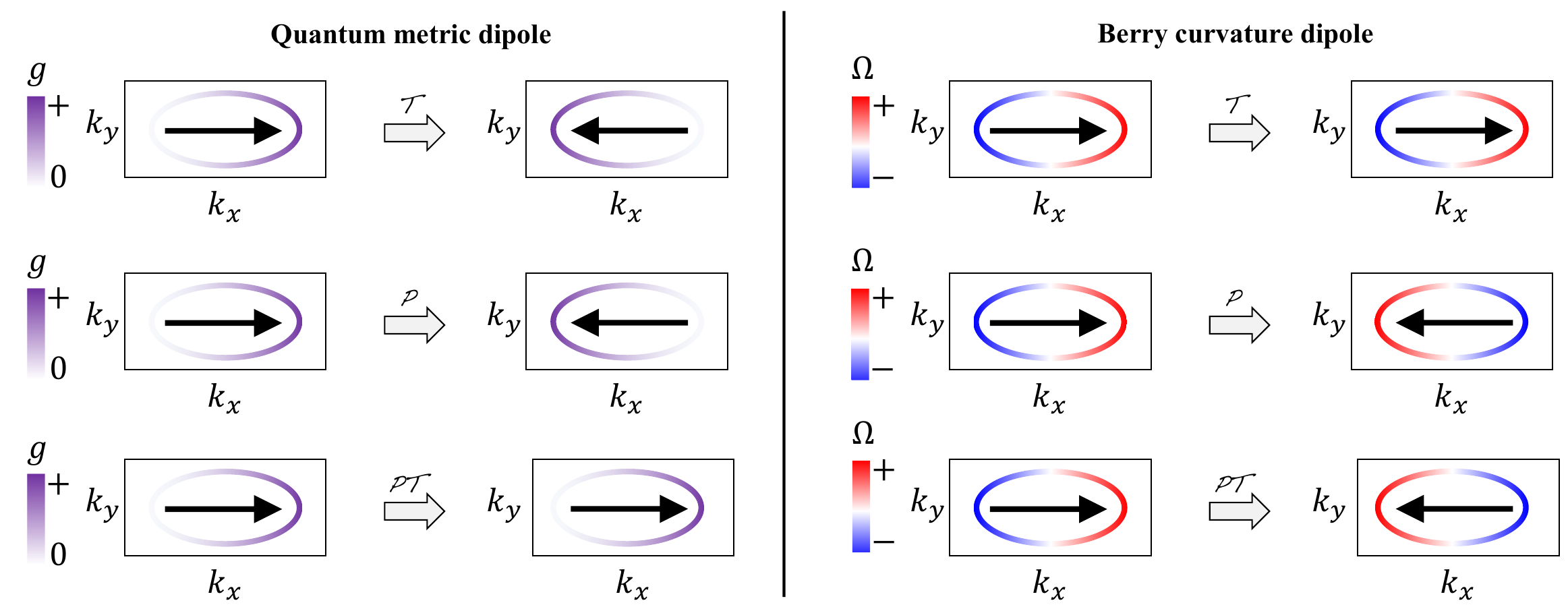}
\caption{\textbf{Symmetry properties of the quantum metric dipole and Berry curvature dipole.} $\mathcal{T}$, $\mathcal{P}$ and $\mathcal{PT}$ are the time-reversal symmetry, space-inversion symmetry and space-time symmetry, respectively.}\label{QMD_BCD}
\end{figure*}

\color{blue} \underline{Symmetry Requirement and Physical Picture:} \color{black} We start by presenting the symmetry requirement and physical picture of the quantum metric nonlinear Hall effect. This effect can be written as 

\begin{equation}
J_j=\sigma_{yxx}E_x^2,
\end{equation}
where the nonlinear Hall conductivity is proportional to the quantum metric dipole (QMD), $\sigma_{yxx}\propto D^{\textrm{QM}}$. The quantum metric dipole is given by 

\begin{equation}
D^{\textrm{QM}}=\int_\mathbf{k} (v_yg_{xx}-v_xg_{yx}) \delta(\varepsilon-\varepsilon_{\mathrm{F}}) 
\end{equation}

Phenomenologically, the quantum metric dipole $D^{\textrm{QM}}$ can be visualized as an uneven distribution of the quantum metric $g$ around the Fermi surface. As shown in Fig.~\ref{QMD_BCD}, the $+k_x$ side of the Fermi surface has a larger quantum metric than the $-k_x$ side (note that the quantum metric, as a distance, is always positive). The quantum metric dipole is odd under time-reversal symmetry $\mathcal{T}$ and space-inversion symmetry $\mathcal{P}$ but it is even under space-time $\mathcal{PT}$. As such, the intrinsic nonlinear Hall effect driven by the quantum metric dipole requires breaking both $\mathcal{T}$ and $\mathcal{P}$, while preserving their combination, $\mathcal{PT}$. Such a symmetry requirement can be realized in $\mathcal{PT}$-symmetric antiferromagnetic materials. As a comparison, the Berry curvature dipole is even under $\mathcal{T}$ but odd under space-inversion symmetry $\mathcal{P}$, which corresponds to noncentrosymmetric materials.

Another crucial but frequently overlooked aspect is that the Hall effect should be intrinsically \textit{antisymmetric}. In the linear order, this means $\sigma_{xy}=-\sigma_{yx}$. In the second order, this means $\sigma_{yxx}=-\sigma_{xyx}$. For the quantum metric dipole, one can easily confirm this because the kernel of the integral, $(v_yg_{xx}-v_xg_{yx})$, is antisymmetric. For the Berry curvature dipole, one can also confirm this because the kernel is $\Omega_{yx}v_x$ and the Berry curvature $\Omega_{yx}$ is inherently antisymmetric.

\begin{figure*}
\centering
\includegraphics[width=15cm]{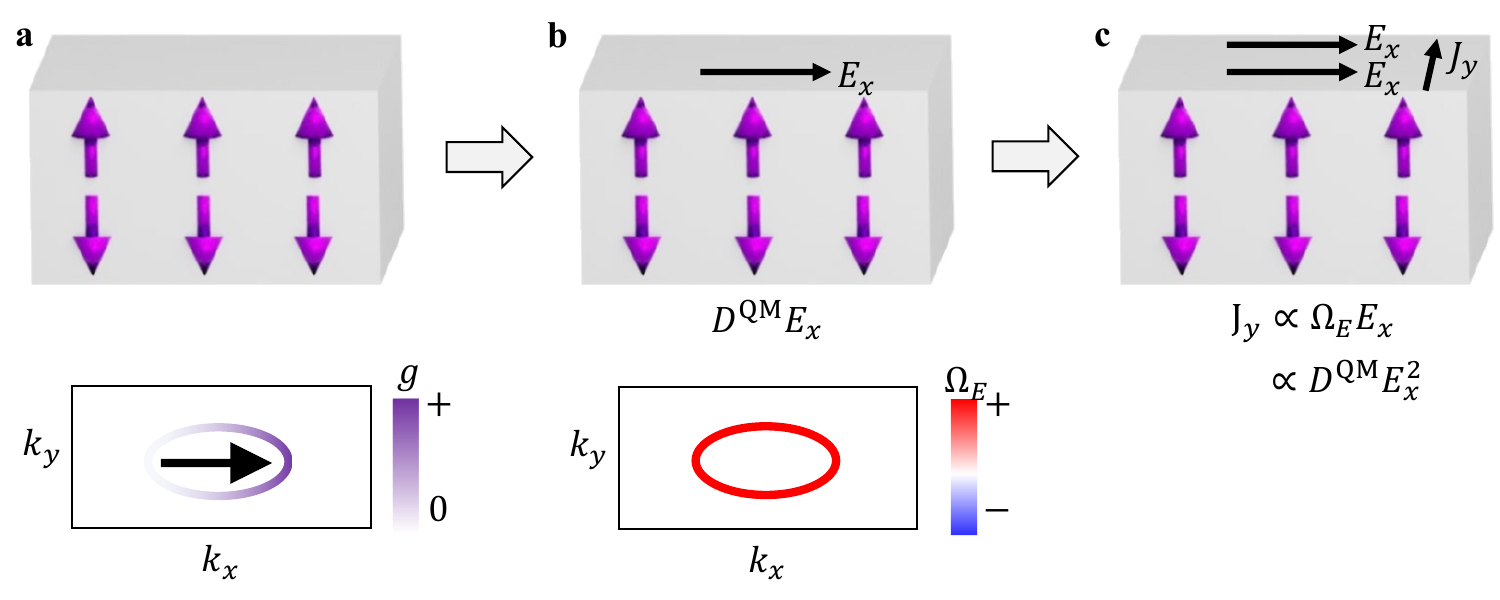}
\caption{\textbf{Physical picture for the quantum metric dipole induced nonlinear Hall effect.} We start from a $\mathcal{PT}$ symmetric antiferromagnetic metal. Because of the $\mathcal{PT}$ symmetry, the system has zero Berry curvature everywhere in $k$ space and no AHE conductivity. However, it does support a nonzero quantum metric dipole $D^{\textrm{QM}}$. The application of an electric field generates a nonzero Berry curvature that is proportional to the quantum metric dipole, $\Omega_E\propto D^{\textrm{QM}}E$. Hence, the electric-field-driven system has nonzero Hall conductivity. As a result, the second application of electric field leads to the Hall current.}\label{NLH_physical}
\end{figure*}

We proceed to explain the physical picture, as shown in Fig.~\ref{NLH_physical}. We start from a $\mathcal{PT}$ symmetric antiferromagnetic metal. Because of the $\mathcal{PT}$ symmetry, the system has zero Berry curvature everywhere in $k$ space and no AHE conductivity. However, it does support a nonzero quantum metric dipole $D^{\textrm{QM}}$. The application of an electric field generates a nonzero Berry curvature that is proportional to the quantum metric dipole, $\Omega_E\propto D^{\textrm{QM}}E$. Hence, the electric-field-driven system has a nonzero Hall conductivity. As a result, the second application of the electric field leads to the Hall current.

\vspace{3mm}
\begin{figure*}
\centering
\includegraphics[width=15cm]{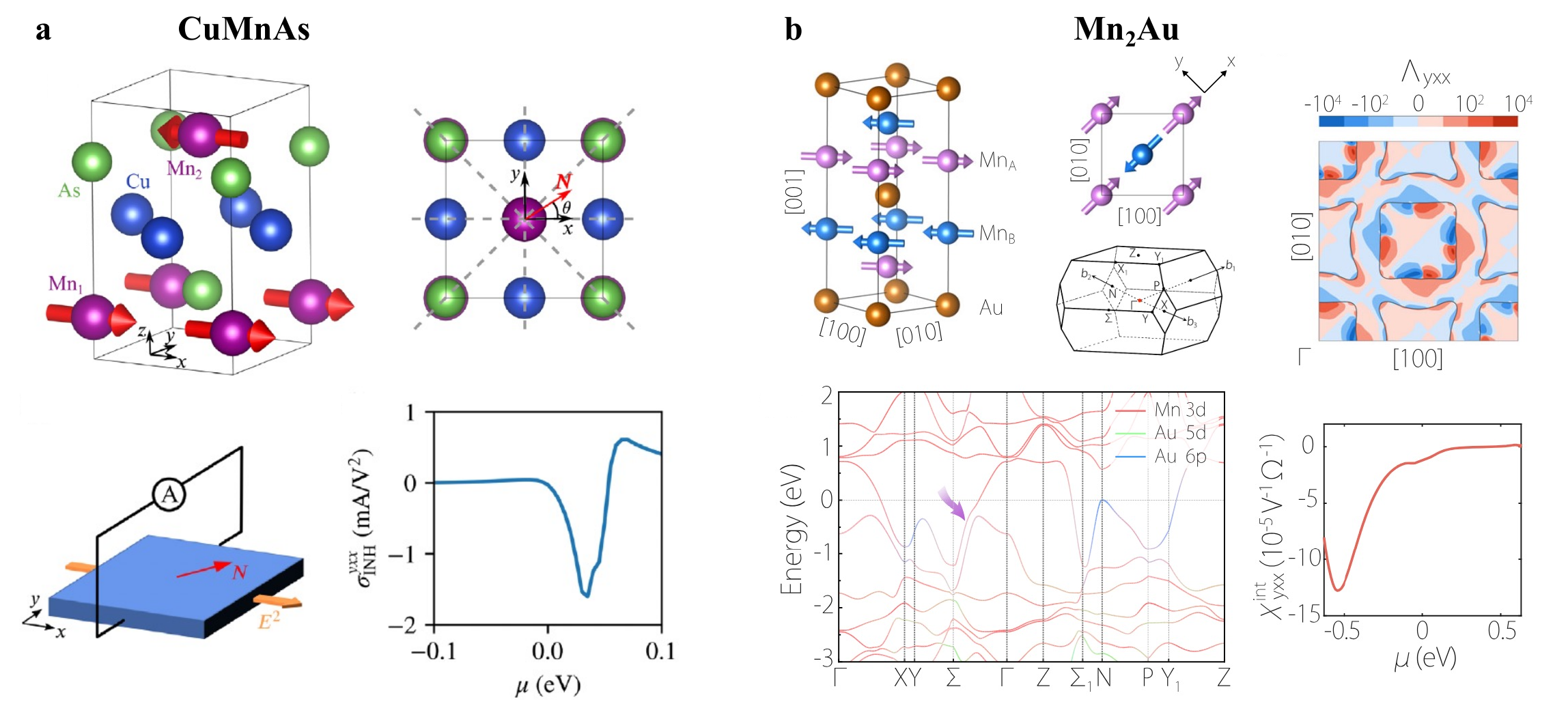}
\caption{\textbf{Theoretical prediction of the quantum metric dipole induced nonlinear Hall effect in CuMnAs and Mn$_2$As.} (a) Prediction and calculation of the nonlinear Hall effect in the $\mathcal{PT}$-symmetric antiferromagnetic metal CuMnAs. From Ref.~\cite{Wang2021Intrinsic}. (b) Prediction and calculation of the nonlinear Hall effect in the $\mathcal{PT}$-symmetric antiferromagnetic metal Mn$_2$As. From Ref.~\cite{liu2021intrinsic}.}\label{NLH_THY}
\end{figure*}

\color{blue} \underline{Theoretical Prediction:} \color{black} In 2014, Ref. \cite{PhysRevLett.112.166601} proposed the theory of the intrinsic nonlinear Hall effect. However, at the time, material platform remained unknown. In 2021, Refs \cite{Wang2021Intrinsic,liu2021intrinsic} proposed that this effect can be realized in CuMnAs and Mn$_2$Au and performed DFT calculations, as shown in Fig.~\ref{NLH_THY}. While the actual experimental observation occurred in a different material, these theoretical works clearly pointed out the $\mathcal{PT}$-symmetric antiferromagnets as the ideal platform. Later, the intrinsic nonlinear Hall effect was theoretically analyzed in Refs. \cite{lahiri2022intrinsic, Smith2022momentum,kaplan2022unification,huang2023nonlinear,huang2023Scaling,atencia2023disorder,kaplan2023general,gong2024nonlinear,wang2023intrinsic,mazzola2023discovery,zhuang2024intrinsic,wang2024intrinsic,xiong2024anomalous, zhang2025logarithmically}.

\vspace{3mm}

\begin{figure*}
\centering
\includegraphics[width=17.2cm]{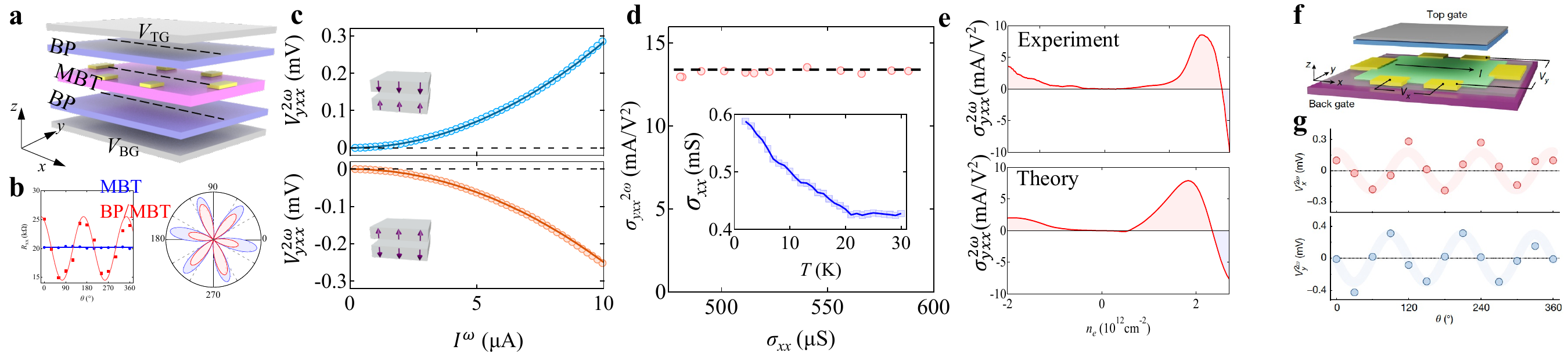}
\caption{\textbf{Experimental observation of the quantum metric dipole induced nonlinear Hall effect.} (a) Schematic illustration of 2L BP/6SL MnBi$_2$Te$_4$/2L BP device. (b) Angle-resolved resistance and optical second-harmonic generation (SHG) measurements of a 6SL MnBi$_2$Te$_4$ before and after interfaced with BP. (c) Measured nonlinear Hall voltage as a function of incident current for the two opposite AFM states. (d) The scaling between the nonlinear Hall conductivity and the Drude conductivity $\sigma_{xx}$. (e) (\textbf{A}) Experimentally measured and theoretically calculated nonlinear Hall conductivity $\sigma^{2\omega}_{yxx}$ as a function of carrier density $n$ for the BP/6SL MnBi$_2$Te$_4$/BP band structure. From Ref.~\cite{gao2023quantum}. (f) Device schematic. (g) Measured nonlinear conductivity as a function of the direction of the current inside the plane. From Ref.~\cite{wang2023quantum}.}\label{NLH_EXY}
\end{figure*}

\begin{figure*}
\centering
\includegraphics[width=17.2cm]{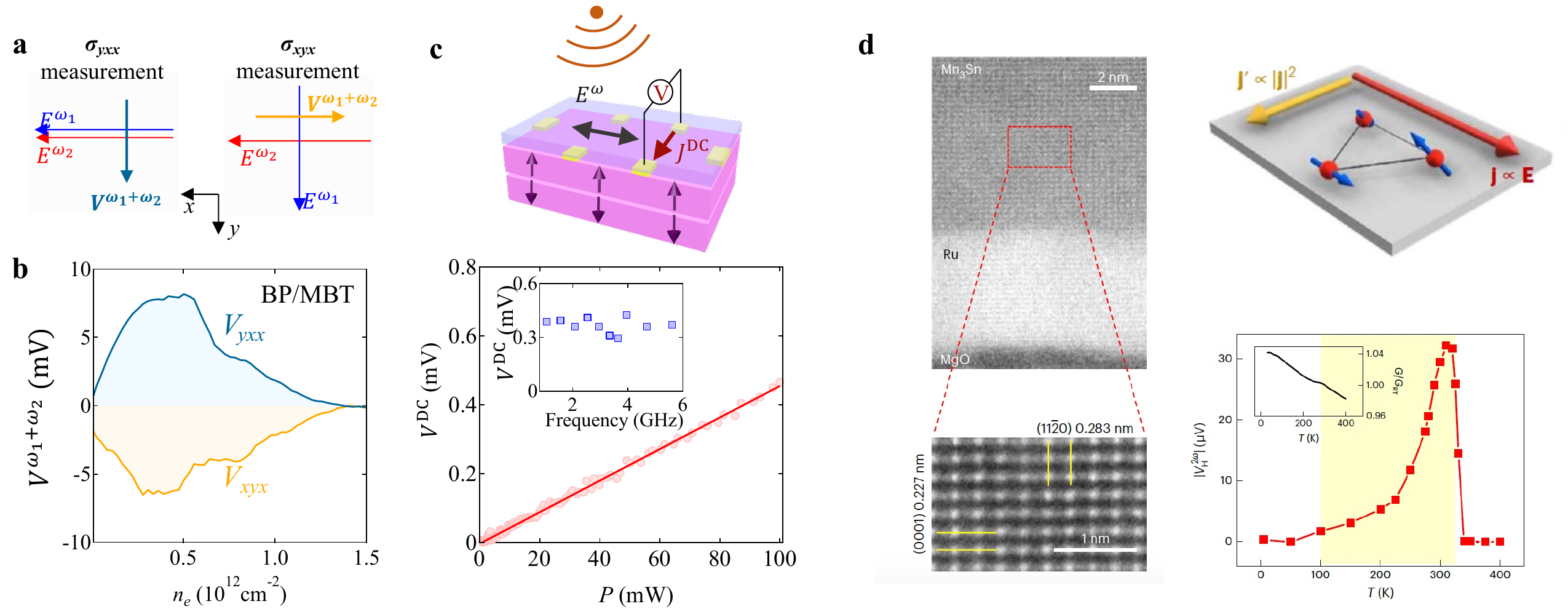}
\caption{\textbf{Systematic studies on the quantum metric dipole induced nonlinear Hall effect.} (a) The schematic of $\sigma _{yxx}$ and $\sigma _{xyx}$ measurements. (b) The experiment results of the $\sigma _{yxx}$ and $\sigma _{xyx}$ as a function of carrier density $n_e$ on a BP/MnBi$_2$Te$_4$ device. (c) Schematic illustration of microwave rectification and measured microwave rectification based on the intrinsic nonlinear Hall effect. Inset is the DC signal $V^{\textrm{DC}}$ as a function of microwave frequency. From Ref.~\cite{gao2023quantum}. (d) Cross-section TEM data of the Mn$_3$Sn on Ru. Schematic illustration of the nonlinear Hall effect in the noncollinear antiferromagnetic state of Mn$_3$Sn. Measured nonlinear Hall voltage as a function of temperature. From Ref.~\cite{han2024room}.}\label{NLH_EXY_2}
\end{figure*}

\color{blue} \underline{Experimental Observations:} \color{black}  In 2023, Refs.~\cite{gao2023quantum, wang2023quantum} reported the experimental observation of the quantum metric dipole induced intrinsic nonlinear Hall effect. The material platform in Ref.~\cite{gao2023quantum} is the heterostructure between six-layer MnBi$_2$Te$_4$ and black phosphorus (BP), as shown in Fig.~\ref{NLH_EXY}\textbf{a}. This is an ideal platform for the following reasons: (1) Even-layer MnBi$_2$Te$_4$ is a $\mathcal{PT}$-symmetric antiferromagnet. It has nontrivial topological electronic structure, suggesting large quantum metric in its low-energy bands. (2) The system is a 2D vdW device. The gate tunability means that one can systematically study the nonlinear Hall effect as a function of both chemical potential and displacement electric field. Moreover, the atomically thin thickness enables the application of a large in-plane electric field, which is highly favorable for the studies of nonlinear effect. (3) The interface with BP allows one to break the MnBi$_2$Te$_4$’s three-fold rotational symmetry $C_{3z}$, because $C_{3z}$ enforces the nonlinear conductivity to be totally symmetric $\sigma_{yxx}=\sigma_{xyx}$, therefore excluding any Hall effect. First, the authors of Ref.~\cite{gao2023quantum} experimentally checked the $C_{3z}$ breaking by the BP interfacing via both transport and optics. For pure MnBi$_2$Te$_4$ without BP, the resistance is isotropic (Fig.~\ref{NLH_EXY}b), consistent with the $C_{3z}$ symmetry. By contrast, the MnBi$_2$Te$_4$/BP heterostructure’s resistance shows clear angular anisotropy with a $180^{\circ}$ periodicity, providing a clear signature of the $C_{3z}$ breaking. First, the authors then performed nonlinear transport by passing a current at frequency $\omega$ ($I^{\omega}$) and use the lock-in technique to detect nonlinear Hall voltage at second-harmonic $V^{2\omega}$. As shown in Fig.~\ref{NLH_EXY}c, they observed clear nonlinear Hall effect. Moreover, the sign of the nonlinear Hall effect flips sign as one flips the sign of the AFM order. This proves that the measured nonlinear Hall signal is time-reversal odd. This excludes the possibility of Berry curvature dipole (time-reversal even) and is consistent with the quantum metric dipole (time-reversal odd). Moreover, the authors have shown that the nonlinear Hall effect vanishes as one increases the temperature above Neel temperature $T_\textrm{N}$. Third, the authors proved that the nonlinear Hall effect is dissipationless, which means that the intrinsic nonlinear Hall conductivity is independent of the scattering time. This was achieved by the scaling method, similar to what has been done in the linear AHE studies \cite{nagaosa2010anomalous}. In particular, they measured temperature dependence of both the nonlinear Hall conductivity $\sigma^{2\omega}_{yxx}$ and the linear Drude conductivity $\sigma_{xx}$. Then they could plot $\sigma^{2\omega}_{yxx}$ as a function of $\sigma_{xx}$ (the Drude conductivity is proportional to the relaxation time $\tau$). As shown in Fig.~\ref{NLH_EXY}d, the result shows that $\sigma^{2\omega}_{yxx}$ is independent of $\sigma_{xx}$, therefore proving the non-dissipative (intrinsic) nature of the nonlinear Hall effect. Fourth, the authors managed to demonstrate the \textit{antisymmetric} nature of the observed nonlinear Hall signal by inventing an electrical sum-frequency generation method (SFG) (Fig.~\ref{NLH_EXY_2}a). This SFG method clearly defines the path direction of each current by separating different currents using different frequencies, therefore allowing them to measure $\sigma_{xyx}$ and showing that  $\sigma_{xyx}\simeq-\sigma_{yxx}$, i.e., the majority of their observed signal is antisymmetric (Fig.~\ref{NLH_EXY_2}b). Fifth, the authors further studied the Fermi level (carrier density) dependence of the nonlinear Hall signal, which they found to be in good agreement with the quantum metric dipole as a function of Fermi level obtained from DFT calculations. Finally, using the observed nonlinear Hall effect, they further demonstrated harvesting of wireless signal, i.e., turning the RF radiation into DC electricity. In Ref.~\cite{wang2023quantum}, the authors performed nonlinear transport measurements on bare MnBi$_2$Te$_4$. The observed nonlinear transport signal in both transverse and longitudinal directions that respect $C_{3z}$ symmetry.

After the initial observation in 2D MnBi$_2$Te$_4$ systems, researchers explored the intrinsic nonlinear Hall effect in a wide range of other systems. For example, in Ref.~\cite{han2024room}, the authors observed the quantum metric dipole nonlinear Hall effect in the noncollinear antiferromagnet Mn$_3$Sn film interfaced with Ru. They showed that the quantum metric effect persists up to room temperature (Fig.~\ref{NLH_EXY_2}d), thanks to the high Neel temperature of Mn$_3$Sn.

\vspace{3mm}

\begin{figure*}
\centering
\includegraphics[width=17.2cm]{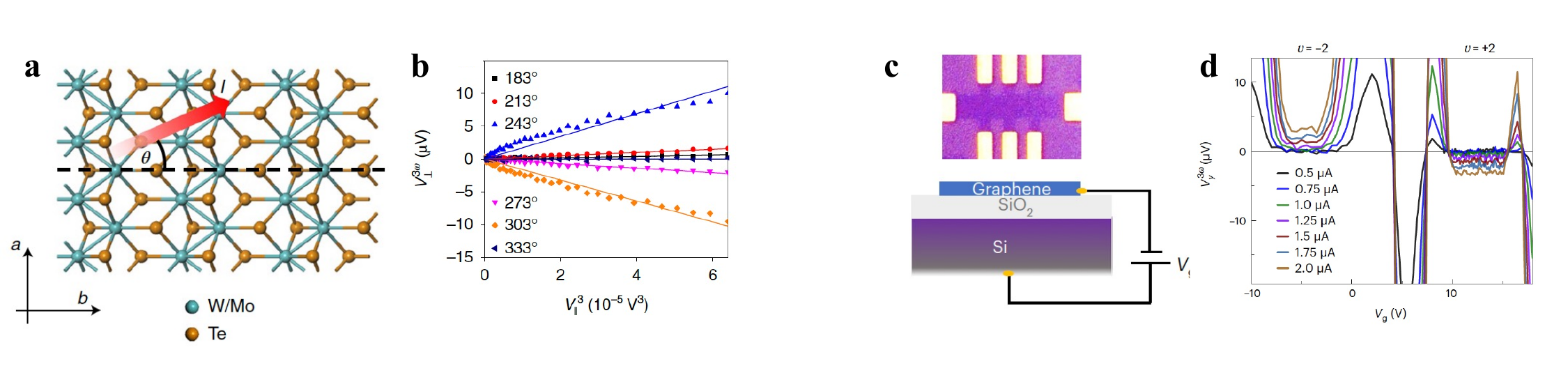}
\caption{\textbf{Third order Hall effects.} (a,b) Third order Hall effect observed in 1T$^\prime$ MoTe$_2$. From Ref. \cite{Lai2021third}. (c,d) Third order Hall effect observed in the quantum Hall states of graphene. From Ref. \cite{he2024third}. }\label{Third_order}
\end{figure*}
\color{blue} \underline{Third-Order Hall Effect:} \color{black} Beyond second-order, the quantum metric can lead to a third-order Hall effect. This effect does not require one to break the time-reversal or space-inversion symmetry. Therefore, it can be realized in nonmagnetic centrosymmetric materials. The process can be understood as follows. Because of the space-inversion symmetry, the system has no Berry curvature dipole. However, the application of an electric field can generate a nonzero Berry curvature dipole that is proportional to the quantum metric quadrupole of the original system. The further consideration of the electric field another two times leads to a nonzero Hall current, i.e., a third-order Hall effect. This third-order Hall effect was first observed in 1T$^\prime$ MoTe$_2$ crystals \cite{Lai2021third}. Later, it has been studied in a wide range of materials \cite{wang2022room, ye2022orbital, nag2023third, ye2023control, zhao2023gate, he2024third, wang2024orbital, liu2025giant}. One interesting recent experiment is the observation of the third-order Hall effect in the quantum Hall state in bilayer graphene. All previously detected nonlinear Hall effects depend on the Fermi surface. By contrast, the quantum Hall state has a full energy gap. The authors proposed some models based on the interactions between edge states. This is a physically reasonable approach. However, how the quantum geometry (such as the Berry curvature and the quantum metric) affects the nonlinear Hall effect in the quantum Hall state remains an open question.

\vspace{3mm}

\color{blue} \underline{Other Nonlinear and Nonreciprocal Effects:} \color{black}  In addition to the nonlinear Hall effect, the quantum metric is proposed to give rise to novel electromagnetic responses. In the low-frequency regime, the quantum metric can contribute to nonlinear magnetoresistance \cite{sala2024quantum}. In the optical regime, the quantum metric dipole is closely related to the nonreciprocal directional dichroism \cite{Malashevich:2010Band, PhysRevLett.122.227402,pozo2023multipole}, which means that the transmission of light is different along forward and backward directions. The quantum metric dipole is also proportional to a novel kind of intrinsic photocurrent, i.e., the magnetic injection current \cite{Ahn2020, PhysRevResearch.2.033100, Watanabe2021chiral}. In the plasmonic regime, the quantum metric dipole can lead to nonreciprocal bulk plasmon propagations in magnetic metals \cite{arora2022quantum}. The quantum metric may also be related to the capacitance of Chern insulators \cite{komissarov2024quantum}, the magnetoelectric coupling of antiferromagnetic metals as well as various spin transport and spin Hall phenomena \cite{xiao2022intrinsic, zheng2024interlayer, feng2024quantum, wang2025intrinsic}.

\vspace{5mm}
\subsection{Momentum-Resolved Measurement of Quantum Geometry}

Because of the many novel responses enabled by quantum geometry, it would be of great interest to directly measure the quantum metric and Berry curvature in the band structure. Recently, Ref.~\cite{kang2024measurements} proposed a framework to measure the quantum geometrical tensor (QGT) in crystalline solids using polarization-, spin- and angle-resolved photoemission spectroscopy. In particular, the authors introduced the quasi-QGT, whose real and imaginary parts are denoted as $qg^{n}_{ij}(k)$ and $q\Omega^{n}_{ij}(k)$. In two-band systems, the quasi-QGT is directly proportional to the QGT. In a multiband system, it is an excellent approximation of the QGT. In particular, $qg^{n}_{ij}(k)$ and $q\Omega^{n}_{ij}(k)$ are defined as follows:

\begin{align}
qg^{n}_{ij}(k) &= \sum_{m\neq n} \textrm{Re} \left[ \frac{ [\langle u_n\arrowvert i\partial_{k_i} H(\mathbf{k}) \arrowvert u_m\rangle \langle u_m\arrowvert i\partial_{k_j} H(\mathbf{k}) \arrowvert u_n\rangle  ] }{\varepsilon_m-\varepsilon_n } \right]\\
q\Omega^{n}_{ij}(k) &= \sum_{m\neq n} \textrm{Im} \left[ \frac{ [\langle u_n\arrowvert i\partial_{k_i} H(\mathbf{k}) \arrowvert u_m\rangle \langle u_m\arrowvert i\partial_{k_j} H(\mathbf{k}) \arrowvert u_n\rangle  ] }{\varepsilon_m-\varepsilon_n } \right]
\end{align}

\begin{figure*}
\centering
\includegraphics[width=17.2cm]{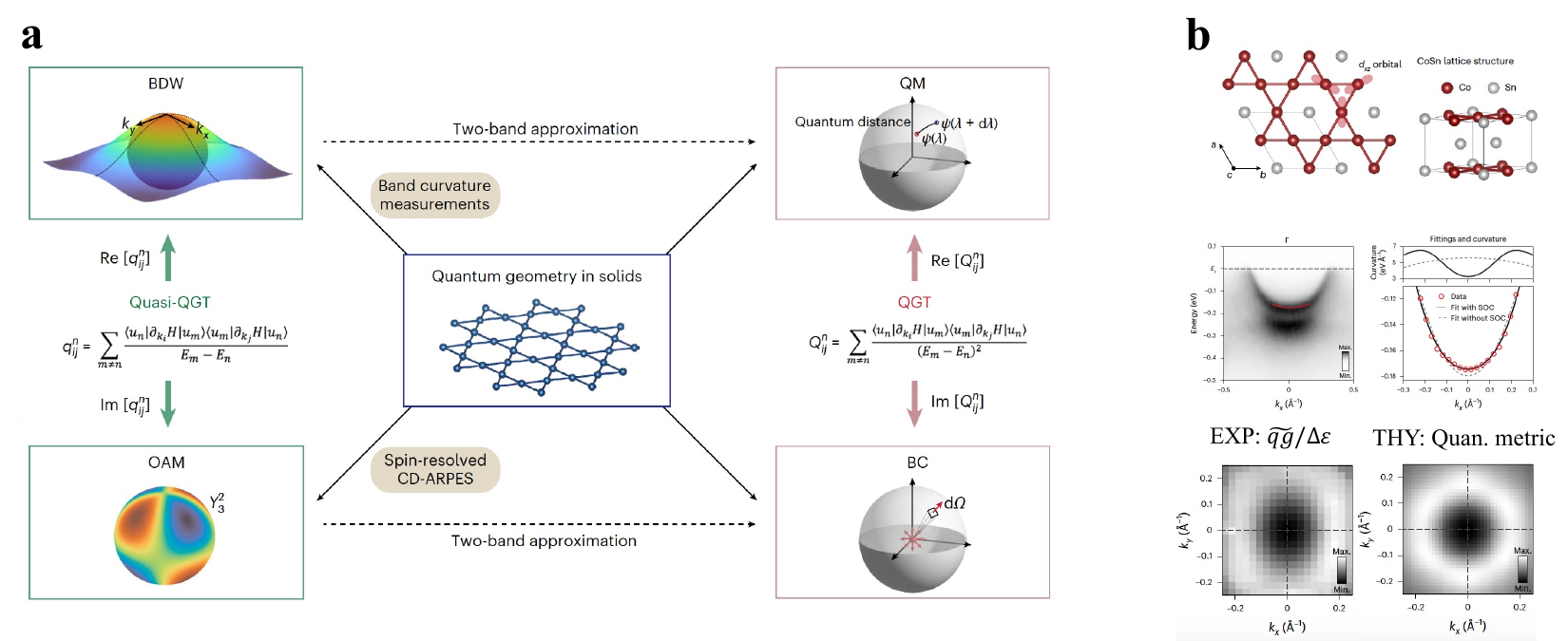}
\caption{\textbf{Momentum-resolved measurement of the quantum geometrical tensor by ARPES.} (a) The quantum geometrical tensor (QGT) of a solid can be measured by introducing the quasi-QGT. The real and imaginary parts of the quasi-QGT are related to the band Drude weight and the orbital angular momentum, which can be measured by ARPES. (b) Band structure of CoSn measured by ARPES allows the extraction of the trace of the real part of the quasi-QGT, $\tilde{qg}$. The ration between the $\tilde{qg}$ and the band gap $\delta \varepsilon$ gives the quantum metric of the band, which can be compared with the theoretical calculations. 
From Ref.~\cite{kang2024measurements}}\label{ARPES_QM}
\end{figure*}

Importantly, this quasi-QGT can be deduced from the ARPES data under some approximations. In particular, the trace of the real part of the quasi-QGT, $qg^{n}_{xx}(k)+ qg^{n}_{yy}(k)= \tilde{qg}$, can be deduced from the ARPES measured band $\varepsilon-k$ dispersion, when the nearest-neighbour hopping is dominant and also when the difference between nearest-neighbour and next-nearest-neighbour distances are not significant.

\begin{align}
\tilde{qg} &= \frac{1}{2} \left( -a^2 (E_n- E_0)- \left(\partial_{k_x}^2+\partial_{k_y}^2 \right) E_n \right)
\label{trace_g}
\end{align}

Using this framework, the authors measured the quasi-QGT in the kagome metal CoSn. As shown in Fig.~\ref{ARPES_QM}, by the measured band structure of CoSn and following Eq.~\ref{trace_g}, they obtained the trace of the quasi-QGT $\tilde{qg}$. The quantum metric then equals to the ratio between the $\tilde{qg}$ and the band gap $\Delta \varepsilon$ under the two-band approximation. Figure~\ref{ARPES_QM} shows the $\tilde{qg}/\Delta\varepsilon$ deduced from data and its comparison with the theoretically calculated quantum metric, which shows a good agreement. Furthermore, the imaginary part of the quasi-QGT can be estimated by circular dichroism (CD)-ARPES experiments, in which one measures the difference in photoemitted electron intensity using right-handed and left-handed circularly polarized light.

\section{Outlook}
\label{sec:sec3}

In this Review, starting from the intuitive pictures of quantum geometry and the formal definition of the quantum geometric tensor (QGT), through theoretical framework, to experimental observations that reveal its physical consequences, we have highlighted how local quantum geometry offers a unifying foundation for understanding a wide range of phenomena in condensed matter systems. Despite the much exciting progresses, the understanding of quantum geometry, especially quantum metric in condensed matter systems is still in its early stage. In the following, we discuss the promising future directions of the researches in quantum geometry.

On the theoretical side, one can generalize the parameters that characterize
the Hamiltonian. While crystal momentum $k$ is central to electronic systems in solids, quantum geometry does not have to be limited to momentum space; extending the parameter spaces to higher-dimensions opens the door to richer physical phenomena. For instance, the real space position $x$ can be added to the phase space, as explored in the wavepacket formalism discussed in section IIA, where the anomalous velocity term naturally emerges, accounting for the anomalous Hall effect. 
Beyond this, additional parameters such as atomic displacements,
strengths of the external fields, or time $t$ can be included, enabling an even broader formulation of quantum geometry and facilitating new predictions of emergent phenomena.
The latter brings geometric structures of space-time into consideration.
Already, we have discussed this in relation to the 
uncertainty relation between the energy and time 
\cite{PhysRevLett.65.1697}.

Another direction is to generalize the QGT, e.g., by
inserting other operators in the definition
of QGT. Originally, QGT measures the response of a quantum state to parameter change. By inserting other operators, one can compute how physical observables respond to parameter change.For instance, as discussed in Section IV.B, such generalizations have proven valuable to describe nonlinear optical responses 
\cite{avdoshkin2024multistategeometryshiftcurrent,
mitscherling2024gaugeinvariantprojectorcalculusquantum}.
In addition, by inserting the operator 
$(H_k-\mu)(I-2P_k)$ where $H_k$ is the Hamiltonian for
the crystal momentum $k$ and $P_k$ is the projection 
operator for the occupied states at $k$,
Resta revealed the relation of this generalized QGT to the
Drude weight and the orbital magnetization
\cite{resta2017geometricalmeaningdrudeweight}.

Another promising direction is to go beyond electrons alone and to study combined systems that couple electrons to other degrees of freedom, such as photons and magnons. In such systems, hybrid quasiparticles with quantum geometric properties in the combined Hilbert space may be engineered,  leading to novel emergent phenomena.
The geometry and topology of photons
have been studied extensively \cite{RevModPhys.91.015006},
paving the way for exploring the quantum geometry of coupled electron-photon systems. Another example is the Berry curvature in a magnon-phonon coupled system \cite{PhysRevLett.117.217205}. A notable example along this line is Floquet engineering of electron-photon 
coupled system, where the periodic photo-radiation drives the system into non-equilibrium steady states with engineered band structure and quantum geometric properties
\cite{rudner2020floquetengineershandbook,
annurev:/content/journals/10.1146/annurev-conmatphys-031218-013423,
LIU2023100705}.

As mentioned above, quantum geometry is closely related to quantum information, which is linked to the holography principle in quantum gravity.
Ryu and Takayanagi derived the formula connecting the
Black hole entropy to the entanglement entropy of the 
many-body quantum systems \cite{PhysRevLett.96.181602}. In this sense, quantum geometry in condensed matter physics is conceptually related to that in high energy physics such as string theory and 
quantum gravity. On the other hand, a key advantage of condensed matter systems is that we can study quantum information in tabletop experiments, such as through measurements of magnetic susceptibility. 
This direction will be further expanded in the future 
to connect physical properties with the information 
that the materials have. In particular, understanding quantum entanglement between a system and its environment, 
especially in non-equilibrium states, remains an important topic for future research.

Quantum geometry also represents one of the most active frontiers in condensed matter physics, offering many exciting opportunities for future experimental exploration. First, in terms of the quantum metric, while it has been observed in the nonlinear Hall effect and the flat band superconductivity, it would be highly interesting to explore additional experimental consequences of the quantum metric in novel quantum materials. For example, the quantum metric is predicted to manifest in magnetic photogalvanic current in $\mathcal{PT}$-symmetric antiferromagnets, as well as in the plasmonic and excitonic responses of time-reversal symmetry-breaking materials \cite{Ahn2020, PhysRevResearch.2.033100, Watanabe2021chiral, arora2022quantum,ying2024flat}. Moreover, the quantum metric may strongly influence the transport properties of flat-band charge orders, Wigner crystals, spin-density waves or other superfluid-like transitions, analogous to the flat-band superconductivity \cite{PhysRevB.102.165118, PhysRevB.105.L140506, ying2024flat, hofmann2023superconductivity, chen2025effect}. The quantum metric can further contribute to the quantum capacitance of Chern insulators, quantum inductance of multiferroics as well as in general the electromagnetic responses of novel quantum materials \cite{komissarov2024quantum, tanaka2025superfluid}. Finally, the spin space group, which is much larger than the magnetic space group, characterizes the magnetic phenomena without SOC. Recent theory based on spin space group has predicted magnetic geometry induced quantum geometry and nonlinear transports without SOC \cite{zhu2025magnetic}, indicating rich quantum geometrical nonlinear responses in helical magnet, altermagnet and other exotic magnetic materials.  Experimental investigation of these possibilities may significantly deepen our understanding of the quantum metric.

Second, in terms of the nonlinear response induced by quantum geometry, it would be highly valuable to systematically explore the energy harvesting and wave-mixing efficiency of the responses. For example, recent theory has predicted that the nonlinear Hall effect can enable $100\%$ energy conversion efficiency for harvesting microwave energies \cite{shi2023berry,onishi2024high}. Similarly, one may expect large efficiency in terms of sum-frequency generation, different-frequency generation as well as multi-wave mixing enabled by quantum metric and Berry curvature multipoles \cite{Morimoto-Nagaosa16, zhang2023higher, Parker19,li2024quantum,liu2025giant}. Moreover, it would be interesting to explore how quantum geometry can enhance the transduction between microwave and visible photons in the quantum limit. Proper material design and device engineering are required to experimentally measure the efficiency and to compare with theory. High efficiency wave mixing at high frequencies or at the quantum limit would lead to a wide range of new applications in terms of sensors, energy harvesters as well as quantum information engineering.

Third, experimental investigation of quantum geometry in strongly correlated systems represents a compelling direction of future research \cite{tai2023quantum,devereaux2023angle,monsen2024supercell, carmichael2025probing, kitamura2025quantum, liu2024quantum, xiao2025effects, marti2025spin, paschen2021quantum}. So far, experiments have been focused on the weakly interacting materials, where single-particle band structure is a good approximation. How quantum geometry manifests in strongly correlated systems like Mott insulators, spin liquids, and unconventional superconductors remains an open question. Both experimental studies and theoretical analyses are urgently needed to provide a proper understanding.

In summary, the study of quantum geometry has already transformed our understanding of condensed matter systems and holds tremendous potential for future discoveries. While current progress is exciting, our exploration of quantum geometry is still in its early stages. With theoretical developments continuing to expand its scope to generalized QGT and hybrid systems, and new material platform being explored from weakly interacting systems to strongly-correlated systems, as well as advanced measurement techniques being developed, the investigation of quantum geometry will not only deepen and broaden the theoretical foundations of condensed matter systems, but also lead to novel physical phenomena, paving the way for new paradigms of functional materials and quantum information science.

\section*{Acknowledgements}
A. Gao was supported by the Shanghai Jiao Tong University 2030 Initiative Grant Number WH510363003/013 and Shanghai Pujiang Program Grant Number 23PJ1404900. N. Nagaosa was supported by JSPS KAKENHI Grant Numbers 24H00197, 24H02231 and 24K00583. N. Nagaosa was supported by the RIKEN TRIP initiative. N. Ni was supported by the U.S. Department of Energy (DOE), Office of Science, Office of Basic Energy Sciences under Award Number DE-SC0021117. S.-Y. Xu was supported by the Center for the Advancement of Topological Semimetals (CATS), an Energy Frontier Research Center (EFRC) funded by the US Department of Energy (DOE) Office of Science, through the Ames National Laboratory under contract DE-AC0207CH11358, the Air Force Office of Scientific Research (AFOSR) grant FA9550-23-1-0040, the Office of Naval Research (ONR) grant the N000142512285, and the National Science Foundation (NSF; Career Grant No. DMR-2143177). S.-Y.X. acknowledges the Sloan Foundation and Corning Fund for Faculty Development.


\bibliography{refs.bib}

\end{document}